\documentclass[floatfix,%
reprint,
 amsmath,amssymb,
 aps,
]{revtex4-2}
\usepackage{array}

\usepackage{graphicx}
\usepackage{dcolumn}
\usepackage{bm}
\usepackage{eqnarray, amsmath, amsthm, amssymb, amsfonts}
\usepackage{siunitx}
\usepackage{chemformula}

\usepackage{ragged2e}
\usepackage{textcase}

\usepackage{booktabs}
\usepackage{algpseudocode}
\usepackage{algcompatible}
\usepackage{algorithmicx}

\usepackage{xcolor} 
\usepackage{tikz}
\usepackage[ruled,linesnumbered]{algorithm2e}
\usepackage{mdframed,float}
\usepackage[most]{tcolorbox}
\usepackage[caption=false]{subfig}
\usepackage[colorlinks=true,
linkcolor=blue,
citecolor=green,
urlcolor=red]{hyperref}
\usepackage{xr-hyper}
\usepackage{dcolumn, longtable}
\DeclareUnicodeCharacter{2212}{-}

\newcommand{\nocontentsline}[3]{}
\newcommand{\tocless}[2]{\bgroup\let\addcontentsline=\nocontentsline#1{#2}\egroup}

\begin{document}
\setlength{\LTpre}{0pt}
\setlength{\LTpost}{0pt}

\preprint{APS/123-QED}

\title{Flux trapping in NbTiN strips}
\author{Ruiheng Bai}
 \affiliation{
Laboratory of Atomic and Solid State Physics,
Cornell University, Ithaca, NY, USA
}
\author{Aliakbar Sepehri}
\affiliation{Department of physics, University of North Dakota, Grand Forks, ND, USA}
\author{Anne-Marie Valente-Feliciano}
\affiliation{Thomas Jefferson National Accelerator Facility, Newport News, VA, USA}
\author{Anna Herr}
\author{Quentin Herr}
\affiliation{imec USA, Kissimmee, FL, USA}
\altaffiliation[Now at ]{Snowcap Compute}
\author{Yen-Lee Loh}
\affiliation{Department of physics, University of North Dakota, Grand Forks, ND, USA}
\author{Katja C. Nowack}
\email{kcn34@cornell.edu}
 \affiliation{
Laboratory of Atomic and Solid State Physics,
Cornell University, Ithaca, NY, USA
}
\affiliation{
 Kavli Institute at Cornell for Nanoscale Science, Ithaca, New York 14853, USA
}
\date{\today}

\begin{abstract}
We use scanning superconducting quantum interference device (SQUID) microscopy to image individual vortices in superconducting strips fabricated from NbTiN thin films. 
By repeatedly field‑cooling strips of different widths in applied magnetic fields, we extract the threshold field at which the first vortex enters a strip, as well as the number and spatial configuration of vortices beyond this threshold. 
We model vortex behavior with and without considering the effect of pinning by numerically minimizing the Gibbs free energy of vortices in the strips. 
Our measurements provide a first experimental benchmark for understanding the flux trapping properties of NbTiN thin films, directly relevant to NbTiN-based superconducting circuits and devices.
    
\end{abstract}

\maketitle

\tocless\section{Introduction}

Understanding and characterizing vortex behavior in fabricated superconducting structures is essential for applications such as superconducting digital circuits (SDCs) \cite{mukhanov2011}, on-chip resonators \cite{song2025}, superconducting quantum information devices \cite{bahrami_vortex_2026}, and radiation detectors across a wide range of wavelengths \cite{Ohkubo_Suzuki_Tanabe_Pressler_Ukibe_2002}. 
In thin-film superconductors, vortices are often trapped or pinned and can persist even at very low fields.

In finite-sized structures, the presence of edges modifies the vortex energy landscape, leading to behavior that deviates from that observed in continuous films. 
For example, the threshold field required for the first vortex to enter, as well as the spatial distribution of subsequent vortices, depends not only on pinning sites and intrinsic material properties but also on the sample geometry. 
A superconducting strip provides a simple yet technologically relevant geometry for exploring these effects. 

Experimentally, vortex behavior in superconducting strips has been studied in a variety of materials, including Nb \cite{Stan2004}, \ch{YBa2Cu3O_y} (YBCO) \cite{Kuit2008, Kuit2009}, amorphous MoSi \cite{Ceccarelli2019}, Pb \cite{Ge2023}, and \ch{NdBa2Cu3O_y} (NBCO) \cite{Suzuki_Li_Utagawa_Tanabe_2000}.
These works employed local magnetometry techniques to image vortex configurations as a function of applied field. 
From these images, one can detemine the number of vortices in a given strip as a function of field. 
Most of these studies \cite{Stan2004,Kuit2008, Kuit2009,Ge2023} predominantly focused on the threshold field for the first vortex and its dependence on strip width, as discussed in more detail below.

Several theoretical models have been proposed to predict the threshold field in superconducting strips in the Pearl limit, where the vortex core size is much smaller than the strip width and the Pearl length is much larger \cite{Clem1998,Likharev1971,Kuit2008}. Much less attention, however, has been given to vortex behavior beyond the threshold or to the role of pinning. Bronson \textit{et al.} \cite{Bronson2006} addressed these issues with a numerical model that incorporates vortex-vortex interactions and pinning to predict vortex configuration in strips. 

Beyond the Pearl limit, Sardella et al. considered the threshold field in tall rectangular pillars \cite{sardella_matching_1999}. Vortex behavior above the threshold in laterally confined superconductors have also been studied with Ginzberg-Landau based models \cite{karapetrov_transverse_2009, mcnaughton_causes_2022}, but with a focus on regimes where vortex-core structure, strong confinement, and nonequilibrium effects are important. De Souza Silva \textit{et al.} \cite{de_souza_silva_flux_2001} investigated vortex penetration in strips for the case of a strip cooled in zero field and then subjected to an increasing magnetic field, in contrast to the field-cooling protocol employed here. A summary of our theoretical frameworks is presented below, with additional details and comparison to previous work provided in the Supplementary Material \cite{supp}. 

In this work, we combine magnetic imaging with theoretical models to study vortex behavior in NbTiN strips. 
We use scanning SQUID microscopy (SSM) to determine the vortex configurations in the strips as a function of applied magnetic field. 
To complement these measurements, we perform simulations that capture both the threshold field and the field dependence of the number of vortices beyond the threshold. 
Our simulations build on the model of Bronson \textit{et al.} \cite{Bronson2006}, with modifications that account for the finite strip length and a revised treatment of the vortex self-energy (see Supplementary Material \cite{supp}, Sec. III.). 

We perform systematic measurements across two samples and multiple strip widths, with high resolution in magnetic field, which allows us to resolve the evolution of vortex configurations in detail near the entry threshold. In particular, we find that close to the threshold field, the vortex number increases more rapidly than the linear dependence commonly assumed in the experimental literature. This behavior is consistent with theoretical expectations as discussed previously \cite{Bronson2006} and is reproduced by our own modeling, but has not been clearly resolved experimentally. In addition, detailed analysis of this regime allows us to extract information about the pinning properties of our samples.

Our results suggest that commonly used methods of extracting threshold fields, based on extrapolating a linear vortex density, can lead to systematic inaccuracies and that careful characterization of the region around the threshold field is meaningful. More generally, they highlight that vortex behavior in strips depends sensitively on pinning and can vary between samples fabricated from the same superconducting material. Overall, we find that the details of the vortex behavior are not simply accounted for by assuming a common coherence length or vortex core radius across our samples.

Our observations are, at least in part, enabled by the use of NbTiN. NbTiN exhibits higher kinetic inductance, which modifies vortex energetics, and differs structurally from e.g. Nb, leading in principle to different pinning characteristics. We find that NbTiN allows us to access a regime in which intrinsic vortex behavior near the threshold field can be resolved, albeit still broadened by pinning. 

In addition to enabling these observations, NbTiN is of practical interest.  It is a promising material for superconducting digital circuits (SDCs) due to its ability to support high critical current densities at small feature sizes, its chemical and thermal robustness, and its compatibility with multilayer fabrication\cite{holmes2021,Iraci2024,Lozano2024,Herr2024,Pokhrel2024}.  Its relatively high critical temperature (up to \SI{17.3}{\kelvin}) and large kinetic inductance are advantageous for device performance and allows dense routing with minimal parasitic magnetic coupling.
Most SDC efforts to date have relied on Nb, an elemental superconductor with a critical temperature of \SI{9.2}{\kelvin}.
Although Nb thin films can be readily deposited by DC sputtering and patterned by conventional etching, achieving reproducible performance in sub-\SI{250}{\nano\meter} circuits remains difficult. 

NbTiN's chemical stability allows processing temperatures approaching \SI{1000}{\degreeCelsius}, depending on composition and deposition conditions, and growth on a variety of dielectrics in multilayer stacks \cite{Valente-Feliciano2015, Farrahi2019, Lozano2024}. NbTiN exhibits a low microwave surface resistance, which reduces microwave losses relative to Nb \cite{valente-feliciano2016,mazin2020,hahnle2021}.

A major challenge for SDCs is uncontrolled flux trapping, which can cause circuit failure and lead to variations in circuit behavior from cooldown to cooldown. 
For NbTiN thin films, direct measurements of flux-trapping properties have not previously been reported. 
Our measurements presented here provide the first characterization of flux trapping in NbTiN thin films, with direct relevance to NbTiN-based SDCs, and allows us to  characterize their pinning properties.

This paper is organized as follows. 
Sections \ref{Exp methods} and \ref{Theo methods} describe the experimental and theoretical methods used in this study. 
In Sec. \ref{Threshold field}, we extract threshold field values from magnetic images of strips cooled in different magnetic field and compare them with theoretical models. 
In Sec. \ref{Above threshold field}, we compare experimental results for vortex trapping above threshold field with simulations that either include or neglect pinning. 
We conclude in Sec. \ref{Conclusion}.

\tocless\section{Samples and experimental methods}\label{Exp methods}

\begin{figure}
    \centering
    \includegraphics[width = 0.45\textwidth]{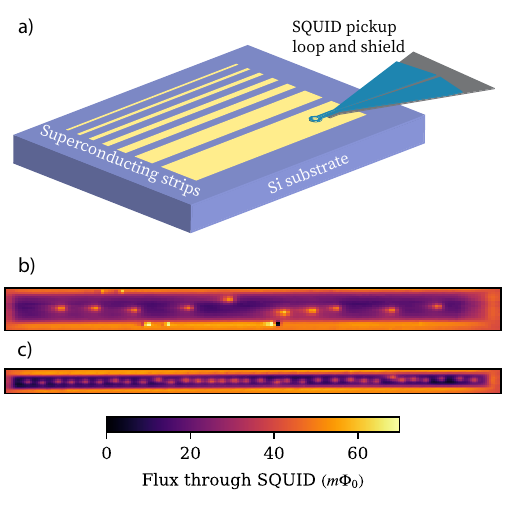}
    \caption{\justifying
    (a) Schematic of the SQUID pickup loop positioned above a NbTiN strip to detect the stray magnetic field produced by trapped vortices. Strips are \SI{200}{\micro\meter} long with width ranging from \SI{2.5}{\micro\meter} to \SI{40}{\micro\meter}. 
    (b, c) Images of strips with widths of (b) \SI{20}{\micro\meter} cooled in a \SI{16}{\micro\tesla} field and (c) \SI{10}{\micro\meter} cooled in an \SI{80}{\micro\tesla} field. 
    Bright spots correspond to individual vortices.
    The color scale (magnetic flux through the SQUID in $m\Phi_0$) is identical for both images.}
    \label{fig:fig1}
\end{figure}

A \SI{50}{\nano\meter}-thick NbTiN film was deposited by physical vapor deposition  (PVD) and patterned using \SI{193}{\nano\meter} immersion lithography \cite{Lozano2024}. 
Two samples, S1 and S2, were studied, both originating from the same \SI{300}{\milli\meter} wafer: S1 was taken from the region between center and edge, and S2 from the center, placing the two samples at least  \SI{67}{\milli\meter} apart. A previous study \cite{Lozano2024}, using the same large-scale deposition tool and recipe, reports significant radial variations in the critical temperature and resistivity across a 300 mm wafer as well as changes in film structure as a function of position on the wafer. Samples from a  wafer deposited under identical conditions as the one studied here, taken at locations corresponding to S1 and S2, yielded room-temperature resistivities of approximately \SI{250}{\micro\ohm\centi\meter} and \SI{290}{\micro\ohm\centi\meter}, with superconducting transition temperatures of \SI{13.5}{\kelvin} and \SI{12.6}{\kelvin}, respectively. 
From these values, the low-temperature London penetration depths $\lambda$ were estimated to be \SI{460}{\nano\meter} and \SI{530}{\nano\meter} for S1 and S2, respectively \cite{Lozano2024, khan_characterization_2023, Bartolf2016}. Reported values for the low-temperature coherence length $\xi$ in NbTiN thin films range from 3 to \SI{5}{\nano\meter} depending on parameters such as thickness, substrate and processing conditions \cite{Sidorova2021,Pratap2023,Lee2024}. Both samples included strips with widths between \SI{2.5}{\micro\meter} to \SI{40}{\micro\meter} and length \SI{200}{\micro\meter}. 

We use scanning SQUID microscopy to image vortices in our samples \cite{Huber2008}. 
The SQUID sensor has a pickup loop with inner and outer diameters of approximately \SI{0.75}{\micro\meter} and \SI{1.6}{\micro\meter}, respectively, and is scanned about \SI{1}{\micro\meter} above the sample (see Fig. \ref{fig:fig1}a). 
A low-field magnet coil surrounding the cryostat provides control over the out-of-plane magnetic field. In principle, the magnetic field applied to a given strip is modified due to screening by neighboring strips and other nearby superconducting structures. Structures other than the strips are located sufficiently far away that their impact is limited. To evaluate the effect of neighboring strips, we numerically solve the London equations using the Python SuperScreen package \cite{bishop-van_horn_superscreen_2022}. These simulations show that, under realistic conditions, the resulting change in magnetic field is less than 1\%. Details of magnetic field calibration and control, as well as the screening simulations, are provided in the Supplementary Material \cite{supp}.

The sample is mounted on a stage with weak thermal coupling to the cryostat. 
A heater and thermometer on this stage allow us to control the sample temperature independently of both the cryostat and SQUID temperatures. Unless noted otherwise, our measurements are performed with the following procedure. 
The sample is first heated above its critical temperature $T_c$, after which the out-of-plane magnetic field is applied. 
The sample is then cooled to approximately \SI{8}{\kelvin} while maintaining the field. 
At this temperature, the vortices are immobile, and we acquire magnetic images to determine their number and positions. 

Figures \ref{fig:fig1}b and \ref{fig:fig1}c show representative images of strips with widths of \SI{20}{\micro\meter} and \SI{10}{\micro\meter}, cooled in an applied field of \SI{16}{\micro\tesla} and \SI{80}{\micro\tesla}, respectively. 
In these images, individual vortices appear as bright dots, while the strip itself appears darker than the surrounding area due to Meissner screening of the applied field. 
The slight tear-drop shape of the vortex images arises from the geometry of the SQUID pickup loop \cite{Huber2008,Kirtley2016}. 
The loop connects to the main SQUID body via two leads, and magnetic flux couples through a small unshielded gap between these leads, distorting the sensitive area from an ideal circular shape. 
By acquiring such images as a function of applied magnetic field, we extract the threshold field, the vortex number as a function of field, and the spatial distribution of vortices in strips of different widths.

The dipole-shaped bright and dark spot pairs, observed at the edges of the 20 µm strips in Figure \ref{fig:fig1}b, are likely due to small magnetic particulates introduced during handling. These features remain visible above the superconducting critical temperature, and we observed a similar feature on the 7 µm strip from the same sample but not on others. We find no evidence that these features affect vortex pinning, as vortices neither preferentially appear near these features nor avoid them, even at high magnetic fields.

\tocless\section{Vortices in an infinite superconducting strip}\label{Theo methods}

\begin{figure}
    \centering
    \includegraphics[width = 0.46\textwidth]{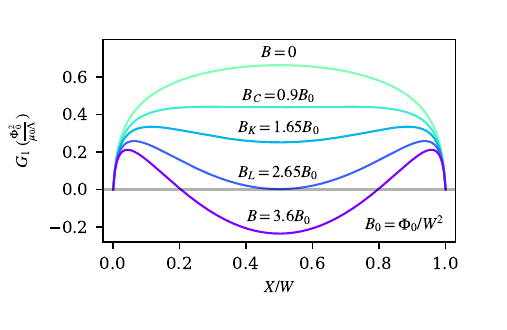}
    \caption{\justifying
    Normalized Gibbs free energy in units of $\Phi_0^2/(\mu_0\Lambda )$ for a single vortex at position $X$ in a strip of width $W$, for $r_c/W = 0.01$. 
    The curves, shown for various applied field values, include those corresponding to the Clem ($B_C$), Kuit ($B_K$), and Likharev ($B_L$) threshold field defined in the main text.
    The characteristic field scale is $B_0=\Phi_0/W^2$.}
    \label{fig:fig2}
\end{figure}

To model vortex behavior in our NbTiN strips, we assume the two-dimensional limit where the London penetration depth $\lambda$ is much larger than the film thickness $d$. 
In this regime, vortices are Pearl vortices \cite{Pearl1964}, characterized by the Pearl length $\Lambda=2\lambda^2/d$. For our samples with $d=\SI{50}{\nano\meter}$ and estimated $\lambda = \SI{460}{\nano\meter}$ (S1) and $\SI{530}{\nano\meter}$ (S2), the corresponding low-temperature values of $\Lambda$ are estimated to be approximately $\SI{8}{\micro\meter}$ and $\SI{11}{\micro\meter}$, respectively. 
A Pearl vortex in an infinite film consists of three regions: (i) a core region ($r\lesssim r_c$), where $r_c$ is the vortex core radius of order the coherence length $\xi$, and the circulating sheet current increases as $K(r)\sim r$; (ii) an intermediate region ($r_c\lesssim r\lesssim \Lambda$) where $K(r)\sim 1/r$; and (iii) an outer region ($r\gg \Lambda$) where $K(r)\sim 1/r^2$.
In our NbTiN films, we expect that $r_c\ll\Lambda$ near $T_c$. 
Under these conditions, vortices can be treated as pointlike particles whose behavior is governed by their Gibbs free energy. 

In a strip of width $W$, the Gibbs free energy of vortices is determined by their individual self-energies and magnetic interactions. For a strip wider than the Pearl length, magnetostatic screening affects the energy of the vortex, and the electrodynamics are described by a self-consistent integral equation (see, e.g., Nakagawa and Kogan \cite{Nakagawa2024}). For an infinite thin-film strip narrower than the Pearl length, Kogan analytically derived the vortex energy and magnetic moment \cite{Kogen1994}.
This condition, $\Lambda\gg W$, is expected to hold in our experiments at the temperature where vortices become immobile or ``freeze,'' which occurs near $T_c$ and sets the relevant temperature for analyzing the vortex behavior \cite{Stan2004}. For example, if freezing occurs at $T/T_c\approx 0.95$, the approximate relevant values of $\Lambda$ are $\SI{170}{\micro\meter}$ and $\SI{220}{\micro\meter}$ for S1 and S2, respectively. Freezing likely occurs even closer to $T_c$.

We consider an infinite strip extending along the $y$-axis, with finite width $W$ along the $x$-axis ($0<X<W$). The result is summarized here; a detailed derivation is provided in Supplementary Material \cite{supp}, Sec. III.A.  
The Gibbs free energy of a single vortex at position $(X,Y)$ is then \cite{Bronson2006,Kogen1994}
\begin{equation}\label{eq:Gv}
    G_v(X,Y)=F_{\text{self}}(X,Y)-m(X,Y)B,
\end{equation}
where $B$ is the applied perpendicular magnetic field, $m(X,Y)$ is the vortex magnetic moment, and $F_{\text{self}}(X,Y)$ is the vortex self-energy, including both core and edge contributions. 

Figure \ref{fig:fig2} shows this Gibbs free energy $G_v$ of a single vortex as a function of applied magnetic field, using the following expression for $m(X,Y)$ and $F_{\text{self}}(X,Y)$:
\begin{eqnarray}\label{eq:magnetic_inf}
    ~m(X,Y)&=&\frac{\Phi_0}{\mu_0 \Lambda} X(W-X),
\end{eqnarray}
\begin{equation}\label{eq:selfEnergy_inf}
    F_{\text{self}}(X,Y)
    =\frac{\Phi_0^2}{2\pi\mu_0 \Lambda}
        \ln\left[
            \frac{2W}{\pi r_c}\sin\left(\frac{\pi X}{W}\right)+1
        \right],
\end{equation}
where $\Phi_0=h/2e$ is the flux quantum \cite{Bardeen1961}, $\mu_0$ the vacuum permeability, and $r_c\sim\xi$ is the vortex core cutoff radius. Because the strip is taken to be infinite along $Y$, $G_v$ is independent of $Y$.

Our expression for $F_{\text{self}}(X,Y)$ includes two modifications relative to previous work.
First, we introduce $r_c$ as an effective vortex core radius rather than identifying it directly with the coherence length $\xi$. 
In thin-film superconductors in the Pearl limit, the vortex core radius $r_c$ used to regularize London or numerical models is not necessarily equal to the Ginzburg-Landau coherence length $\xi$. In our modeling, the vortex core enters the Gibbs free energy only as a short-distance logarithmic cutoff. As emphasized by Kuit et al. \cite{Kuit2008}, different choices of this cutoff correspond to different effective core radii, affecting numerical prefactors but not the characteristic $1/W^2$ scaling of the threshold fields discussed in the next section. Similarly, Nakagawa et al. \cite{Nakagawa2024} use $\xi$ as an effective cutoff in London self-energy calculations without identifying it with a uniquely defined vortex core radius. More generally, the equations governing Pearl vortices do not contain an intrinsic coherence-length scale, and therefore require the introduction of an effective vortex kernel to regularize divergences in current and phase gradients \cite{Semenov2016}.

A physically motivated alternative identifies the vortex core boundary as the location where the circulating current reaches the depairing current, leading to an effective thin-film core radius $r_c\sim (12\Lambda\xi^2)^{1/3}$, which can be parametrically larger than  in the Pearl limit \cite{Bronson2006}. We therefore distinguish $r_c$ from $\xi$ in the Pearl limit, consistent with prior theoretical and experimental work.
Second, the additional ``$+1$" inside the logarithm ensures that $F_{\text{self}}(X,Y)$ vanishes at the strip edges ($X=0,W$) (see Supplementary Material \cite{supp}, Sec. III.A.1.).

To describe vortex behavior beyond the threshold field at which the first vortex enters, we need to include the interaction energy between vortices. 
For two vortices of the same polarity at positions $(x,y)$ and $(X,Y)$, the interaction energy is
\begin{equation}\label{eq:F12_inf}
    \resizebox{0.43\textwidth}{!}{$
    F_{vv}(x,y;X,Y)=
    \frac{\Phi_0^2}{2\pi\mu_0 \Lambda}
    \ln\left[
    \frac{\cos\frac{\pi(x+X)}{W}-\cosh\frac{\pi(y-Y)}{W}}
         {\cos\frac{\pi(x-X)}{W}-\cosh\frac{\pi(y-Y)}{W}}
    \right].
    $}
\end{equation}
In Ref. \cite{Kuit2008}, vortex-vortex screening is incorporated through a mean-field replacement $B\rightarrow B_a-n\Phi_0$, whereas in the formulation here vortex density effects are treated explicitly via the interaction energy $F_{vv}$. This distinction allows us to capture nonuniform vortex configurations and pinning effects beyond a uniform mean-field description.
Because our NbTiN strips are finite in length, we derived modified forms of $m,F_{\text{self}}$ and $F_{vv}$ [Eqs. \eqref{eq:magnetic_inf}, \eqref{eq:selfEnergy_inf}, and \eqref{eq:F12_inf}] for semi-infinite strips.  This is equivalent to adding corrections to account for the influence of the strip ends
(see Supplement \cite{supp}, Sec. III.B.).

To study the effect of pinning, we include energy terms that model localized pinning sites. A common approach assigns random positions in the strip where the vortex free energy is locally reduced. These pinning terms lower the threshold field for vortex entry and compete with vortex-vortex interactions, which favor ordered vortex arrangements. To capture these effects, we implement a toy model (see Supplementary Material \cite{supp}, Sec. III.E.), following the approach of Bronson \textit{et al.} \cite{Bronson2006}, which we solve using numerical simulations, and discussed further in Sec. \ref{Above threshold field}. 

\tocless\section{Threshold field for inducing the first vortex in superconducting \NoCaseChange{NbTiN} strips}\label{Threshold field}

\begin{figure}
    \centering
    \includegraphics[width = 0.47\textwidth]{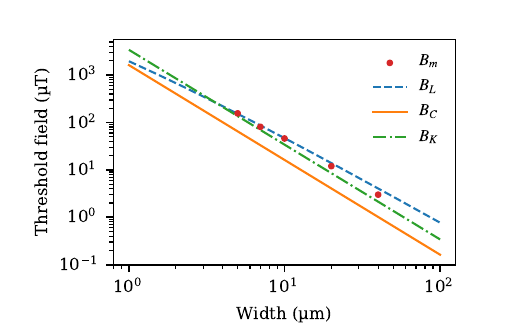}
    \caption{\justifying
    Measured threshold field $B_m$ (red dots) extracted from SQUID images and compared with the model predictions $B_C$, $B_K$ and $B_L$. 
    A fit of the Likharev model $B_L$ to the data yields a best-fit vortex core radius of $r_c=\SI[parse-numbers=false]{92\pm 5}{\nano\meter}$}.
    \label{fig:fig3}
\end{figure}

Figure \ref{fig:fig2} shows that, in the absence of pinning, the first vortex enters at the strip center, where the Gibbs free energy develops a local minimum as the applied magnetic field increases. Several models for the threshold field have been proposed \cite{Maksimova_1998, Kuznetsov_Eremenko_Trofimov_1999, Likharev1971, Clem1998, Kuit2008, Kuit2009}. Likharev \cite{Likharev1971} argued that a vortex becomes trapped once it is absolutely stable, corresponding to the field at which the  Gibbs free energy $G_v$ in Eq. \eqref{eq:Gv} becomes zero or negative. For a vortex at the strip center, this condition gives
\begin{equation}
    B_L = \frac{2\Phi_0}{\pi W^2}
        \ln{\left(\frac{2 W}{\pi r_c}+1\right)}.
    \label{eq:likharev}
\end{equation}
The threshold field $B_L$ reflects the competition between the magnetic energy and self-energy of a vortex in the strip. It is independent of the Pearl length $\Lambda$, since both contributions scale identically with $\Lambda$.

Clem \cite{Clem1998} defined the threshold field as the field where a \emph{local} minimum first appears in the Gibbs free energy, rendering the vortex metastable. This leads to 
\begin{equation}
    B_C = \frac{\pi \Phi_0}{4 W^2}
    \label{eq:clem}
\end{equation}

Kuit \textit{et al.} \cite{Kuit2008, Kuit2009} introduced a model that incorporates vortex-antivortex pair generation just below the critical temperature, in addition to the contributions considered above. Their analysis yields
\begin{equation}
    B_K = 1.65 \frac{\Phi_0}{W^2}
    \label{eq:kuit}
\end{equation}

Notably, $B_K$ and $B_C$ depend only on the strip width $W$, whereas $B_L$ also depends logarithmically on the ratio $W/r_c$. 

In Fig. \ref{fig:fig3}, we compare the measured threshold fields $B_m$ of strips with varying width $W$ from sample S2 to the theoretical predictions. Here, we define the threshold field $B_m$ as the highest applied field at which cooling a strip does not nucleate a vortex. We focus on S2 because sample S1 exhibits vortex entry at fields well below the expected thresholds, likely due to a few strong pinning sites. Our method for determining $B_m$ differs from previous studies \cite{Stan2004,Kuit2008,Kuit2009,Ge2023}, which estimate the threshold by extrapolating the high-field linear dependence of vortex number to zero. Such extrapolation underestimates the threshold field \cite{Bronson2006}, as we will show below. 

In Fig. \ref{fig:fig3}, $B_m$ decreases with increasing width $W$, reflecting the reduced influence of edge barriers in wider strips. Comparing $B_m$ to the three models, we find that Clem's criterion $B_C$ consistently underestimates the threshold for all widths, while both $B_K$ and $B_L$ provide reasonable agreement. For $B_L$, fitting to Eq. \eqref{eq:likharev} yields an effective core radius of $r_c=\SI[parse-numbers=false]{92 \pm 5}{\nano\meter}$. This value is significantly larger than the low temperature coherence length of NbTiN, typically reported to be \SI[parse-numbers=false]{3-5}{\nano\meter}. We attribute this difference mainly to two factors. First, similar to the Pearl length, the relevant $r_c$ corresponds to its value at the temperature at which vortices become immobile and the configuration freezes out, which typically occurs within a few percent of the critical temperature \cite{Stan2004}. At these temperatures, $r_c$ is substantially larger. Second, even with the definition of $B_m$ used here, its value is lowered by the presence of pinning compared to the pinning-free case, which leads to an overestimate of $r_c$. In the next section, we discuss estimates of $r_c$ from modeling that explicitly includes pinning and considers the overall shape of the curve beyond the threshold field. Finally, as explained above we stress that we treat $r_c$ as an effective parameter, which is not straightforward to compare quantitatively to the the coherence length at low temperature.

\tocless\section{Number of vortices above the threshold field}\label{Above threshold field}

\begin{figure}
    \centering
    \includegraphics[width = 0.47\textwidth]{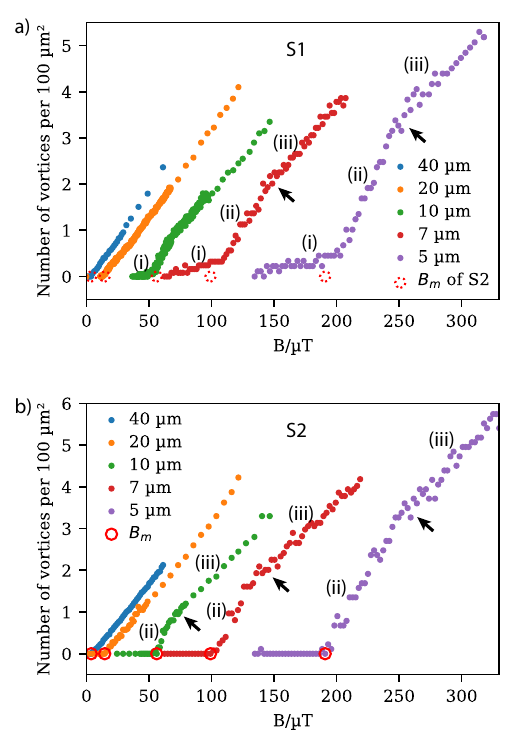}
    \caption{\justifying
    Number of vortices per \SI{100}{\micro\meter\squared} trapped in superconducting strips as a function of applied magnetic field for a) S1 and b) S2, with widths ranging from 5 to $\SI{40}{\micro\meter}$. 
    In S1 strips, stronger pinning sites produce shallow tails extending to lower magnetic fields. 
    In S2 strips, the threshold fields $B_m$ (from Fig. \ref{fig:fig3}) are marked by circles in b.
    }
    \label{fig:fig4}
\end{figure}

In this section, we examine the behavior of vortices as a function of applied field beyond $B_m$ as additional vortices are trapped in the strips during cooling. Figure \ref{fig:fig4} shows the vortex number, extracted from images, for strips with widths between 5 and \SI{40}{\micro\meter} in samples S1 and S2. As labeled in Figure \ref{fig:fig4}, the data exhibit several regimes (i) a shallow tail at low fields, where vortices appear gradually, (ii) a steep increase in vortex number, and (iii) a slower linear increase in vortex number at higher fields. A shoulder occurs when transitioning from (ii) to (iii), where the slope softens marked with black arrows in a few curves in Figure \ref{fig:fig4}. Regime  (i) is only observed for sample S1, and a kink is visible when transitioning from (i) to (ii). The slope in regime (iii) is smaller than the slope expected for adding one vortex per $\Phi_0$ threading the area $\mathcal{A}$, i.e., $N = (\mathcal{A}/\Phi_0) B$. Aside from the shallow tail in regime (i) in S1, the vortex numbers in both samples are nearly identical.


As discussed in the modeling section below, in the absence of pinning, a steep initial rise in vortex number as seen in regime (ii) at the strip center is expected.  Near the threshold field, the intervortex spacing is large. The minimum of the Gibbs free energy confines vortices to the strip center, and while  vortex-vortex interactions enforce a uniform spacing, their contribution to the total energy remains small due to the large spacing. As the vortex spacing decreases with increasing field and becomes comparable to the width of the strip $W$, vortex-vortex interactions become increasingly important, slowing down the addition of further vortices (regime (iii)). This softening of the slope produces the observed shoulder. This shoulder is thus consistent with basic expectations for the vortex behavior, although it has not been clearly resolved in previous studies. Pinning further modifies this simple picture.

As further discussed below, we attribute regime (i) which is observed for S1 but not S2 to differences in the pinning landscape, which is plausible given that S1 and S2 originate from two locations on a wafer more than 67 mm apart. In the Supplemental Material \cite{supp}, we show a zoom in and individual images corresponding to regime (i) in S1. The curve shows discrete steps and inspection of the corresponding images shows that vortices repeatedly occupy a few selected locations within the strip, which we associate with strong pinning sites.

Previous studies \cite{Stan2004,Kuit2008,Kuit2009,Ge2023} also report a linear dependence of vortex number on applied field at high fields (regime (iii)). Kuit \textit{et al.} \cite{Kuit2008,Kuit2009} provide data for only two strip widths with relatively sparse field sampling, limiting direct comparison. The data of Ge \textit{et al.} \cite{Ge2023} show signatures of the shoulder that we observe, although it is not clearly resolved and was not discussed. Stan \textit{et al.} \cite{Stan2004} observe a transition from a linear dependence to a shallower slope, producing a kink similar to our S1 data. However, they do not resolve a shoulder, which Bronson \textit{et al.} attribute to pinning effects masking the shoulder. 



\begin{figure*}[hbt!]
     \includegraphics{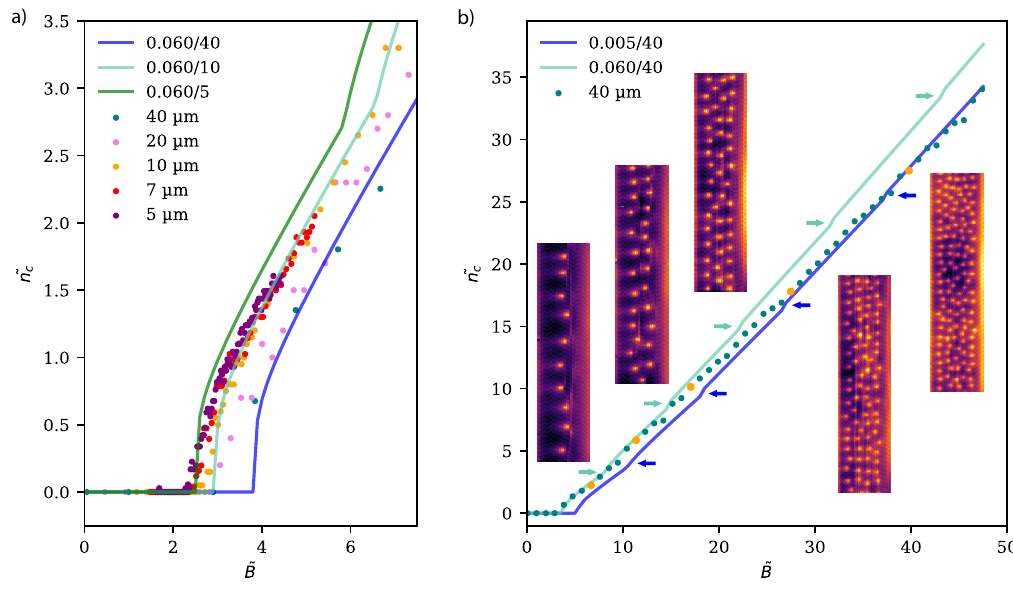}
     \caption{\justifying
     Number of vortices in an area $W^2$ obtained from the vortex lattice (VL) model without pinning (solid lines) and from experiment (symbols). 
     Steps in the simulated curves, highlighted by arrows, correspond to the addition of a new column of vortices. 
     The magnetic field is expressed in units of $B_0 = \Phi_0/W^2$, i.e., $\tilde{B}=B/B_0$.
     a) Results for S2 strips with widths $W=40,20,10,7,\text{ and }5\,\mu$m, using the dimensionless vortex core radius $\tilde{r}_c=r_c/W$ with $r_c =\SI{60}{\nano\meter}$}.
     b) Results for the \SI{40}{\micro\meter} strip on S2 across an extended range of $\tilde{B}$. 
      Images at the fields marked by orange circles are consisted with the formation of additional vortex columns at approximately the fields indicated by arrows.
     \label{fig:nvsB}
\end{figure*}

We next compare our experimental data with two theoretical models that capture the interplay of vortex-vortex interactions, finite strip size, and pinning, closely following the approach of Bronson \textit{et al.} \cite{Bronson2006}. Both models are based on the Gibbs free energy introduced earlier in Eq. \eqref{eq:Gv} (see Supplementary Material \cite{supp}, Secs. II.C-D. for details). 

In the first model, vortices form an ordered lattice (vortex lattice, VL) in the absence of pinning within infinite-length strips. As the applied field increases, the number of vortex columns increases accordingly. 

To explore the influence of pinning near $B_m$, we introduce a second, one-dimensional model. Here, vortices occupy discrete sites along the strip center in semi-finite strips. Pinning is incorporated by lowering the free energy at randomly selected sites by amounts drawn from a distribution. A Monte Carlo simulated annealing algorithm is then used to obtain the equilibrium vortex configuration for a given applied field, and we refer to this model as simulated annealing (SA) below. 

For the VL model, we use an optimization algorithm to minimize the Gibbs free energy for different vortex numbers and periodic vortex arrangements, including vortex-vortex interaction (see Supplementary Material \cite{supp} Sec. III.C.). The solid lines in Figure \ref{fig:nvsB} show the results for infinite strips of width $W$ and several values of $r_c/W$. 

The horizontal axis is normalized by $B_0 = \Phi_0/W^2$, which corresponds to one flux quantum per area $W^2$. The vertical axis shows the average number of vortices, $\tilde{n}_c$, in a section of length $W$ (equivalently to a square area $W^2$) within the strip. With this rescaling, the area law becomes a straight line of slope 1, and the threshold field in the VL simulations appears at $B_L/B_0$. At this threshold, vortices first align along the strip center with spacing $a > W$. As the field increases, the vortex number grows steeply until $a \approx W$ (or equivalently $\tilde{n}_c \approx 1$), when the  vortex-vortex interactions become significant and the curve rounds off. This behaviour follows from the assumption $\Lambda \gg W$. At still higher field, additional vortex columns enter with regular spacings, producing subtle shoulders in the simulated curves at each column addition (see Figure \ref{fig:nvsB}b ).

The experimental data for strips from S2 are shown as dots in Fig. \ref{fig:nvsB}a), and nearly collapse onto a single curve when plotted on the rescaled axes. The simulated curves use  $\tilde{r}_c=r_c/W$ ranging from $0.060/5$ to $0.060/40$, corresponding to $r_c=\SI{60}{\nano\meter}$ and $W=\SI[parse-numbers=false]{5-40}{\micro\meter}$. The value $r_c=\SI{60}{\nano\meter}$ is smaller than the value estimated from the threshold fields in Fig. \ref{fig:fig3}. This can be explained by the fact that threshold fields are decreased by pinning, which results in a larger estimated $r_c$ value. Overall, the simulations reproduce the experimental data well, though discrepancies appear both near the threshold and at higher fields. Close to the threshold, pinning effects are expected to play an important role, as investigate below.

One might expect that a single value of $r_c$ characterizes the thin film, such that the data for a strip of a given width $W$ would follow a curve corresponding to the ratio $r_c/W$, and wider strips should agree with the corresponding lower $r_c/W$ ratio. This qualitative trend is observed. For example, the data for the \SI{20}{\micro\meter} strip lie below those for the \SI{10}{\micro\meter} strip. However, none of the experimental curves follows a curve corresponding to a fixed value of $r_c/W$. In particular, for $B/B_0 \gg 1$ (see Fig. \ref{fig:nvsB}b), the data shifts to an decreasing effective $r_c/W$, implying an decreasing effective $r_c$ given that the strip width is fixed by lithography. Despite this field-dependent shift of $r_c$, the number of vortex columns observed in images approximately matches the predicted number.

A full understanding of this discrepancy is beyond the scope of this work, but we briefly offer a few speculations on its origin. Deviations near the threshold field may arise from pinning in the strips, however at higher fields we expect pinning to be less important. A key assumption in our models is that a single value of $r_c$ and $\Lambda$ characterizes the strips. Both parameters are strongly temperature dependent and diverge as the temperature approaches $T_c$. We assume that the relevant values for $\Lambda$ and $r_c$ are those at the temperature where vortices become immobile, which the literature \cite{Stan2004} suggests is close to $T_c$. This implies that $\Lambda$ and $r_c$ significantly exceed their low-temperature values, supporting the assumption $\Lambda \gg W$ used in deriving the Gibbs free energy. At the same time, their strong temperature dependence makes the relevant values sensitive to the cooling process.

Because vortex capture and freezing is a dynamic process that occurs near $T_c$, we speculate that the effective $r_c$ depends on the applied magnetic field. At a fixed field, the Gibbs free energy evolves as the sample cools through the temperature-dependent changes of $r_c$ and $\Lambda$. The observation that higher-field data align with smaller $r_c/W$ ratios suggests that vortices freeze further below $T_c$ as the field increases. This trend is consistent with the notion that vortices fills strong pinning sites first. At low magnetic fields, few vortices are trapped in a strip, filling only strong pinning sites. As magnetic field increases, more vortices get trapped and they start to occupy pinning sites with weaker pinning potential. Vortices in such sites can stay mobile until a lower temperature, lowering the freezing temperature for higher field.

\begin{figure*}[hbt!]
     \includegraphics[width=\textwidth]{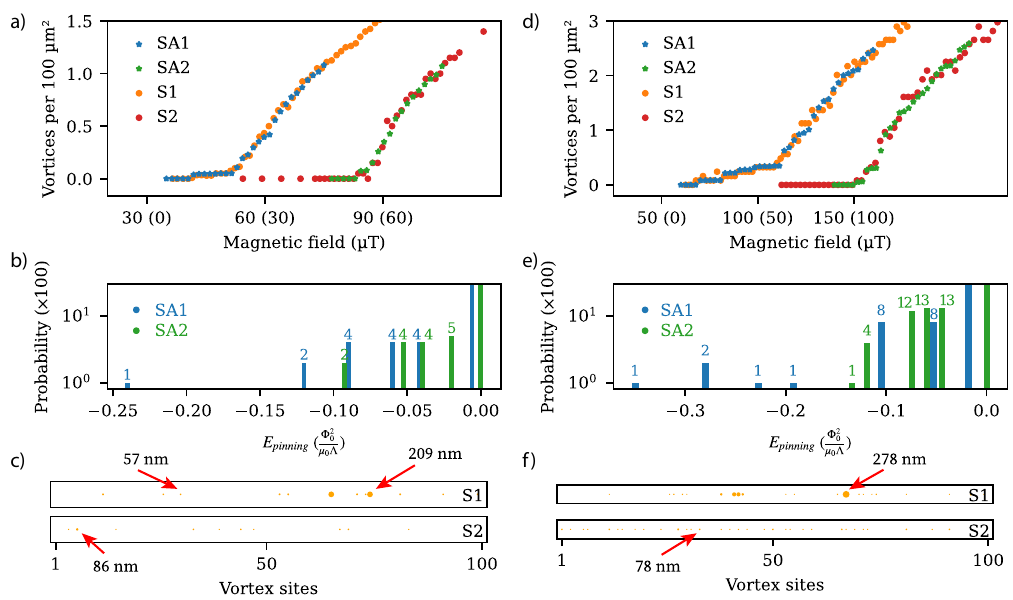}
     \caption{\justifying
     (a, d) Comparison of SA results with experimental data for (a) \SI{10}{\micro\meter}  and (d) \SI{7}{\micro\meter}  strips from samples S1 and S2.
     Results for S2 are shifted in field by \SI[parse-numbers=false]{30(50)}{\micro\tesla}, as indicated by the axis labels in brackets.
     (b, e) Histograms of pinning energy distributions for 100 allowed sites along the (b) \SI{10}{\micro\meter} and (e) \SI{7}{\micro\meter} strips for S1 and S2.
     Bars show the probability of pinning sites for a given pinning energy (expressed in units of $\Phi_0/(\mu_0\Lambda)$); numbers above bars indicate counts.
     (c, f) Pinning energy configurations for S1 (top) and S2 (bottom) from a representative simulation run. Dot sizes are calculated with the model described in the main text that correlates pinning energies to effective hole sizes on a superconducting film and then enlarged $\times5$ for visibility. 
     For S1, radii range from 57-\SI{209}{\nano\meter}} (\SI{10}{\micro\meter} strip) and 42-\SI{278}{\nano\meter} (\SI{7}{\micro\meter} strip).
     For S2, radii range from 43-\SI{86}{\nano\meter} (\SI{10}{\micro\meter} strip) and 48-\SI{78}{\nano\meter} (\SI{7}{\micro\meter} strip).
     \label{fig:MC_simu}
\end{figure*}


The second model, simulated annealing (SA), captures vortex behavior near the threshold field. In this approach, vortices are constrained to discrete sites along the strip center, with the number of available sites chosen to exceed by far the maximum number of vortices expected at the highest field. Although vortex-vortex interactions are included, the model is by construction only applicable in the regime where a single column of vortices is expected. In the absence of pinning, the SA model produces the same result as the VL model at low fields (see Supplementary Material \cite{supp}, Sec. III.D.). 

To incorporate pinning, we add energy terms that randomly lower the free energy of a vortex at some sites along the strip. These pinning terms reduce the threshold field for the first vortex to enter, broaden the steep initial rise in vortex number, and produce a shallow tail. The magnitude of these effects depends on both the energy scale and the density of the pinning sites. 
To obtain equilibrium configurations in the presence of pinning, we employ SA with exponential cooling—a procedure distinct from experimental cooling—to minimize the system’s energy.


Figures \ref{fig:MC_simu}a and \ref{fig:MC_simu}d compare simulations with experimental data for the \SI{10}{\micro\meter} and \SI{7}{\micro\meter} strips from samples S1 and S2, respectively. Core radii of $r_c = \SI{60}{\nano\meter}$ were used for the \SI{10}{\micro\meter} strips, and $r_c=\SI{40}{\nano\meter}$ for the \SI{7}{\micro\meter} strips. Details of how $r_c$ is chosen are provided in the Supplementary Material. Briefly, we use the $r_c$ primarily to shift the entire curve horizontally through its effect on the single-vortex threshold field, while the pinning energies primarily to affect the curve near the threshold field. 
The simulations show good agreement with the data up to the field at which a second column of vortices begins to form -- around \SI{180}{\micro\tesla} for the \SI{7}{\micro\meter} strip and approximately \SI{110}{\micro\tesla} for the \SI{10}{\micro\meter} strip. The corresponding pinning site densities and energy distributions are shown in Figs. \ref{fig:MC_simu}b and \ref{fig:MC_simu}e on a common axis for both samples. To capture the shallow tails in S1, the model requires sparse but strong pinning sites. In contrast, in S2 the broadened shoulder arises from weaker pinning sites with higher density. These differences reflect spatial variations in the pinning landscape across the wafer from which both samples originated.

In superconducting digital circuits, long openings-referred to as "moats"-have been fabricated around the circuits to guard the active circuits from trapping vortices \cite{Bermon_1983}. Early designs employed extended trenches, whereas more recent approaches use much smaller square or circular holes patterned  into the ground plane \cite{Golden_Parmar_Semenov_Tolpygo_2025, Herr_Talanov_Herr_2020}. For moats to be effective, they must energetically compete with naturally occurring pinning sites in the film. Consequently, the pinning energies are of central importance in design of moat structures.

To estimate the moat size for effectice pinning, we model intrinsic pinning sites as circular holes in the thin film and map pinning energies to equivalent hole diameters. The reduction in Gibbs free energy for a vortex trapped at the center of a circular hole of radius $r_h$ is evaluated assuming $\Lambda\gg r_c$. In the presence of an out-of-plane magnetic field, the total Gibbs free energy of a vortex pinned at the hole is
\begin{equation}\label{eq:GTotal}
    G_{\text{tot}}(X,Y)=G_{\text{pin}}+G_v(X,Y),
\end{equation}
where $G_{\text{pin}}$ accounts for both the reduction in the vortex self-energy ($\Delta F_{\text{self}}$) and the modification of magnetic energy ($\Delta F_{mB}$) due to the hole, such that 
\begin{equation}
    G_{\text{pin}}=\Delta F_{\text{self}} - \Delta F_{mB}. 
\end{equation}
Approximate analytical expressions for these terms are given in the Supplementary Material \cite{supp}, Sec. III.E, and are used to numerically convert the pinning energies in Figs. \ref{fig:MC_simu}b,e into equivalent hole sizes. Figures \ref{fig:MC_simu}c,f show representative realizations of the pinning configurations, where each dot corresponds to an equivalent hole size (enlarged by a factor of five for visibility).

We find that the largest effective pinning site diameters required to reproduce the observed pinning energies are on the order of several hundred nanometers in S1, but significantly smaller in S2 (see Fig. \ref{fig:MC_simu}). In future work, it would be valuable to compare these extracted sizes from strip measurements with the performance of engineered moats patterned into the same films. In addition, identifying the microscopic origin of the pinning sites is desirable; however, linking individual pinning sites observed in magnetic imaging to their microscopic origin remains challenging, as it requires combining nanoscale structural probes with techniques that access vortex configurations under comparable conditions.

A limitation of the SA approach is that it is based on a one-dimensional toy model. In contrast, the images show that even before a second vortex column forms, some vortices appear off-center. A fully two-dimensional time-dependent Ginzburg-Landau simulation with a realistic pinning landscape could capture such behavior more accurately and also account for vortex dynamics during cooling. However, such modeling lies beyond the scope of the present work.

\tocless\section{Conclusion}\label{Conclusion}

We investigated vortex trapping in NbTiN superconducting strips cooled in magnetic field using a combination of scanning SQUID microscopy and modeling. In addition to determining the threshold field for the first vortex entry, we presented detailed measurements and modeling of vortex behavior above this threshold. Our results clearly show the expected features: an initially steep increase in vortex number that softens once the vortex spacing approaches the strip width. We also identified distinct signature of pinning, which differ between the two samples and are captured by a one-dimensional Monte Carlo simulated annealing model. 

Our Gibbs free energy-based models, which treat vortices as point-like particles and include localized pinning sites, capture many features of the vortex behavior in NbTiN strips. Nonetheless, discrepancies remain. For instance, the optimal vortex core radius $r_c$ appears to vary with the applied magnetic field. Moreover, the measured threshold field for our \SI{10}{\micro\meter} strips is lower than that reported for a  Nb strip with the same width fabricated from a 200 nm thick film \cite{Stan2004}, despite Nb having a longer low-temperature coherence length than NbTiN. 
This comparison highlights that the threshold field is sensitive to the pinning landscape, the temperature at which vortices become immobile and kinetic effects during field cooling, which depend on details of the film, its microstructure and the cooling protocol. As a result, threshold fields can vary significantly even between nominally similar samples, and are not uniquely determined by intrinsic material parameters alone. This sample-to-sample variability is also reflected in comparing different studies on Nb strips \cite{Stan2004, kapur2026fluxtrappingcharacterizationsuperconducting} which  show threshold fields different from each other.

Quantitative comparison between samples of different materials and even the same material therefore remains challenging. Despite the above caveats, we believe that the type of measurements presented here can provide useful insights into the pinning landscape and flux trapping properties of thin films, although they should not be interpreted as a measurement of the vortex core radius.

Future work will extend our models to include kinetic effects and a more refined treatment of pinning. Experiments that directly probe vortex freezing, quantify pinning strength, and test the influence of cooling protocols will be essential for guiding these efforts. In addition, systematic comparisons of strips fabricated from different superconducting films (e.g., Nb and NbTiN) as well as strips fabricated from the same superconducting material under identical experimental conditions will help clarify the connection between the effective vortex core radius $r_c$ used in modeling and the intrinsic low-temperature coherence length $\xi$.
\begingroup
\let\oldaddcontentsline\addcontentsline
\renewcommand{\addcontentsline}[3]{} 
\begin{acknowledgments}
Research was partially sponsored by the U.S. Department of Energy, Office of Science, Offices of Nuclear Physics and Advanced Scientific Computing Research under contract DE-AC05-06OR23177, and by the Army Research Office under Grant Number W911NF-24-1-0150. Initial scanning SQUID measurements were also supported by the Cornell Center for Materials Research with funding from the NSF MRSEC program (DMR-1719875). Sample design and fabrication at imec and imec USA were also supported by imec INVEST+ and by Osceola County. 
\end{acknowledgments}

\endgroup

\widetext
\newpage
\begin{center}
\textbf{\large Supplemental Materials: Flux trapping in NbTiN strips}
\end{center}
\setcounter{equation}{0}
\setcounter{section}{0}
\setcounter{figure}{0}
\setcounter{table}{0}
\setcounter{page}{1}
\makeatletter
\renewcommand{\theequation}{S\arabic{equation}}
\renewcommand{\thefigure}{S\arabic{figure}}
\renewcommand{\bibnumfmt}[1]{[S#1]}
\renewcommand{\citenumfont}[1]{S#1}

\SetCommentSty{mycommfont}
\SetKwComment{Comment}{/* }{ */}
\algrenewcommand\algorithmiccomment[1]{\hfill{/* #1 */}}

\newcommand{\mycommfont}[1]{\footnotesize\itshape #1}
\SetCommentSty{mycommfont}
\SetKwComment{Comment}{/* }{ */}
\newtcolorbox{summarybox}[1][]{%
  enhanced,
  breakable,
  colback=white,
  colframe=black,
  boxrule=0.8pt,
  arc=3pt,
  before skip=10pt plus 2pt minus 1pt,
  after skip=10pt plus 2pt minus 1pt,
  title=#1,
}
\newcommand{\panel}[2]{%
  \begin{tikzpicture}
    \node[inner sep=0pt, outer sep=0pt] (img)
      {\includegraphics[width=\linewidth]{#2}};
    \node[anchor=north west, xshift=2pt, yshift=-2pt,
          font=\bfseries\small, fill=white, inner sep=1pt,
          rounded corners=1pt] at (img.north west) {#1};
  \end{tikzpicture}%
}

\setcounter{secnumdepth}{4}
\setcounter{tocdepth}{4}
\renewcommand{\thesection}{\Roman{section}}
\makeatletter
\renewcommand\section{\@startsection{section}{1}{0pt}
   {3.5ex plus 1ex minus .2ex}
   {2.3ex plus .2ex}
   {\normalfont\large\bfseries}}
\renewcommand{\thesubsection}{\thesection.\Alph{subsection}}
\renewcommand\subsection{\@startsection{subsection}{2}{0pt}%
   {3.25ex plus 1ex minus .2ex}%
   {1.5ex plus .2ex}%
   {\normalfont\normalsize\bfseries}}
\renewcommand{\thesubsubsection}{\thesubsection.\arabic{subsubsection}}
\renewcommand\subsubsection{\@startsection{subsubsection}{3}{0pt}%
   {3.25ex plus 1ex minus .2ex}%
   {1.5ex plus .2ex}%
   {\normalfont\normalsize}}%
\newcounter{subsubsubsection}[subsubsection]
\renewcommand{\thesubsubsubsection}{\thesubsubsection.\alph{subsubsubsection}} 
\newcommand\subsubsubsection{\@startsection{subsubsubsection}{4}{0pt}%
  {3.25ex plus 1ex minus .2ex}%
  {1.5ex plus .2ex}%
  {\normalfont\normalsize\bfseries}}
\newcommand\l@subsubsubsection{\@dottedtocline{4}{7.1em}{3.8em}}
\renewcommand\p@section{}
\renewcommand\p@subsection{}
\renewcommand\p@subsubsection{}
\renewcommand\p@subsubsubsection{}
\makeatother
\makeatletter
\renewcommand\l@section{\@dottedtocline{1}{0em}{2.3em}}
\renewcommand\l@subsection{\@dottedtocline{2}{2.3em}{3.2em}}
\renewcommand\l@subsubsection{\@dottedtocline{3}{5.5em}{4.1em}}
\renewcommand\l@subsubsubsection{\@dottedtocline{4}{9.6em}{4.1em}}
\makeatother
\makeatletter
\providecommand*{\toclevel@subsubsubsection}{4}
\makeatother

\title{Supplementary Material for ``Flux trapping in NbTiN strips''}
\maketitle

\tableofcontents

\section{Scanning SQUID Magnetometry}

Measurements were carried out in a Montana Instruments Workstation  operating at a base temperature of ~\SI{4.2}{\kelvin}. 
A home-built piezoelectric scanner with an in-plane range of approximately 
\SI{200}{\micro\meter} was used to scan the sample.
During the measurement, the SQUID sensor was positioned about \SI{1}{\micro\meter} above the sample surface. 
Larger lateral and vertical displacements were achieved using a stack of cryogenic coarse-positioning stages. 
The scanning SQUID sensor employed the same gradiometric design described in \cite{Huber2008}, featuring a pickup loop with an inner diameter of \SI{1.5}{\micro\meter}. 
The SQUID was coupled to a SQUID-array amplifier mounted on the cold plate of the cryostat, which provided low-noise amplification. 
Further details of the readout scheme are given in \cite{Huber2008}. 
The sample was mounted on a stage equipped with a separate heater and thermometer. 
Because of the weak thermal connection between the stage and the cryostat, the sample could be heated to \SI{20}{\kelvin} without significantly raising the cryostat temperature.

\subsection{Feedback control of applied magnetic field}
Magnetic field is applied to samples using a resistive solenoid wound on an aluminum cylinder that fits around the cryostat. The current $I$ through the coil is controlled using a current source. In addition to the field generated by the coil, several other sources of magnetic field contribute to the total field at the sample. Static contributions include the Earth’s magnetic field and nearby magnetic components, such as a stainless-steel tabletop and mounting hardware. Time-dependent background fields arise from superconducting magnets operating in neighboring labs on the same floor. For this reason, we use closed-loop control of the out-of-plane magnetic field to actively compensate for the variations in the out-of-plane background field.

To enable closed-loop control, we use a fluxgate sensor mounted on top of the cryostat centered approximately 10 cm above the sample to monitor the magnetic field in real time. The sensor is aligned such that it detects the magnetic field along the out-of-plane direction. In the following, we will refer to the out-of-plane magnetic field as magnetic field for brevity.

To control the magnetic field at the sample, we need to relate the magnetic field read at the sensor and sample which we describe with the following equations,
\begin{equation}
    B_{sample} = \alpha I+B_{static\_sample}+B_{dynamic}
\end{equation}
\begin{equation}
    B_{sensor} = \beta I+B_{static\_sensor}+B_{dynamic}
\end{equation}
where $\alpha$/$\beta$ and $B_{static\_sample}$/$B_{static\_sensor}$ denote the magnet constant and static background field at the sample/sensor location, and $I$ is the current in the coil. $B_{dynamic}$ is the time-varying contribution to the background field, which we assume to be the same at the sensor and sample location given that the sources for this contribution are at least 10 m away. The difference between $B_{static\_sample}$ and $B_{static\_sensor}$ is caused by magnetic parts in or near the cryostat. With these equations, we can rewrite $B_{sample}$ as,
\begin{equation}
\begin{split}
\label{eq:gammaDelta}
    B_{sample} & = B_{sensor}+(\alpha-\beta)I+B_{static\_sample}-B_{static\_sensor}\\
    & = B_{sensor}+\gamma I+\Delta
\end{split}
\end{equation}
with $\gamma$ defined as $\alpha-\beta$ and $\Delta$ defined as $B_{static\_sample}-B_{static\_sensor}$. Note that we expect $\alpha>\beta$ because the sample is approximately at the center of the solenoid. The calibration of $\gamma$ and $\Delta$ is described in the following paragraphs. With $B_{sensor}$ updating every \SI{10}{\second}, the feedback controller adjusts the magnet current to maintain the applied magnetic field at the sample at the target value.

\subsection{Calibration of magnet constants}
Eq.~\ref{eq:gammaDelta} can be rewritten as
\begin{equation}
    B_{sample}-B_{sensor} = \gamma I+\Delta.
\end{equation}
To determine $\gamma$ and $\Delta$, we measure $B_{sensor}$ and $B_{sample}$ as a function of the coil current $I$. The sensor directly provides $B_{sensor}$. We determine the magnetic field at the sample $B_{sample}$ by counting the number of vortices in a superconducting film, because the vortex density, $n$, is related to the applied magnetic field by,
\begin{equation}
    n = \frac{B_{sample}}{\Phi_0}
\end{equation}
where $\Phi_0$ is the magnetic flux quantum.

\begin{figure}[h]
    \centering
    \includegraphics[width=0.65\linewidth]{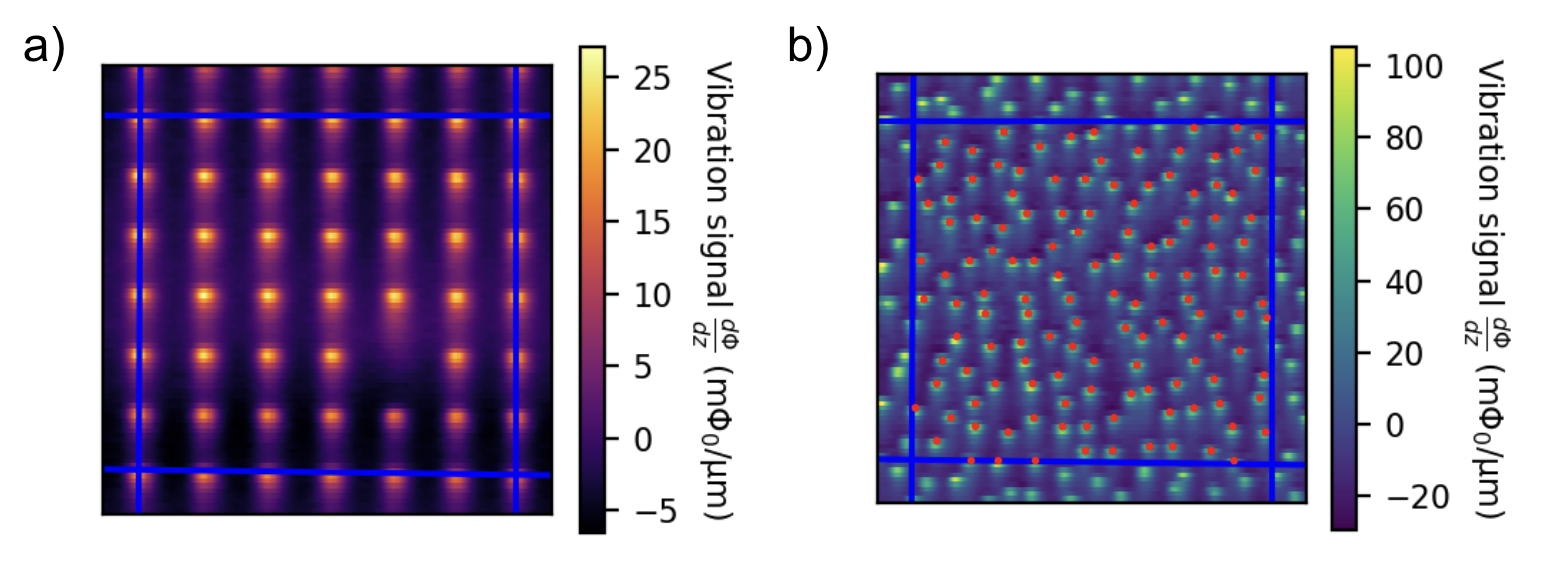}
    \caption{a) Image of vortices pinned at an array of anti-dots patterned in a NbTiN film. The square anti-dots are \SI{0.5}{\micro\meter} wide and have an edge-to-edge spacing of \SI{20}{\micro\meter} such that the distance between the blue lines, which are drawn through the center of the vortices trapped at the anti-dots, is \SI{123}{\micro\meter}, and hence the  area defined by the blue lines corresponds to $(\SI{123}{\micro\meter})^2$. b) Image of the \SI{2}{\micro\meter} thick NbTiN film cooled in a magnetic field and imaged below $T_c$. We determine that 129 vortices are within the $(\SI{123}{\micro\meter})^2$ area, labelled by red dots.}
    \label{fig:areacalibration}
\end{figure}

For this calibration measurement, we mount a \SI{2}{\micro\meter} thick NbTiN superconducting film next to a patterned NbTiN structure on the sample holder with a center to center distance of $\sim\SI{9}{\milli\meter}$. Experimentally, we control the voltages applied to the piezoelectric scanners, we therefore first image the patterned structure, an array of anti-dots, to calibrate the scan area. Fig.~\ref{fig:areacalibration} a) shows an image of the anti-dot array. The anti-dots are \SI{0.5}{\micro\meter} wide with an edge-to-edge spacing of \SI{20}{\micro\meter}. Vortices pinned at the anti-dots therefore define a grid, and the blue lines drawn through vortex centers are spaced \SI{123}{\micro\meter}. The square enclosed by these lines therefore defines a calibrated area of $(\SI{123}{\micro\meter})^2$.

\begin{figure}[h]
    \centering
    \begin{minipage}{0.45\textwidth}
    \includegraphics[width=\linewidth]{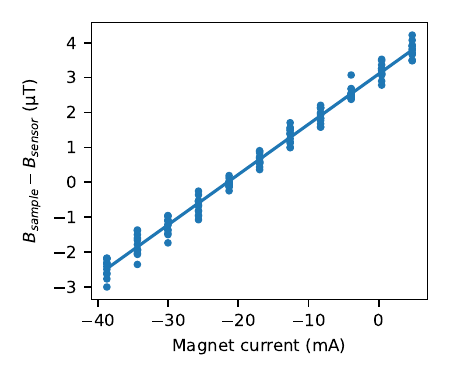}
    \end{minipage}
    \hfill
    \begin{minipage}{0.5\textwidth}
    \caption{\justifying
    Combined data from current series measurements on the 13 different locations. The fitted slope is $\gamma$ and the y interception is $\Delta$.
    }
    \label{fig:alphafitting}
    \end{minipage}
\end{figure}

After calibrating the scan area, we image the unpatterned NbTiN film after cooling it through $T_c$ in a series of currents applied to the coil, corresponding to fields between approximately \SI{-20}{\micro\tesla} and \SI{20}{\micro\tesla} at the sample. The scanning parameters, including the size of the total scan window, the number of lines and scanning speed, are kept identical to those used for the patterned sample to ensure that the calibrated scan area is valid. Using identical scan parameters avoids small variations in scan size that can arise from piezo hysteresis and creep when different voltage ranges and scan rates are used.

For each image, vortices whose centers fall within the calibrated $(\SI{123}{\micro\meter})^2$ area are counted to determine the vortex density and hence $B_{sample}$. The total scan window is chosen to be slightly larger than the counting region to allow identifying whether a vortex is inside it. A representative image from the current series is shown in Fig.~\ref{fig:areacalibration} b), where 129 vortices are identified within the counting region.

To reduce the effect of local variations in vortex density due to defects or film inhomogeneity, the measurement series is repeated at 13 different locations on the film. The combined data from all measurements are shown in Fig.~\ref{fig:alphafitting}. Fitting Eq.~\ref{eq:gammaDelta} yields $\gamma=\SI[parse-numbers=false, per-mode=symbol]{144\pm9}{\tesla\per\ampere}$, and a background field difference, $\Delta=\SI[parse-numbers=false]{3.11\pm0.06}{\micro\tesla}$. The uncertainty is dominated by  the uncertainty in determining the counting area (limited by pixelation) and by uncertainty of the fit.

We also determine $\gamma$ by individually measuring $\alpha$ and $\beta$ and taking their difference. To obtain the magnet constant $\beta$ at the sensor, we vary the coil current and record the corresponding sensor reading. The resulting linear dependence yields a value of \SI[per-mode=symbol]{690}{\micro\tesla\per\ampere}. The measurement is completed in under one minute, and performed while the background field is stable. The magnet constant at the sample is measured in an analogous manner by placing the fluxgate sensor at the sample position, yielding \SI[parse-numbers=false, per-mode=symbol]{835\pm5}{\micro\tesla\per\ampere}. The uncertainty mainly comes from the uncertainty of the sensor's position along the z direction. From these measurements, we obtain $\gamma=\SI[parse-numbers=false, per-mode=symbol]{145\pm5}{\micro\tesla\per\ampere}$, in good agreement with the value determined by vortex counting.

Although $\gamma$ determined by vortex counting has a relative uncertainty of ~6\%, it accounts for only a fraction of the total value of  $B_{sample}$. The larger contribution comes from $B_{sensor}$, which is measured with high precision. Using $\alpha=\beta+\gamma$ yields $\alpha=\SI[parse-numbers=false, per-mode=symbol]{834\pm9}{\micro\tesla\per\ampere}$ where the uncertainty is dominated by the uncertainty in $\gamma$. This corresponds to an approximate 1\% relative uncertainty in the magnetic field at the sample for a fixed current in the coil.

\section{Magnetic field screening between strips}
The strips are arranged with a spacing of \SI{10}{\micro\meter} (see Fig.~1 a in the main text). Each strip screens the applied magnetic field and therefore modifies the field experienced by neighboring strips. This screening is described by the London equations and persists up to the threshold field. Once vortices form in a strip, vortices carry part of the applied flux, reducing the Meissner screening currents. To estimate the impact of screening, we perform simulations using a software package that solves the London equations in two dimensions (Superscreen, \cite{bishop-van_horn_superscreen_2022}).
\begin{figure}[h]
    \centering
    \begin{minipage}{0.75\textwidth}
    \includegraphics[width=\linewidth]{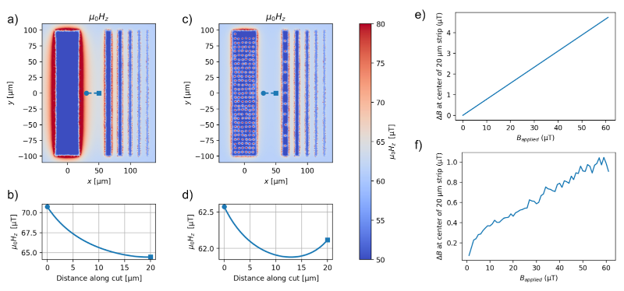}
    \end{minipage}
    \hfill
    \begin{minipage}{1\textwidth}
    \caption{\justifying
    Finite-element analysis of the screening effect of neighboring strips using the superscreen Python package \cite{bishop-van_horn_superscreen_2022}. A low temperature London penetration depth of \SI{460}{\nano\meter} is assumed in all simulations. a) Simulated magnetic field for an applied field of \SI{61}{\micro\tesla} with the \SI{20}{\micro\meter} strip removed and no vortices in the remaining strips. b) Magnetic field profile across the width of the \SI{20}{\micro\meter} strip along the dotted line in a). c, d) Simulations analogous to a, b) but including vortices in the strips. The number of vortices in each strip is chosen to match that observed experimentally at \SI{61}{\micro\tesla}, and for simplicity the vortices are assumed to form a square lattice. e, f) Magnetic field at the center of the \SI{20}{\micro\meter} strip as a function of applied field, calculated without and with vortices in the neighboring strips. In f), the initial steep slope arises from the threshold field of the \SI{40}{\micro\meter} strip; the \SI{40}{\micro\meter} strip screens more strongly below this value than above.
    }
    \label{fig:Screen460}
    \end{minipage}
\end{figure}

\begin{figure}[h]
    \centering
    \begin{minipage}{0.75\textwidth}
    \includegraphics[width=\linewidth]{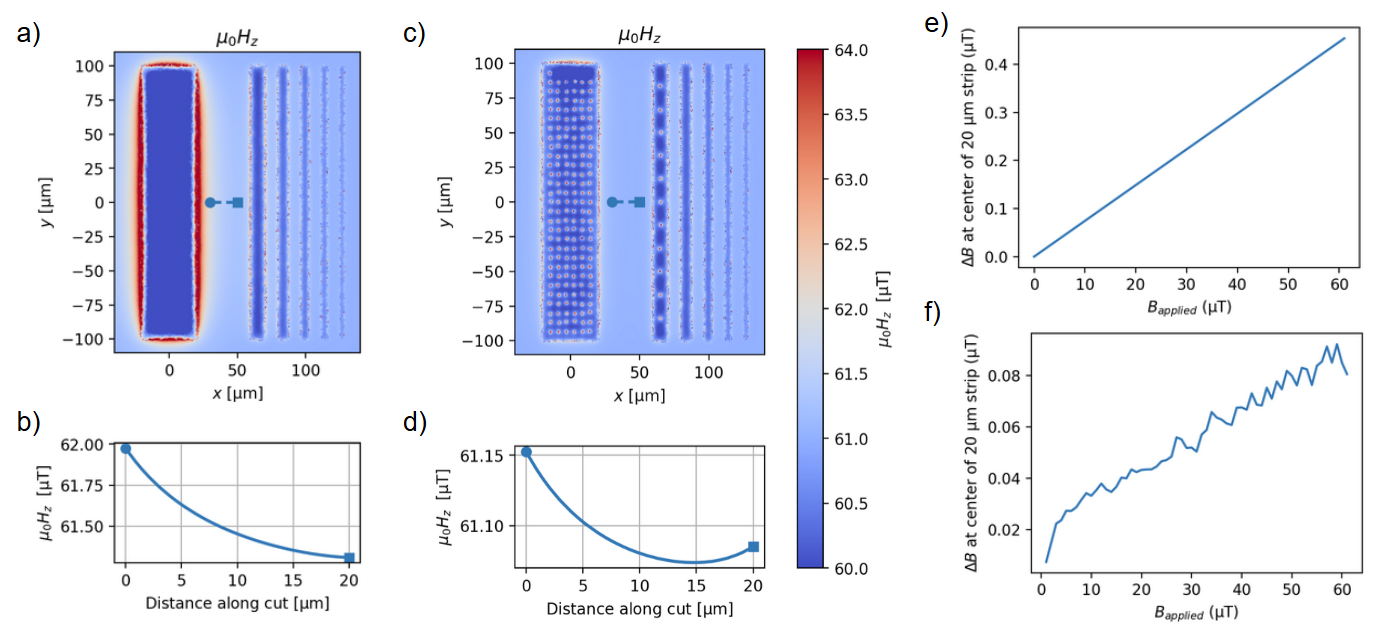}
    \end{minipage}
    \hfill
    \begin{minipage}{1\textwidth}
    \caption{\justifying
    Same simulations as in Fig.~\ref{fig:Screen460} but assuming $\lambda\sim\SI{2060}{\nano\meter}$, corresponding to the estimated London penetration depth at $T=0.95T_c$.
    }
    \label{fig:Screen2060}
    \end{minipage}
\end{figure}

In Figs.~\ref{fig:Screen460} and \ref{fig:Screen2060}, we show simulations of how the field at the position of the \SI{20}{\micro\meter} strip is modified by the presence of all neighboring strips. To isolate this effect, we remove the \SI{20}{\micro\meter} strip from the simulation, and extract the field values along its width. Fig.~\ref{fig:Screen460} shows the results at \SI{61}{\micro\tesla} applied field, assuming a London penetration of $\lambda=\SI{460}{\nano\meter}$, corresponding to a Pearl length of $\Lambda=\SI{8464}{\nano\meter}$ for a thin film with thickness $d=\SI{50}{\nano\meter}$ using $\Lambda=2\lambda^2/d$. This value corresponds to the low temperature penetration depth, $\lambda_0$, estimated from resistivity and $T_c$. We consider two cases: (i) all strips in the Meissner state (no vortices), and (ii) a more realistic configuration in which neighboring strips host vortices with densities matching those observed in the SSM measurements. In Fig.~\ref{fig:Screen460}, we repeat the same simulations for a larger penetration depth, which is more relevant, as the field at the temperature at which vortices become immobile determines the trapped vortex configuration. For example, at $T=0.95T_c$, we estimate $\lambda\sim\SI{2060}{\nano\meter}$ assuming $\lambda_0=\SI{460}{\nano\meter}$ and Ginzburg-Landau temperature dependence $\lambda(T)=\lambda_0(1-\frac{T}{T_c})^{-0.5}$.

In all cases, the field across the strip width is slightly non-uniform and enhanced compared to the applied field of \SI{61}{\micro\tesla}. The correction is appreciable ($\sim 8\%$) only when assuming $\lambda_0$ and the absence of vortices, which represents an unrealistic limit. In all other cases, the correction is below 2\% and falls below 0.3\% for the most realistic scenario, which accounts  for the increased penetration depth at the temperature where vortices become immobile and includes vortices in neighboring strips. We obtain similar results for other strip widths, with even smaller corrections due to the geometry: while the spacing is fixed at \SI{10}{\micro\meter}, the neighboring strips become narrower.

We also note that the correction is largely linear in the applied field. The only noticeable non-linearity in Figs.~\ref{fig:Screen460}f and \ref{fig:Screen2060}f occurs when crossing the threshold field for vortex entry into the wider (\SI{40}{\micro\meter}) neighboring strip, which happens at a lower field lower than vortex entry into the \SI{20}{\micro\meter} strip. This behavior repeats systematically for strips with smaller width.

When the \SI{20}{\micro\meter} strip is included in the simulation, it screens the field onto neighboring strips, which in turn modified the screened field on the \SI{20}{\micro\meter} strip. These higher-order screening effects are small and neglected in the analysis shown in Figs.~\ref{fig:Screen460} and \ref{fig:Screen2060}.

In summary, we conclude that the effect of screening is small and does not significantly impact the shape of the curves or the conclusions drawn from them.

\section{Analytical and Numerical Studies of Pearl Vortex Trapping}
\label{app:InfiniteStrip}

In this section, we derive 
the two theoretical models used in the main text to analyze the experimental data and extract key properties of NbTiN.
All superconducting strips considered here are in the two-dimensional limit, with film thickness $d\ll\lambda$ (the London penetration depth).
In this regime, magnetic flux penetrates in the form of Pearl vortices \cite{Pearl1964}. 
The vortex core radius $r_c$ is of order $r_c\sim\xi$ (the coherence length), while the strip width $W$ and the Pearl length $\Lambda=2\lambda^2/d$ satisfy $r_c\ll W\ll\Lambda$.
These conditions allow vortices to be modeled as point-like particles. 

We remark on similarities and differences with previous theoretical works.
The problem of finding Pearl vortex configurations in thin films ($L\times W\times d$ with $d\rightarrow0$) has similarities to the problem of Abrikosov vortex configurations in very tall rectangular pillars ($a\times b\times \infty$), as studied in Sardella \textit{et al.} \cite{sardella_matching_1999}.
We compare our $W\times\infty$ analytic threshold field, Eq. (5), with the finite-$W\times L$ London result of Sardella \textit{et al.}, Eq. (22). For $L=\SI{200}{\micro\meter}$, $W=\SI{20}{\micro\meter}$, $r_c=\SI{200}{\nano\meter}$, and $\lambda=\SI{100}{\micro\meter}$, these expressions yield $B_L\approx\SI{13.7}{\micro\tesla}$ and $B_c\approx\SI{14.04}{\micro\tesla}$ with $\sim2\%$ difference, indicating that finite-length corrections are negligible for the threshold field at $L/W=10$.
In other words, the threshold field for a $\SI{20}{\micro\meter}\times\infty\times\SI{0.05}{\micro\meter}$ film is practically the same as threshold field for a $\SI{20}{\micro\meter}\times\SI{200}{\micro\meter}\times\infty$ pillar.
However, \textit{Sardella et al.} did not find solutions with three or more rows of vortices, possibly due to an overly restrictive ansatz.  
For pinning-free systems, we employ an ansatz that allows multiple rows/columns of vortices following Bronson \textit{et al.} \cite{Bronson2006}, and we find that the number of rows/columns increases with increasing magnetic field, whereas Sardella \textit{et al.} focus on linear-chain configurations.
In addition, Sardella \textit{et al.} regularize the London self-energy divergence using a short-distance cutoff equal to $\xi$, while we use the value $r_c\approx0.966\xi$. Finally, Sardella \textit{et al.} do not consider pinning effects, which play a central role in our trapping/freezing analysis.

de Souza Silva \textit{et al.} \cite{de_souza_silva_flux_2001},  investigated vortex penetration dynamics controlled by surface barriers using overdamped Langevin equations, where the first-entry field is defined via a local force-balance criterion at distances of order $\xi$.  Their model was appropriate for studying hysteresis in $M(H)$ curves, where the superconducting strip is held at a fixed temperature and the field is slowly cycled. In contrast, our experiments involved field cooling, i.e., the field was held constant while temperature was reduced. We assume that cooldown is so slow that the vortex configuration remains in thermal equilibrium until the temperature drops below a certain "vortex freezing temperature".

Previous numerical studies based on self-consistent Ginzburg-Landau and time-dependent Ginzburg-Landau simulations have investigated vortex configurations, row instabilities, and driven dynamics in laterally confined superconductors [e.g., Karapetrov \textit{et al.} \cite{karapetrov_transverse_2009}; McNaughton \textit{et al.} \cite{mcnaughton_causes_2022}]. These approaches explicitly resolve vortex-core structure and nonequilibrium dynamics and are particularly relevant for nanoscale systems and strong confinement regimes. In contrast, our work focuses on micrometer-scale strips deep in the London-Pearl limit ($\Lambda\gg W$), where vortices can be treated as pointlike objects and a quasi-static free-energy description is appropriate. By combining analytical Pearl-vortex energetics with Monte Carlo simulated annealing in the presence of pinning, we isolate the thermodynamic landscape governing vortex trapping, row formation, and stability during field cooling, providing a complementary perspective to GL-based simulations.

In thin strips, vortices are influenced by both the applied magnetic field and edge screening effects \cite{Kogen1994}. 
We extend earlier work on infinitely long strips \cite{Kogan2007,Bronson2006} to the case of semi-infinite strips bounded by three straight edges. We stress here that the exact value of Pearl length $\Lambda$ is not relavent to the results of our models. Although the Pearl length sets the overall energy scale, it cancels from the threshold condition and the dimensionless formulation used here; stability and trapping are therefore governed primarily by the strip geometry, consistent with recent London-limit analyses.

We provide detailed derivations of the vortex self-energy $F_{\text{self}}$, vortex magnetic moment $m$, and vortex-vortex interaction energy $F_{vv}$ in both geometries, which may be valuable to some readers.
For clarity, all notation---covering material parameters, energies, fields, and simulation variables---is summarized in Table \ref{tab:Notations}.

\clearpage
\setlength{\LTpre}{0pt}
\setlength{\LTpost}{0pt}
\noindent
\begin{longtable}{l@{\hspace{1.cm}}l}
    \caption{Notations used in the paper and this supplement.} 
    \label{tab:Notations} \\
    \hline
    \hline
    \textbf{Symbol} & \textbf{Meaning} \\
    \hline
    \endfirsthead
    
    \multicolumn{2}{c}{\tablename~\thetable{} -- continued from previous page}\\
    \hline
    \hline
    \textbf{Symbol} & \textbf{Meaning}\\
    \hline
    \endhead

    \hline 
    \multicolumn{2}{r}{\textit{Continued on next page}}\\
    \endfoot

    \hline
    \hline
    \endlastfoot
    
    \multicolumn{2}{l}{\textbf{Experimental samples}} \\
    \hline
        S1 & Sample taken from middle region between center and edge of the \SI{300}{\milli\meter} wafer; \\
            & $~~~\rho\approx\SI{250}{\micro\ohm\centi\meter}$, $T_c\approx \SI{13.5}{\kelvin}$\\
        S2 & Sample taken from center of the \SI{300}{\milli\meter} wafer; \\ 
            & $~~~\rho\approx\SI{290}{\micro\ohm\centi\meter}$, $T_c\approx \SI{12.6}{\kelvin}$\\
    \hline
    \multicolumn{2}{l}{\textbf{Fields and normalizations}} \\
    \hline
        $B$ & Applied magnetic field (normal to film, \unit{\micro\tesla}) \\
        $B_L=\dfrac{2\Phi_0}{\pi W^2}\ln\left(\dfrac{2W}{\pi r_c}+1\right)$ & Threshold field (Likharev criterion \cite{Likharev1971}: vortex Gibbs free energy $G_v\le0$ at strip center) \\
        $B_{C}=\dfrac{\pi\Phi_0}{4W^2}$ & Threshold field (Clem criterion \cite{Clem1998}: first local minimum of $G_v$) \\
        $B_K=1.65\,\dfrac{\Phi_0}{W^2}$ & Threshold field (Kuit \emph{et al.} model \cite{Kuit2008}) \\
        $B_m$ & Experimentally measured maximum field at which cooling yields no trapped vortices \\
        $\mathbf{A}$ & Vector potential \\
        $\mathbf{K}$ & Sheet current \\
        $\mathcal{M}=m/\mathcal{A}$ & Areal magnetization \\
        $\Omega$ & Areal vorticity \\
        $m$ & Vortex magnetic moment \\
        $B_0=\Phi_0/W^2$ & Characteristic field scale set by strip width $W$ \\
        $\tilde{B}=B/B_0$ & Dimensionless magnetic field \\
    \hline
    \multicolumn{2}{l}{\textbf{Theoretical models}} \\
    \hline
        VL & Vortex lattice without pinning, modeled for \textbf{infinite length strips}\\
        SA & One-dimensional Monte Carlo simulated annealing with pinning, \\
        & modeled for \textbf{semi-infinite length strips}\\
    \hline
    \multicolumn{2}{l}{\textbf{Energies and free energies}} \\
    \hline
        $F_{\text{self}}$ & Self-energy of a single vortex, including core and edge effects \\
        $F_{mB} = mB$ & Magnetic moment-field coupling energy \\
        $G_v = F_{\text{self}} - F_{mB}$ & Gibbs free energy of a single vortex \\
        $F_{vv}$ & Vortex–vortex interaction free energy \\
        $\Delta F_{\text{self}}=F_{\text{self}}^{\text{hole}}-F_{\text{self}}^{\text{film}}$ & Self-energy change due to a circular hole of radius $r_h$\\
        $\Delta F_{mB}=F_{mB}^{\text{hole}}-F_{mB}^{\text{film}}$ & Change in moment-field coupling due to a hole/defect\\
        $G_{\text{pin}} = \Delta F_{\text{self}} - \Delta F_{mB}$ & Pinning Gibbs free energy, defined as the energy shift caused by defect or hole \\
        $G_{\text{tot}} = G_v + F_{vv}$ & Total Gibbs free energy of a vortex (no pinning) \\
        $G_{\text{tot+pin}} = G_{\text{pin}} + G_v + F_{vv}$ & Total Gibbs free energy of a vortex (with pinning) \\
    \hline
    \multicolumn{2}{l}{\textbf{Auxiliary functions}} \\
    \hline
        $\mathrm{H}(Y)$ & Heaviside step function: $\mathrm{H}(Y)=0$ for $Y\leq 0$, $\mathrm{H}(Y)=1$ for $Y>0$ \\
        $\mathrm{sgn}(Y)$ & Sign function: $\mathrm{sgn}(Y)=-1,0,+1$ \\
    \hline
    \multicolumn{2}{l}{\textbf{Geometry and material parameters}} \\
    \hline
        $X,Y$ & In-plane coordinates of a vortex\\
        $W$ & Strip width (\unit{\micro\meter}) \\
        $L$ & Strip length (\unit{\micro\meter}) \\
        $\mathcal{A} = WL$ & Total area of the superconducting strip (\unit{\micro\meter\squared}) \\
        $d$ & Strip thickness (\unit{\nano\meter}) \\
        $\lambda$ & London penetration depth (\unit{\micro\meter})\\
        $\Lambda=2\lambda^2/d$ & Pearl length, i.e., the thin-film screening length (\unit{\micro\meter}) \\
        $\xi$ & Coherence length (\unit{\nano\meter})\\
        $r_c$ & Vortex core radius, treated as a fit parameter (typically of order $\xi$) \\
        $r_h$ & Radius of a circular hole/defect (pinning site, \unit{\nano\meter}) \\
        $\theta$ & Phase of the superconducting order parameter \\
        $\Upsilon_{2\text{D}}=\frac{2}{\mu_0\Lambda}$ & 2D superfluied stiffness \\
        $T_c$ & Critical temperature (\unit{\kelvin}) \\
        $\rho$ & Resistivity at room-temperature (\unit{\micro\ohm\centi\meter}) \\
    \hline
    \multicolumn{2}{l}{\textbf{Dimensionless variables and lattice geometry}} \\
    \hline
        $X_i,Y_i$ & Physical coordinates of vortices within the unit cell (motif) \\
        $\tilde{X}=X/W,~\tilde{Y}=Y/W$ & Dimensionless coordinates \\
        $u\in\mathbb{Z}$ & Unit-cell index in periodic sums \\
        $\tilde{g}_c=\tilde{G}/\tilde{a}=(\tilde{G}_v+\tilde{F}_{vv})/\tilde{a}$ & Dimensionless Gibbs free energy density (dimensionless unit cell area is $(aW)/W^2\Rightarrow\tilde{a}$).\\
        $E_0=\Phi_0^2/(\mu_0 W)$ & Energy scale \\
        $G_c=\tilde{g}_cE_0$ & Gibbs free energy per unit cell \\
        $a$ & Physical lattice period along strip length, i.e., the unit-cell size in the $y$ direction \\
        $\tilde{a}=a/W$ & Dimensionless lattice spacing normalized by strip width \\
        $c$ & Number of vortex columns across the width strip \\
        $\tilde{n}_{\tilde{a}}=1/\tilde{a}=W/a$ & Dimensionless \emph{unit cell density} (used for counting vortices along the strip; $n_a=1/(aW)$) \\
        $\tilde{n}_c=c/\tilde{a}=cW/a$ & Dimensionless \emph{vortex density}, normalized by strip width squared $W^2$ \\
        $n_c=c/(aW)$ & \emph{vortex number per unit cell} (vortex density in the strip) \\
        $n_v=\tilde{n}_c L/W$ & Total number of vortices in a finite strip of width $W$ and lengtn $L$ \\
    \hline
    \multicolumn{2}{l}{\textbf{Simulated annealing and simulation parameters}} \\
        \hline
        $l_0=\SI{1}{\micro\meter}$ & Reference length scale \\
        $B_1=\Phi_0/l_0^2=\SI{2067.834}{\micro\tesla}$ & Magnetic field scale \\
        $E_0=\Phi_0^2/(\mu_0 l_0)$ & Characteristic energy scale \\
        $N$ & Number of discretized sites along the strip centerline\\
        $E_{\text{pin}}$ & Pinning energy per vortex per site, drawn from a distribution (unit: $\Phi_0^2/(\mu_0\Lambda)$) \\
        $p$ & Probability weight assigned within the pinning-energy distribution \\
        $kT$ & Thermal energy used in Metropolis updates \\
         & ~~~ (with annealing schedule uses $k_BT\to 0.997~k_BT$ per step) \\
        $\Delta G$ & Change in local Gibbs free energy used in Metropolis acceptance updates \\
        $n_y[i]\in\{0,1\}$ & Binary site-occupation variable (0 empty, 1 occupied) \\
        $t_{\text{max}}$ & Total number of Monte Carlo steps per run\\
        $n_v$ & Total number of vortices in a given realization (MC run) \\
    \hline
    \multicolumn{2}{l}{\textbf{Constants}} \\
    \hline
        $\Phi_0 = h/2e$ & Flux quantum, with $h$ the Planck constant and $e$ the elementary charge \\
        $\mu_0$ & Vacuum permeability (\unit{\henry\per\meter}) \\
        $\alpha\approx0.4123$ & Near-core slope coefficient of the GL order parameter, evaluated in the limit $(\Lambda/\xi)\to\infty$\\
    \hline
\end{longtable}

\subsection{Gibbs Free Energy in an Infinite Length Strip}\label{sec:InfiniteStrip}

\begin{summarybox}[Summary --- Gibbs Free Energy in an Infinite Length Strip]
In an infinitely long superconducting strip of width $W$ ($0<x<W$), the Gibbs free energy of a Pearl vortex at this level of modeling is 
\begin{eqnarray}
    G_{\text{tot}} \;=\; G_v + F_{vv} 
        = F_{\text{self}} - mB + F_{vv}\nonumber
\end{eqnarray}
where 
    the \textbf{self-energy} $F_{\text{self}}(X)$ is
    \begin{eqnarray}
        F_{\text{self}}(X)
        =\frac{\Phi_0^2}{2\pi\mu_0\Lambda} 
            \ln\left[
                \frac{2W}{\pi r_c}
                \sin\left(\frac{\pi X}{W}\right) 
                + 1 
            \right],\nonumber
    \end{eqnarray}
    the \textbf{vortex magnetic moment} $m(X)$ is 
    \begin{eqnarray}
        m(X)=\frac{\Phi_0}{\mu_0\Lambda}X(W - X),\nonumber
    \end{eqnarray}
    and the \textbf{vortex-vortex interaction free energy} $F_{vv}(x,y;X,Y)$ is
    \begin{eqnarray}
        F_{vv}(x,y;X,Y)
        =\frac{\Phi_0^2}{2\pi\mu_0\Lambda} 
            \ln\left[
                \frac{\cos\left(\tfrac{\pi(x+X)}{W}\right)
                     -\cosh\left(\tfrac{\pi(y-Y)}{W}\right)}
                     {\cos\left(\tfrac{\pi(x-X)}{W}\right)
                     -\cosh\left(\tfrac{\pi(y-Y)}{W}\right)} 
            \right].\nonumber
    \end{eqnarray}
$G_\text{tot}$ includes vortex core free energy, edge screening, magnetic coupling to the external field $B$, and vortex-vortex interactions. 
This expression is our starting point for analyzing field thresholds, vortex stability, and interactions in infinite-length strips.
\end{summarybox}

\subsubsection{Single-Vortex Self-Energy \texorpdfstring{$F_{\text{self}}$}{Fself} (Edges and Core)}\label{sec:InfiniteStripSelfEnergy}

\begin{summarybox}[Summary — Single-Vortex Self-Energy]
We compute the position-dependent self-energy $F_{\text{self}}(X)$ of a Pearl vortex in an \emph{infinitely long} superconducting strip of width $W$
using an electrostatic analogy:
\begin{eqnarray}
    F_{\text{self}}(X)=\frac{\Phi_0^2}{2\pi\mu_0\Lambda} 
    \ln\left[ 
    \frac{2W}{\pi r_c}\,\sin\left(\frac{\pi X}{W}\right) + 1 \right],\nonumber
\end{eqnarray}
This is the free energy of the vortex, including the logarithmic core contribution and the edge-screening term.  The ``+1'' is added to prevent spurious negative logarithmic divergences at the strip edges.  The expression vanishes smoothly at $X=0$ and $X=W$, and grows logarithmically away from the edges. 
\end{summarybox}


We will assume that the superconducting film obeys the London relation
$
    \mathbf{K}=\Upsilon_{2\mathrm{D}}
    \left(\frac{\Phi_0}{2\pi}\nabla\theta-\mathbf{A}\right)
$,
where $\mathbf{K}$ is sheet current, $\theta$ is the phase, $\Upsilon_{2\mathrm{D}} = \frac{2}{\mu_0\Lambda}$ is the 2D superfluid stiffness, and $\Lambda$ is the Pearl length.
If there is no screening and no applied field ($\mathbf{A}=\mathbf{0}$),
\begin{eqnarray}
    \mathbf{K} 
    =\frac{\Phi_0}{\pi\mu_0\Lambda}\, \nabla \theta.
\end{eqnarray}
By analogy with fluid dynamics, we define the areal vorticity as the curl of the sheet current, $\Omega\hat{\mathbf{z}}=\nabla\times\mathbf{K}$.  Suppose a vortex is located at 2D position $\mathbf{S}$, so that the phase increases by $\Delta\theta=2\pi$ going around the vortex core.  Integrating over a region containing the vortex and applying Stokes' law gives
\begin{eqnarray}
 \iint d^2s~ \Omega(s) = \oint \mathbf{ds}\cdot\mathbf{K} 
= \frac{\Phi_0}{\pi\mu_0\Lambda} \Delta\theta 
= \frac{2\Phi_0}{\mu_0\Lambda}.   
\end{eqnarray}
Therefore we may write
\begin{eqnarray}
    \Omega(\mathbf{s}) 
    =\frac{2\Phi_0}{\mu_0\Lambda}\,
    \delta^{(2)}(\mathbf{s}-\mathbf{S}).
\end{eqnarray}

Instead of the phase $\theta$, it is often more convenient to work with the areal magnetization $\mathcal{M}$ (also known as ``current potential'' or ``stream function'' \cite{Kogen1994,Kogan2007}) such that the current density is $\mathbf{K}=\nabla\times (\mathcal{M} \hat{\mathbf{z}})$.  Taking the curl of both sides then gives
$\Omega\hat{\mathbf{z}} = \nabla\times(\nabla\times \mathcal{M} \hat{\mathbf{z}})$.
Vector calculus identities reduce this to $-\nabla^2 \mathcal{M} = \Omega$.
Therefore, in a superconducting film with a vortex at position $(X,0)$, the areal magnetization satisfies Poisson's equation
\begin{eqnarray}
-\nabla^2 \mathcal{M} =\frac{2\Phi_0}{\mu_0\Lambda} \delta(x-X) \delta(y)
\label{eq:InfiniteStripPoissonEq}
\end{eqnarray}
with Dirichlet boundary conditions $\mathcal{M}=0$ along the film boundary.  In other words, $\mathcal{M} = \frac{2\Phi_0}{\mu_0\Lambda} G(x,y;X,0)$ where $G(x,y;X,0)$ is the Dirichlet Green function in the infinite strip.  This is analogous to a steady-current-flow problem with current injected at point $(X,0)$ in a resistive strip.  It is also analogous to an electrostatic problem with a line charge of strength $+\omega$ at $(X,0)$ confined between two grounded conducting planes at $x=0$ and $x=W$.  In both these situations, the potential $V(x,y) \equiv G(x,y;X,0)$ obeys
\begin{eqnarray}
    -\nabla^2V(x,y)=\delta(x-X)\,\delta(y),
    \qquad V(x,y)=0 \;\; \text{if } x=0 \text{ or } x=W.
\label{eq:InfiniteStripPoissonEqDimless}
\end{eqnarray}

\begin{figure}[h]
    \centering
    \begin{minipage}{0.45\textwidth}
    \includegraphics[width=\linewidth]{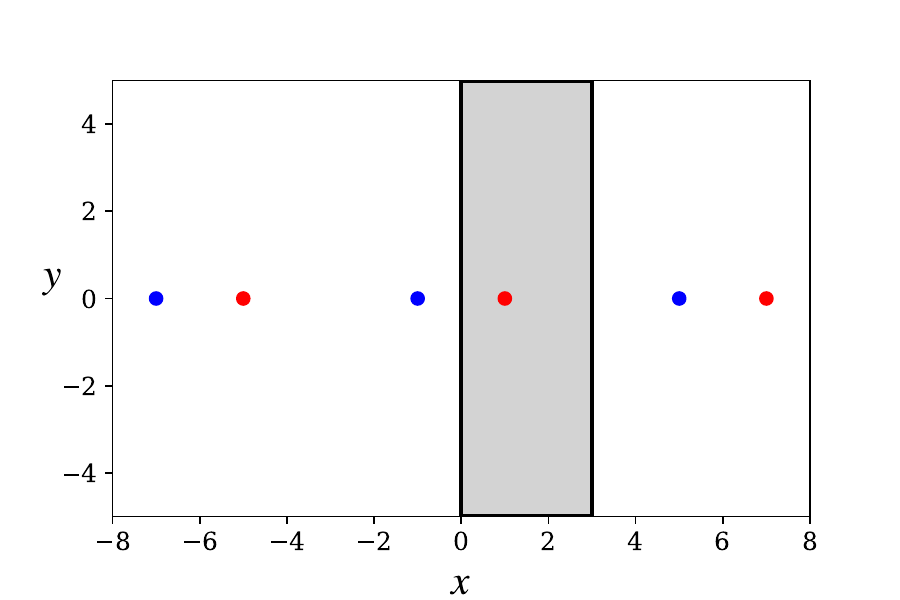}
    \end{minipage}
    \hfill
    \begin{minipage}{0.5\textwidth}
    \caption{\justifying
    Illustration of the method of image charges used to solve the 2D Poisson equation in a strip geometry. 
    The infinite strip of width $W=3$ is shown as the shaded region between $x=0$ and $x=3$.
    Red dots represent positive charges, and blue dots denote negative charges of equal magnitude. 
    This alternating construction guarantees that the potential vanishes at both strip boundaries.
    }
    \label{fig:InfiniteStrip}
    \end{minipage}
\end{figure}

The method of image charges provides a systematic way to satisfy the Dirichlet boundary conditions at the strip edges. 
For each source at position $X$, an infinite sequence of image charges is generated outside the interval $0<x<W$. 
Two families of images appear
\begin{eqnarray}
    X_n^{\pm}=2nW\pm X,\qquad
    \omega_n^{\pm}=\pm1,\qquad n\in\mathbb{Z},
\end{eqnarray}
where $\omega_n^{\pm}=\pm1$ denotes the sign of the each image charge. 
This construction produces a periodic array of alternating positive and negative charges along the $x$-axis, ensuring that the potential vanishes at both boundaries ($x=0$ and $x=W$). 
The potential inside the strip is then obtained by summing the contributions from the original source at $(X,0)$ together with all of its images. 
An illustration is shown in Fig. \ref{fig:InfiniteStrip} for $W=3$ and $X=1$.

Using the 2D Green function $\frac{1}{2\pi}\ln{\frac{1}{s}}$, the total potential is 
\begin{eqnarray}
    V(x,y)&&=\sum_{n=-\infty}^{\infty}\sum_{\alpha=\pm1}\frac{\alpha}{2\pi}
    \ln\frac{1}{\sqrt{\left(x-2nW-\alpha X\right)^2+y^2}}
    =\sum_{n=-\infty}^{\infty}\frac{1}{2\pi}
    \ln\frac{\sqrt{\left(x-2nW+X\right)^2+y^2}}
            {\sqrt{\left(x-2nW-X\right)^2+y^2}}\nonumber\\
    &&=\sum_{n=-\infty}^{\infty}\frac{1}{2\pi}\operatorname{Re}\left\{
    \ln\left[\frac{\left(x-2nW+X\right)+iy}{\left(x-2nW-X\right)+iy}\right]
    \right\}
    =\frac{1}{2\pi}\operatorname{Re}\left\{
    \ln\left[\prod_{n=-\infty}^{\infty}\frac{z-2nW+X}{z-2nW-X}\right]
    \right\}
    \label{eq:InfiniteStripV_Total}
\end{eqnarray}
where $z=x+iy$, $W$ is the width of the strip, and $X$ is the $x$-coordinate of the source charge.

Equation \eqref{eq:InfiniteStripV_Total} expresses the potential as an infinite product. 
To simplify this into a trigonometric form, we consider the auxiliary function
\begin{equation}
    g(z)=\prod_{\substack{k=-\infty\\ \,k\neq 0}}^\infty(1-\frac{z}{k}).
\end{equation}
By construction, $g(z)$ is an entire function with zeros at $z=\pm 1,\pm 2,\pm 3,...$ and with normalization $g(0)=1$. 
It is therefore equivalent to the well-known sine-product formula \cite{Whittaker_Watson_modern_2021}
\begin{eqnarray}
    g(z)=\text{sinc}(\pi z)=\frac{\sin(\pi z)}{\pi z}.
\end{eqnarray}

Using this identity, the infinite product in Eq. \eqref{eq:InfiniteStripV_Total} can be reduced using the sine formula. 
Splitting off the $n=0$ term and rescaling each factor by $2nW$, we obtain
\begin{eqnarray}\label{eq:sinc} 
    \prod_{n=-\infty}^\infty\frac{z-2nW+X-iY}{z-2nW-X-iY}
    &&=\frac{z+X-iY}{z-X-iY}
    \prod_{\substack{n=-\infty\\ n\neq 0}}^\infty
    \frac{z-2nW+X-iY}{z-2nW-X-iY}
    =\frac{z+X-iY}{z-X-iY}
    \prod_{\substack{n=-\infty\\ n\neq 0}}^\infty
    \frac{1-\frac{z+X-iY}{2nW}}{1-\frac{z-X-iY}{2nW}}\nonumber\\
    &&=\frac{z+X-iY}{z-X-iY}\times
    \frac{\text{sinc}\left(\frac{\pi(z+X-iY)}{2W}\right)}
         {\text{sinc}\left(\frac{\pi(z-X-iY)}{2W}\right)}
    =\frac{\sin\left(\frac{\pi(z+X-iY)}{2W}\right)}
          {\sin\left(\frac{\pi(z-X-iY}{2W}\right)}.
\end{eqnarray}
Substituting Eq. \eqref{eq:sinc} into Eq. \eqref{eq:InfiniteStripV_Total}, the total potential becomes
\begin{eqnarray}
    V(x,y)=\frac{1}{2\pi}\operatorname{Re}\left\{
    \ln\left[
    \frac{\sin\left(\frac{\pi(z+X)}{2W}\right)}
         {\sin\left(\frac{\pi(z-X)}{2W}\right)}
         \right]
    \right\}
    =\frac{1}{2\pi}\ln\left|
        \frac{\sin\left(\frac{\pi(z+X)}{2W}\right)}
             {\sin\left(\frac{\pi(z-X)}{2W}\right)}
        \right|,
    \qquad\qquad z=x+iy.
    \label{eq:InfiniteStripMethodOfImages}
\end{eqnarray}


The total potential can be split into the self- and the image-contribution. 
Subtracting the self-term removes the logarithmic divergence and isolates the edge screening part. Introducing a cutoff length $R$, one finds
\begin{eqnarray}
    V_{\text{images}}(x,y)=V(x,y)-V_{\text{self}}(x,y)
    =\sum_{n=-\infty}^{\infty}\sum_{\alpha=\pm}\frac{\alpha}{2\pi}
    \ln
    \frac{R}{\sqrt{\left(x-2nW-\alpha X\right)^2+y^2}}
    -\frac{1}{2\pi}
    \ln\frac{R}{\sqrt{\left(x-X\right)^2+y^2}}.
\end{eqnarray}
Using the closed form for $V(x,y)$ [Eq. \eqref{eq:InfiniteStripMethodOfImages}] and writing the real parts explicitly, we obtain
\begin{eqnarray}
    V_{\text{images}}(x,y)&&=\frac{1}{2\pi}
    \operatorname{Re}\left[
    \ln\frac{\sin{\left(\frac{\pi\left(z+X\right)}{2W}\right)}}
            {\sin{\left(\frac{\pi\left(z-X\right)}{2W}\right)}}
    \right]
    -\frac{1}{2\pi}\operatorname{Re}\left[\ln\left(\frac{R}{z-X}\right)\right]
    =\frac{1}{2\pi}\operatorname{Re}\left\{
    \ln\left[
    \frac{\sin{\left(\frac{\pi\left(z+X\right)}{2W}\right)}}
         {\left(\frac{R}{z-X}\right)\sin{\left(\frac{\pi\left(z-X\right)}{2W}\right)}}
    \right]
    \right\}\nonumber\\
    &&=\frac{1}{2\pi}\operatorname{Re}\left\{
    \ln\left[\frac{\sin{\left(\frac{\pi\left(z+X\right)}{2W}\right)}}
            {\left(\frac{\pi R}{2W}\right)
                ~\text{sinc}\left(\frac{\pi\left(z-X\right)}{2W}\right)}
        \right]
    \right\}.
\end{eqnarray}

Evaluating the image contribution of potential at the position of the physical source $(x,y)=(X,Y=0)$ gives
\begin{eqnarray}\label{}
    V_{\text{images}}(X,0)=\frac{1}{2\pi}
    \ln\left[
    \frac{\sin{\left(\frac{\pi\left(X+X\right)}{2W}\right)}}
         {\left(\frac{\pi R}{2W}\right)~\text{sinc}\left(\frac{\pi\left(X-X\right)}{2W}\right)}
    \right]
    =\frac{1}{2\pi}\ln\left[
    \frac{2W}{\pi R}\sin{\left(\frac{\pi X}{W}\right)}
    \right].
\end{eqnarray}

Therefore, the potential energy per unit length of a line charge $\omega=+1$ at position $x=X$ due to the image line charges is 
\begin{eqnarray}\label{}
    \tilde{U}(X)=\omega V_{\text{images}}(X,0)
    =\frac{1}{2\pi}\ln\left[
        \frac{2W}{\pi R}\sin{\left(\frac{\pi X}{W}\right)}
    \right].
\end{eqnarray}

Using the image method, the energy is counted twice if we take $\omega V_{\mathrm{images}}$, so the electrostatic interaction energy of the real line charge with the grounded conducting planes is
\begin{eqnarray}\label{eq:InfiniteStripPotentialEnergyLineCharge}
    U(X)=\frac{1}{4\pi}\ln\left[\frac{2W}{\pi R}\sin{\left(\frac{\pi X}{W}\right)}\right].
\end{eqnarray}
For the electrostatic problem (a line charge density $\omega$ in a medium of permittivity $\varepsilon_0$), restoring physical units would give
$
    U(X)=\frac{\omega^2}{4\pi \varepsilon_0}
    \ln\left[\frac{2W}{\pi R}\sin{\left(\frac{\pi X}{W}\right)}\right]
$.
By comparing the expression for electrostatic energy density $\left| \mathbf{E} \right|^2/(2\varepsilon_0)$ with superconducting free energy density $\left| \mathbf{K} \right|^2/(2\Upsilon_\text{2D})$,
and by comparing Eqs.~\eqref{eq:InfiniteStripPoissonEq} and \eqref{eq:InfiniteStripPoissonEqDimless}, one may deduce that 
the free energy of a vortex at position $(X,0)$ due to edge screening in an infinite strip $(0<x<W)$ is
\begin{eqnarray}\label{}
    F_{\text{images}}(X)=\frac{\Phi_0^2}{2\pi\mu_0 \Lambda}
    \ln\left[\frac{2W}{\pi R}\sin{\left(\frac{\pi X}{W}\right)}\right].
\end{eqnarray}

Unlike the electrostatic analogy, in a superconducting thin film the core self-energy of a vortex is physically meaningful. 
This energy diverges logarithmically and therefore requires regularization at both short and long distances. 
We write the \emph{core contribution} as
\begin{eqnarray}\label{}
    F_{\text{core}}=\frac{\Phi_0^2}{2\pi\mu_0 \Lambda}\ln\left(\frac{R}{r_c}\right),
\end{eqnarray}
where $r_c\sim\xi$ is the effective vortex core radius and $R$ is a long-distance cutoff determined by the system size or geometry. 

Adding the core contribution to the edge-screening term $F_{\text{images}}(X)$ yields the total single-vortex self-energy in an infinite strip
\begin{eqnarray}
    \label{eq:InfiniteStripF_Self1}
    F_{\text{self}}(X)=F_{\text{core}}+F_{\text{images}}(X)
    =\frac{\Phi_0^2}{2\pi\mu_0 \Lambda}
    \ln\left[\frac{2W}{\pi r_c}\sin{\left(\frac{\pi X}{W}\right)}\right].
\end{eqnarray}

\begin{figure}[h]
    \centering
    \subfloat[][ ]{%
        \includegraphics[width=0.48\textwidth]{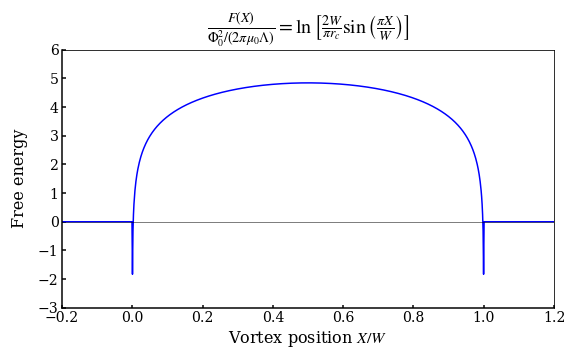}%
        \label{fig:InfiniteStripF_Self1}}
    \hfill
    \subfloat[][ ]{%
        \includegraphics[width=0.48\textwidth]{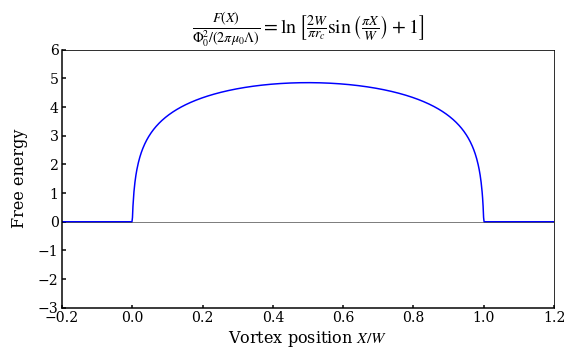}%
        \label{fig:InfiniteStripF_Self2}}
  \caption{\justifying 
        Comparison of the free energy profiles of a vortex in a strip of width $W=1$ with core cutoff $r_c/W=0.005$. 
        (a) The unregularized expression [Eq. \eqref{eq:InfiniteStripF_Self1}] diverges to $-\infty$ at the strip edges, predicting an unphysical infinitely deep potential well. 
        (b) The regularized expression [Eq. \eqref{eq:InfiniteStripF_Self2}] eliminates this divergence and ensures $F_{\text{self}}(0)=F_{\text{self}}(W)=0$, yielding a physically reasonable free energy profile.}
    \label{fig:F_infinite_finite}
\end{figure}

The above derivation treated vortices as point particles. The above expression for $F_{\text{self}}(X)$ becomes invalid within $r_c$ of the strip edges; it becomes negative and diverges to $-\infty$ as the vortex approaches the strip edges ($X\to 0,W$), leading to unphysical infinitely deep trapping potentials (see Fig. \ref{fig:InfiniteStripF_Self1}). 
Thus, in Monte Carlo simulations, Eq. \eqref{eq:InfiniteStripF_Self1} poses a danger  that vortices may become permanently trapped in \emph{bottomless wells}. 
Physically, when a vortex approaches the edge within a distance of $\sim r_c$, its current is nearly canceled by that of its image antivortex, so the energy cost must vanish smoothly at the boundary.
To incorporate this behavior and eliminate the divergence in an empirical way, we introduce a small regularization shift $\delta$, chosen such that $F_{\text{self}}(0)=F_{\text{self}}(W)=0$:
\begin{equation}\label{}
    F_{\text{self}}(X)=\frac{\Phi_0^2}{2\pi\mu_0\Lambda}
    \ln\left[\frac{2 W}{\pi r_c}\sin\left(\frac{\pi X}{W}\right)+\delta\right]=0,\qquad\therefore\quad\delta=1.\nonumber
\end{equation}
Thus, for the purpose of Monte Carlo simulations, we use the following model for the free energy of a vortex in a strip (see Fig. \ref{fig:InfiniteStripF_Self2}):
\begin{equation}\label{eq:InfiniteStripF_Self2}
    \boxed{
    F_{\text{self}}(X)
    =\frac{\Phi_0^2}{2\pi\mu_0\Lambda}
    \ln\left[
        \frac{2 W}{\pi r_c}\sin\left(\frac{\pi X}{W}\right)+1
    \right].}
\end{equation}
More accurate regularization requires solving the Ginzburg-Landau equation in the vicinity of the vortex core. This will be presented in a future paper.

\subsubsection{Single-Vortex Magnetic Moment \texorpdfstring{$m$}{m} with Edge Screening}\label{sec:InfiniteStripMagneticMoment}

\begin{summarybox}[Summary --- Single-Vortex Magnetic Moment]
We compute the magnetic moment $m(X)$ of a single Pearl vortex in an infinitely long strip $0 < x < W$ with edge screening.  
We show that the Dirichlet Green function obtained via the method of images [Eq. \eqref{eq:InfiniteStripMethodOfImages}] and via the direct-integration method [Eq. \eqref{eq:InfiniteDirectIntegration}] are mathematically equivalent [Eq. \eqref{eq:InfiniteDirectIntegrationExplicit}], providing two alternative but identical representations of the Green function [Eq. \eqref{eq:InfiniteStrip_2FormGF}]:
\begin{eqnarray}
    G(x,y;X,0)=\frac{1}{2\pi}\operatorname{Re}\left\{\ln\left[\frac{\sin{\left(\frac{\pi}{2W}\left(x+iy+X\right)\right)}}{\sin{\left(\frac{\pi}{2W}\left(x+iy-X\right)\right)}}\right]\right\}
    =\sum_{n=1}^{\infty}\frac{1}{n\pi}\sin{\left(\frac{n\pi X}{W}\right)}\sin{\left(\frac{n\pi x}{W}\right)}\exp\left(-\frac{n\pi}{W}|y|\right).\nonumber
\end{eqnarray}
Using the second representation in Eq. \eqref{eq:InfiniteStrip_2FormGF}, we derived the areal magnetization field $\mathcal{M}(x,y;X,0)$ [Eq. \eqref{eq:InfiniteStripMagnetization}], multiplied it by prefactor $2\Phi_0/(\mu_0\Lambda)$, and integrated it to obtain the global vortex magnetic moment:
\begin{eqnarray}
    m(X,0)=\frac{\Phi_0}{\mu_0\Lambda}X\left(W-X\right).\nonumber
\end{eqnarray}
Thus, the vortex magnetic moment is a simple quadratic function of the vortex position $X$, vanishing at the strip edges $X=0,W$ and reaching a maximum at mid-strip.
\end{summarybox}



If a vortex is located at $(X,Y)$ in a patterned superconducting film, the areal magnetization $\mathcal{M}(x,y)$ satisfies
\begin{equation}
    -(\partial_x^2 + \partial_y^2)\,\mathcal{M}(x,y) = 
    2\pi\,\delta(x-X)\,\delta(y-Y),
    \qquad\,\mathcal{M}=0\text{ on the strip edges}.
\end{equation}
The magnetic moment of the vortex is then defined as
\begin{equation}
    m = \int_{\mathcal{S}}\mathcal{M}(x,y)\,dx\,dy.
\end{equation}

In principle, one could directly employ the Dirichlet Green function already derived in Sec. \ref{sec:InfiniteStripSelfEnergy}, using the method of images [Eq. \eqref{eq:InfiniteStripMethodOfImages}], to evaluate the magnetization $\mathcal{M}$ and thereby obtain the magnetic moment $m$.
Here, however, we adopt an alternative approach---the \emph{method of direct integration}---which constructs the same Green function by expanding it in Fourier modes along the $x$-direction, thereby reducing the 2D PDE to an ODE in $y$.
Readers primarily interested in the final expression may skip ahead to Eq. \eqref{eq:InfiniteDirectIntegrationExplicit}, where the derivation of the vortex magnetic moment $m$ begins.

We therefore consider the Dirichlet Green function in an infinite strip $0<x<W$, satisfying
\begin{eqnarray}\label{eq:InfiniteStripPoisson}
    -\nabla^2G(x,y)=\delta(x-X)\,\delta(y),
    \qquad\qquad G=0\,\,\text{ for $x=0,W$}.
\end{eqnarray}
Expanding in Fourier modes along the $x$-direction, we obtain
\begin{eqnarray}
    &&[(\frac{n\pi}{W})-\partial_y^2]\tilde{G}_n(y)=\sqrt{\frac{2}{W}}\sin\frac{n\pi X}{W}\delta(y),
    \qquad\therefore\quad
    (\kappa^2-\partial_y^2)g(y)=\delta(y),
    \label{eq:1DScreenedCoulomb}
    \\&&\qquad\qquad\text{where }
    g(y)\equiv\frac{\tilde{G}_n(y)}{\sqrt{\frac{2}{W}}\sin\left(\frac{n\pi x}{W}\right)}
    \text{ and }
    \kappa=\frac{n\pi}{W}.\nonumber
\end{eqnarray} 
Equation \eqref{eq:1DScreenedCoulomb} is the 1D screened Coulomb equation where $\kappa>0$ .

For $y\neq 0$, the delta term vanishes and Eq. \eqref{eq:1DScreenedCoulomb} reduces to
\begin{eqnarray}
    (\kappa^2-\partial_y^2)\,g(y)=0
    \quad\Rightarrow\quad 
    g''(y)=\kappa^2 g(y).
\end{eqnarray}
The general solution is $g(y)=A e^{\kappa y}+B e^{-\kappa y}$. 
Boundedness as $|y|\to\infty$ requires $g(y)=B e^{-\kappa|y|}$.
To fix $B$, we integrate Eq. \eqref{eq:1DScreenedCoulomb} across a vanishing interval around $y=0$:
\begin{eqnarray}
    \int_{-\varepsilon}^{\varepsilon}
    \bigl(\kappa^2 g - g''\bigr)\,dy
    =\kappa^2\int_{-\varepsilon}^{\varepsilon}g\,dy-\bigl[g'(y)\bigr]_{-\varepsilon}^{+\varepsilon}
    =1.
\end{eqnarray}
The contribution from $\kappa^2g$ vanishes as $\varepsilon\to0$, leaving only the discontinuity in $g'(y)$.
From $g(y)=B e^{-\kappa|y|}$,
\begin{eqnarray}
    g'(0^+)=-\kappa B,
    \qquad g'(0^-)=+\kappa B 
    \quad\Rightarrow\quad
    \bigl[g'(y)\bigr]_{0^-}^{0^+}=-2\kappa B=-1,
\end{eqnarray}
Thus, the normalization constant is
\begin{eqnarray}
    B=\frac{1}{2\kappa},
    \qquad\qquad
    g(y)=\frac{1}{2\kappa}e^{-\kappa|y|}.
\end{eqnarray}
This defines the 1D screened-Coulomb Green function, which exhibits exponential decay with screening length $1/\kappa$.
Substituting back into the Fourier expansion, the coefficients are
\begin{eqnarray}
    \tilde{G}_n(y)
    =\sqrt{\frac{2}{W}}\sin\left(\frac{n\pi X}{W}\right)
    \frac{1}{2\kappa}\exp\left(-\kappa|y|\right),
\end{eqnarray}
and hence the full Dirichlet Green function in the strip is
\begin{eqnarray}
    G(x,y)&&
    =\sum_{n=1}^\infty\sqrt{\frac{2}{W}}
        \sin\left(\frac{n\pi x}{W}\right)\tilde{G}_n(y)
    =\sum_{n=1}^\infty\sqrt{\frac{2}{W}}
        \sin\left(\frac{n\pi x}{W}\right)
        \sqrt{\frac{2}{W}}\sin\left(\frac{n\pi X}{W}\right)
        \frac{1}{2\kappa}\exp\left(-\kappa|y|\right)\nonumber\\
    &&=\sum_{n=1}^\infty\frac{1}{n\pi}
        \sin\left(\frac{n\pi X}{W}\right)
        \sin\left(\frac{n\pi x}{W}\right)
        \exp\left(-\frac{n\pi}{W}|y|\right).
    \label{eq:InfiniteDirectIntegration}
\end{eqnarray}

The series sum may be evaluated in closed form as follows. Assume $y\geq0$, then
\begin{eqnarray}\label{}
    \partial_yG(x,y)&&=-\frac{1}{W}\sum_{n=1}^{\infty}
    \sin{\left(\frac{n\pi X}{W}\right)}
    \sin{\left(\frac{n\pi x}{W}\right)}
    \exp\left(-\frac{n\pi}{W}y\right)\nonumber\\
    &&=-\frac{1}{W}\sum_{n=1}^{\infty}
    \left(
        \frac{\exp\left(i\frac{n\pi X}{W}\right)
              -\exp\left(-i\frac{n\pi X}{W}\right)}
              {2i}\right)
    \left(
        \frac{\exp\left(i\frac{n\pi x}{W}\right)
              -\exp\left(-i\frac{n\pi x}{W}\right)}
              {2i}\right)
    \exp\left(i\frac{n\pi}{W}(iy)\right)\nonumber\\
    &&=\frac{1}{4W}\sum_{n=1}^{\infty}
    \left\{
        \exp\left(\frac{in\pi\left(X+x+iy\right)}{W}\right)-
        \exp\left(\frac{in\pi\left(X-x+iy\right)}{W}\right)
        \right.\nonumber\\&&\qquad\qquad\qquad\left.-
        \exp\left(\frac{in\pi\left(-X+x+iy\right)}{W}\right)+
        \exp\left(\frac{in\pi\left(-X-x+iy\right)}{W}\right)
    \right\}\nonumber\\
    &&=\frac{1}{2W}\operatorname{Re}\left\{
    \sum_{n=1}^{\infty}
    \left[
        \exp\left(\frac{in\pi\left(X+x+iy\right)}{W}\right)-
        \exp\left(\frac{in\pi\left(X-x+iy\right)}{W}\right)
    \right]
    \right\}.
    \label{eq:dG_FourierSum}
\end{eqnarray}
The infinite sum may be evaluated using the standard series identity
\begin{eqnarray}\label{}
    \sum_{n=1}^{\infty}e^{in\theta}
    =\sum_{n=1}^{\infty}\left(e^{i\theta}\right)^n
    =\frac{e^{i\theta}}{1-e^{i\theta}}
    =\frac{i}{2}\cot\frac{\theta}{2}-\frac{1}{2}.
\end{eqnarray}
Substituting into Eq. \eqref{eq:dG_FourierSum} yields
\begin{eqnarray}\label{eq:dG_FourierSum2}   
    \partial_yG(x,y)
    =\frac{1}{2W}\operatorname{Re}\left\{
        \frac{i}{2}\left[
            \cot\left(\frac{\pi\left(X+x+iy\right)}{2W}\right)
            -\cot\left(\frac{\pi\left(X-x+iy\right)}{2W}\right)
        \right]
    \right\}.
\end{eqnarray}
To reconstruct the Green function from its derivative in Eq. \eqref{eq:dG_FourierSum2}, we integrate with respect to $y$.
Using
\begin{eqnarray}
    \int d\theta\;\;\cot\theta=\ln(\sin{\theta}),
\end{eqnarray}
we obtain
\begin{eqnarray}\label{eq:InfiniteDirectIntegrationExplicit}
    G(x,y;X,0)&&
    =\frac{1}{4W}\operatorname{Re}\left\{
        \frac{2W}{i\pi}i
            \left[\ln\left(
                \sin{\frac{\pi\left(X+x+iy\right)}{2W}}
            \right)
            -\ln\left(
                \sin{\frac{\pi\left(X-x+iy\right)}{2W}}
            \right)
        \right]
    \right\}\nonumber\\
    &&=\frac{1}{2\pi}\operatorname{Re}\left\{
        \ln\left[
            \frac{\sin\left(\frac{\pi\left(X+x+iy\right)}{2W}\right)}
            {\sin\left(\frac{\pi\left(X-x+iy\right)}{2W}\right)}
        \right]
    \right\},
\end{eqnarray}
where $0<x<W$. 
Equation \eqref{eq:InfiniteDirectIntegrationExplicit} provides a closed-form expression for the Dirichlet Green function, which is exactly equivalent to the result obtained earlier by the method of images [Sec. \ref{sec:InfiniteStripSelfEnergy}, Eq. \eqref{eq:InfiniteStripMethodOfImages}]. 
This confirms that the method of direct integration and the method of images are mathematically equivalent, both yielding the same Dirichlet Green function for the infinite strip:
\begin{eqnarray}\label{eq:InfiniteStrip_2FormGF}
    G(x,y;X,0)
    =\frac{1}{2\pi}\operatorname{Re}\left\{
        \ln\left[
            \frac{\sin{\left(\frac{\pi\left(x+iy+X\right)}{2W}\right)}}
                 {\sin{\left(\frac{\pi\left(x+iy-X\right)}{2W}\right)}}
        \right]
    \right\}
    =\sum_{n=1}^{\infty}\frac{1}{n\pi}
    \sin{\left(\frac{n\pi X}{W}\right)}
    \sin{\left(\frac{n\pi x}{W}\right)}
    \exp\left(-\frac{n\pi}{W}|y|\right).
\end{eqnarray}
This dual representation Eq. \eqref{eq:InfiniteStrip_2FormGF} will be central in evaluating the vortex magnetic moment $m$.

Restoring physical factors (by multiplying by $2\Phi_0/(\mu_0\Lambda)$) gives the areal magnetization for a vortex:
\begin{eqnarray}\label{eq:InfiniteStripMagnetization}
    \mathcal{M}(x,y;X,0)
    =\frac{\Phi_0}{\pi\mu_0\Lambda}
    \operatorname{Re}\left\{
        \ln\left[
            \frac{\sin{\left(\frac{\pi\left(x+iy+X\right)}{2W}\right)}}
                 {\sin{\left(\frac{\pi\left(x+iy-X\right)}{2W}\right)}}
            \right]
    \right\}
    =\frac{2\Phi_0}{\mu_0\Lambda}
    \sum_{n=1}^{\infty}\frac{1}{n\pi}
    \sin{\left(\frac{n\pi X}{W}\right)}
    \sin{\left(\frac{n\pi x}{W}\right)}
    \exp\left(-\frac{n\pi}{W}|y|\right).
\end{eqnarray}

To go from the ``local'' magnetization field $\mathcal{M}(x,y;X,0)$ to the ``global'' magnetic moment $m(X,0)$ of a vortex, integrate Eq. \eqref{eq:InfiniteStripMagnetization} over the whole strip:
\begin{eqnarray}
    m(X,0)&&
    =\int_0^W dx\int_{-\infty}^{\infty}dy\;\;\mathcal{M}(x,y;X,0)\
    =\int_0^Wdx\int_{-\infty}^{\infty}dy\;\;\frac{2\Phi_0}{\mu_0\Lambda}
    \sum_{n=1}^{\infty}\frac{1}{n\pi}
        \sin{\left(\frac{n\pi X}{W}\right)}
        \sin{\left(\frac{n\pi x}{W}\right)}
        \exp\left(-\frac{n\pi}{W}|y|\right)
    \nonumber\\
    &&=\frac{2\Phi_0}{\mu_0\Lambda}
    \sum_{n=1}^{\infty}\frac{1}{n\pi}
        \sin{\left(\frac{n\pi X}{W}\right)}
        \int_0^Wdx\;\sin{\left(\frac{n\pi x}{W}\right)}
        \;\;2\int_0^{\infty}dy\;\exp\left(-\frac{n\pi}{W}y\right)
    \nonumber\\
    &&=\frac{4\Phi_0}{\mu_0\Lambda}
    \sum_{n=1}^{\infty}\frac{1}{n\pi}
        \sin{\left(\frac{n\pi X}{W}\right)}
        \;\left[\frac{W}{n\pi}\left(1-\cos{n\pi}\right)\right]
        \;\left(\frac{W}{n\pi}\right)
    =\frac{4\Phi_0W^2}{\mu_0\Lambda}
    \sum_{n=1}^{\infty}
        \frac{1-\cos(n\pi)}{\left(n\pi\right)^3}
        \sin{\left(\frac{n\pi X}{W}\right)}.
    \label{eq:InfiniteStripMagneticMomentSeries}
\end{eqnarray}
To evaluate the series in Eq. \eqref{eq:InfiniteStripMagneticMomentSeries} in closed form, expand the dimensionless function $f(x)=\frac{x}{W}\left(1-\frac{x}{W}\right)$ in orthonormal basis functions $\phi_n(x)=\sqrt{2/W}\sin(n\pi x/W)$ on $(0,W)$:
\begin{eqnarray}
    \int_0^Wdx\;
        \left[\sqrt{\frac{2}{W}}\sin{\left(\frac{n\pi x}{W}\right)}\right]
        \frac{x}{W}\left(1-\frac{x}{W}\right)
    =\frac{\sqrt{8W}\left(1-\cos{n\pi}\right)}{n^3\pi^3}.
    \label{eq:InfiniteStripFourierSine1}
\end{eqnarray}
This gives $f(x)$ as a Fourier sine series
\begin{eqnarray}
    \sum_{n=1}^{\infty}
        \left[\sqrt{\frac{2}{W}}\sin{\left(\frac{n\pi X}{W}\right)}\right]
        \frac{\sqrt{8W}\left(1-\cos{n\pi}\right)}{n^3\pi^3}
    =\frac{X}{4W}\left(1-\frac{X}{W}\right)
    \label{eq:InfiniteStripFourierSine2}
\end{eqnarray}
Hence,
\begin{eqnarray}
    \boxed{
    m(X,0)
        =\frac{4\Phi_0W^2}{\mu_0\Lambda}
            \frac{X}{4W}
            \left(1-\frac{X}{W}\right)
        =\frac{\Phi_0}{\mu_0\Lambda}
            X\left(W-X\right).
    }
    \label{eq:InfiniteStripMagneticMoment}
\end{eqnarray}
We have chosen $Y=0$ for simplicity, but the system is translationally invariant in the $y$-direction, so it is obvious that $m(X,Y)=m(X,0)$.  The vortex magnetic moment is a quadratic function of the vortex position across the strip, $X$; it   vanishes at the edges ($X=0$ and $X=W$) and peaks in the middle ($X=W/2$).

\subsubsection{Vortex–Vortex Interaction \texorpdfstring{$F_{vv}$}{Fvv} with Edge Screening}

\begin{summarybox}[Summary — Vortex–Vortex Interaction]
In an infinitely long strip $0<x<W$, the interaction free energy of two Pearl vortices located at $(x,y)$ and $(X,Y)$ is given by
\begin{eqnarray}
    F_{vv}(x,y;X,Y)
    =\frac{\Phi_0^2}{2\pi\mu_0\Lambda}\,
        \ln\left[
            \frac{\cos\left(\tfrac{\pi(x+X)}{W}\right)
                  -\cosh\left(\tfrac{\pi(y-Y)}{W}\right)}
                 {\cos\left(\tfrac{\pi(x-X)}{W}\right)
                  -\cosh\left(\tfrac{\pi(y-Y)}{W}\right)}
        \right].\nonumber
\end{eqnarray}
This expression is symmetric under exchange of the two vortices $(x,y)\leftrightarrow (X,Y)$, reduces to the usual logarithmic repulsion if both vortices are far from the edges, and captures the influence of \emph{edge screening} through the explicit $W$ dependence on the strip width $W$.
\end{summarybox}

Recall that for an infinite strip ($0<x<W$), if a source is located at $(X,Y)$, the potential in the strip is [Eq. \eqref{eq:InfiniteStripMethodOfImages}]
\begin{eqnarray}
    V(x,y;X,Y)
    =\frac{1}{2\pi}
        \ln\left|
            \frac{\sin{\left[\frac{\pi\left(z+X-iY\right)}{2W}\right]}}
                 {\sin{\left[\frac{\pi\left(z-X-iY\right)}{2W}\right]}}
            \right|,\qquad\qquad z=x+iy.
\end{eqnarray}
The energy experienced by a second source at $(x,y)$ is simply
\begin{eqnarray}
    U(x,y;X,Y)=\omega V(x,y;X,Y).
\end{eqnarray}
Line charges repel with energy $U=\frac{\omega^2}{2\pi\varepsilon_0}\ln\frac{1}{\left|\textbf{S}_1-\textbf{S}_2\right|}$, whereas vortices repel with free energy $F_{vv}=\frac{\Phi_0^2}{\pi\mu_0\Lambda}\ln\frac{1}{\left|\textbf{S}_1-\textbf{S}_2\right|}$. 
By the electrostatic-superconducting correspondence introduced earlier, the electrostatic interaction of line charges maps to the vortex-vortex free energy, with the replacement $\frac{\omega^2}{2\pi\varepsilon_0}\;\mapsto\; \frac{2\Phi_0^2}{\mu_0\Lambda}$.
Thus,
\begin{eqnarray}
    F_{vv}(x,y;X,Y)
        =\frac{\Phi_0^2}{\pi\mu_0\Lambda}
            \operatorname{Re}\left\{
                \ln\left[
                    \frac{\sin{\left(\pi\frac{z+X-iY}{2W}\right)}}  
                         {\sin{\left(\pi\frac{z-X-iY}{2W}\right)}}
                \right]
            \right\},
\end{eqnarray}
In terms of real-valued quantities only,
\begin{eqnarray}\label{eq:InfiniteStripFvv}
    \boxed{
    F_{vv}(x,y;X,Y)
        =\frac{\Phi_0^2}{2\pi\mu_0\Lambda}
            \ln\left[
                \frac{\cos{\left(\frac{\pi\left(x+X\right)}{W}\right)}- 
                      \cosh{\left(\frac{\pi\left(y-Y\right)}{W}\right)}}
                     {\cos{\left(\frac{\pi\left(x-X\right)}{W}\right)}- 
                      \cosh{\left(\frac{\pi\left(y-Y\right)}{W}\right)}}
            \right].
    }
\end{eqnarray}
Equation \eqref{eq:InfiniteStripFvv} is symmetric under the exchange of $(x,y)$ and $(X,Y)$, as required for an interaction energy. 

\subsection{Gibbs Free Energy in a Semi-Infinite Length Strip}\label{sec:SemiInfiniteStrip}

\begin{summarybox}[Summary --- Gibbs Free Energy in a Semi-Infinite Length Strip]
Here we show that the Gibbs free energy for a Pearl vortex in a semi-infinite superconducting strip of width $W$ ($0<x<W$, $0<y<\infty$) is
\[
G_{\text{tot}} = G_v + F_{vv} = F_{\text{self}} - mB + F_{vv},
\]
where
the \textbf{self-energy $F_{\text{self}}(X,Y)$} is
  \[
  F_{\text{self}}(X,Y) 
  =\frac{\Phi_0^2}{2\pi\mu_0\Lambda}\,
    \ln\left[
        \frac{2W}{\pi r_c}
        \frac{\sin\left(\tfrac{\pi X}{W}\right)
              \sinh\left(\tfrac{\pi Y}{W}\right)}
             {\sqrt{\sin^2\left(\tfrac{\pi X}{W}\right)+
              \sinh^2\left(\tfrac{\pi Y}{W}\right)}}+1
    \right],
  \]
the \textbf{vortex magnetic moment $m(X,Y)$} is
  \[
  m(X,Y) 
  =\frac{\Phi_0}{\mu_0\Lambda}
    \left[
        X(W-X) - 
        \frac{8W^2}{\pi^3}
        \sum_{n=1,3,5,\dots}^\infty 
            \frac{1}{n^3}
            \sin\left(\tfrac{n\pi X}{W}\right)
            e^{-n\pi Y/W}
    \right],
  \]
and the \textbf{vortex–vortex interaction free energy $F_{vv}(x,y;X,Y)$} is
  \[
  F_{vv}(x,y;X,Y) 
  =\frac{\Phi_0^2}{2\pi\mu_0\Lambda}
    \ln\left[
        \frac{\cos\left(\tfrac{\pi(x+X)}{W}\right)-
              \cosh\left(\tfrac{\pi(y-Y)}{W}\right)}
             {\cos\left(\tfrac{\pi(x-X)}{W}\right)-
             \cosh\left(\tfrac{\pi(y-Y)}{W}\right)}
        \cdot
        \frac{\cos\left(\tfrac{\pi(x-X)}{W}\right)-
              \cosh\left(\tfrac{\pi(y+Y)}{W}\right)}
             {\cos\left(\tfrac{\pi(x+X)}{W}\right)-
              \cosh\left(\tfrac{\pi(y+Y)}{W}\right)}
  \right].
  \]
$G_\text{tot}$ includes core energy, edge screening at all three boundaries ($x=0$, $x=W$, and $y=0$), magnetic coupling to the external field $B$, and vortex–vortex interactions.
%
%
\end{summarybox}

\subsubsection{Single-Vortex Self-Energy \texorpdfstring{$F_{\text{self}}$}{Fself} (Edges and Core)}\label{sec:SemiInfiniteSelfEnergy}

\begin{summarybox}[Summary --- Single-Vortex Self-Energy in a Semi-Infinite Length Strip]
We compute the position-dependent self-energy $F_{\mathrm{self}}(X,Y)$ of a Pearl vortex in a semi-infinite strip ($0 < x < W$, $0 < y < \infty$),
\begin{equation}
    F_{\mathrm{self}}(X,Y)
    =\frac{\Phi_0^2}{2\pi\mu_0\Lambda}
        \ln\left[
            \frac{2W}{\pi r_c}\,
            \frac{\sin\left(\tfrac{\pi X}{W}\right)
                  \sinh\left(\tfrac{\pi Y}{W}\right)}
                 {\sqrt{\sin^2\left(\tfrac{\pi X}{W}\right)
                 +\sinh^2\left(\tfrac{\pi Y}{W}\right)}}+ 1
        \right],\nonumber
\end{equation}
where $r_c$ is the effective vortex core radius. 
The ``+1'' is added to prevent spurious negative logarithmic divergences at the strip edges. 
Compared to an infinite strip, a semi-infinite strip has an additional boundary at $y=0$, 
which leads to extra factors in $F_{\mathrm{self}}$ containing $\sinh(\pi Y/W)$.
\end{summarybox}


We now consider a semi-infinite strip.
Compared to the infinite strip studied in Sec. \ref{sec:InfiniteStrip}, there is now an additional edge at $y=0$, so the system is no longer translationally invariant in the $y$-direction, and the vortex position $Y$ must be treated explicitly. 
Let $G(x,y;X,Y)\equiv V(x,y)$ denote the 2D Dirichlet Green function of the Poisson equation in the semi-infinite strip, 
\begin{eqnarray}
    -\nabla^2 V(x,y)=\delta(x-X)\delta(y-Y)~~
    \text{ where }\;
    0<x<W,\; 0<y<\infty,\;
    \text{ and }\;
    V(0,y)=V(W,y)=V(x,0)=0.
\end{eqnarray} 

\begin{figure}[h]
    \centering
    \begin{minipage}{0.45\textwidth}
    \includegraphics[width=\linewidth]{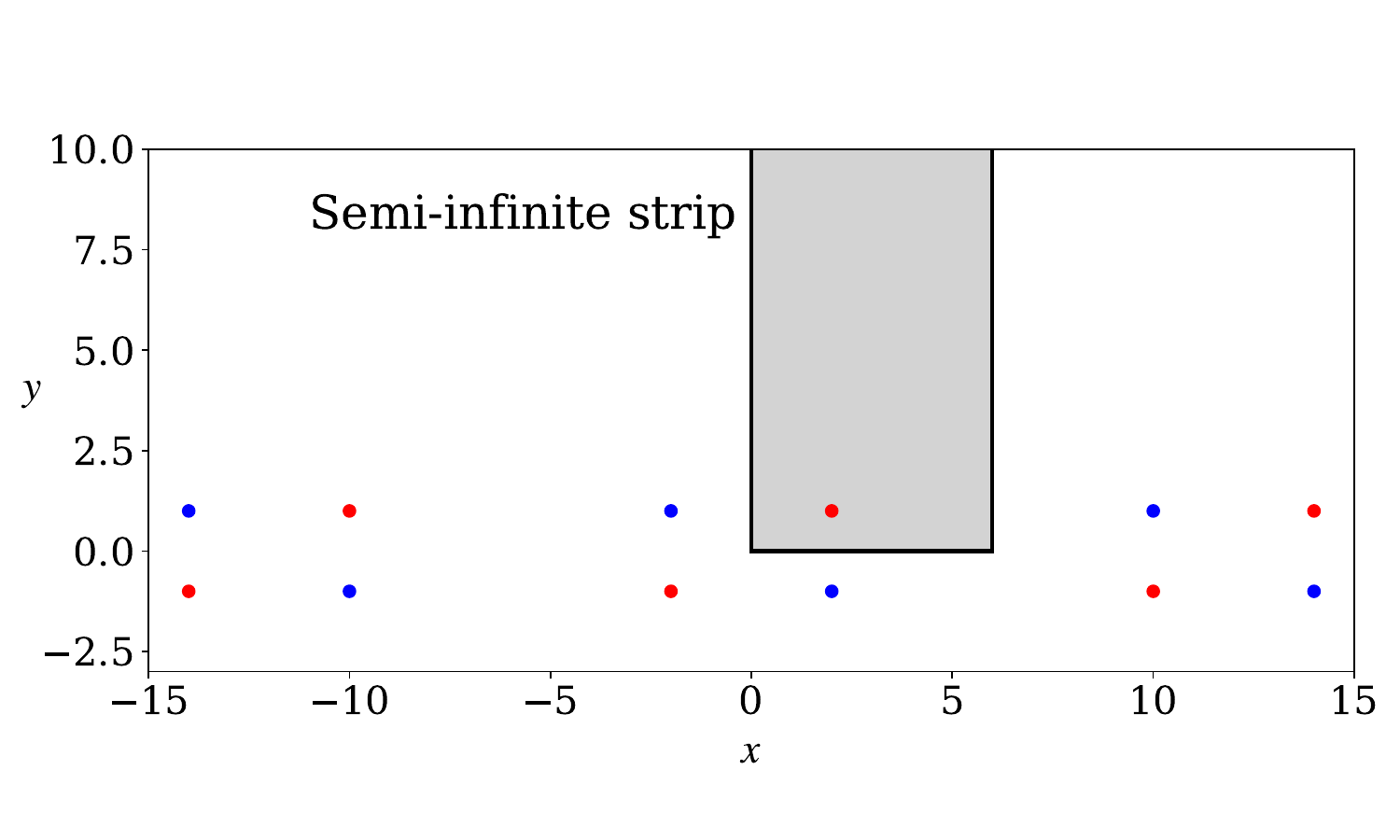}
    \end{minipage}
    \hfill
    \begin{minipage}{0.5\textwidth}
    \caption{\justifying
    A source at $(X,Y)$ in a semi-infinite strip $(0<x<W,\;0<y<\infty)$
    behaves like an infinite array of quadrupoles consisting of sources (red) and sinks (blue) located at $(\pm X+2nW,\pm Y)$, with integer $n$.
    This periodic image arrangement enforces Dirichlet boundary conditions at $x=0$, $x=W$, and $y=0$.
    }
    \label{fig:1DQudrupoles}
    \end{minipage}
\end{figure}
 In the language of 2D electrostatics, $V(x,y)$ is the potential due to a point source at $(X,Y)$ with Dirichlet boundary conditions along the three strip edges.  In 3D electrostatics, $V(x,y)$ is the potential due to a line charge near three grounded conducting planes.

A suitable configuration of image charges is a 1D array of quadrupoles, each composed of two positive and two negative charges, and extending infinitely along the $\pm x$ direction.  See Fig. \ref{fig:1DQudrupoles}.
Each quadrupole is centered at $(nW,0)$ ($n\in\mathbb{Z}$).  The charges are located at  ($\pm X+2nW,\pm Y$).  The total potential $V(x,y)$ is thus
\begin{eqnarray}
    V(x,y)&&=\sum_{n=-\infty}^\infty\sum_{\alpha=\pm1}\sum_{\beta=\pm}
    \frac{\alpha\beta}{2\pi}
    \ln\left[
        \frac{1}{\sqrt{(x-2 n W-\alpha X)^2+(y-\beta Y)^2}}
    \right]\nonumber\\
    &&=\sum_{n=-\infty}^\infty\frac{1}{2\pi}
    \ln\left[
        \frac{\sqrt{(x-2nW+X)^2+(y-Y)^2}}
             {\sqrt{(x-2nW-X)^2+(y-Y)^2}}\times
        \frac{\sqrt{(x-2nW-X)^2+(y+Y)^2}}
             {\sqrt{(x-2nW+X)^2+(y+Y)^2}}
    \right]\nonumber\\
    &&=\sum_{n=-\infty}^\infty\frac{1}{2\pi}
    \operatorname{Re}\left\{
        \ln\left[
            \frac{(x-2nW+X)+i(y-Y)}{(x-2nW-X)+i(y-Y)}\times
            \frac{(x-2nW-X)+i(y+Y)}{(x-2nW+X)+i(y+Y)} 
        \right]
    \right\}\nonumber\\
    &&=\frac{1}{2\pi}
    \operatorname{Re}\left\{
        \ln\prod_{n=-\infty}^\infty
            \left[
                \frac{z-2nW+X-iY}{z-2nW-X-iY}\times
                \frac{z-2nW-X+iY}{z-2nW+X+iY}
            \right]
    \right\},\qquad\qquad z=x+iy.
    \label{eq:quadrupolesPotential}
\end{eqnarray}

Using the sine-product identity introduced earlier in Eq. \eqref{eq:sinc}, the infinite product in Eq. \eqref{eq:quadrupolesPotential} reduces to
\begin{equation}\label{eq:SemiInfiniteTotalV}
    V(x,y)=\frac{1}{2\pi}
    \operatorname{Re}\left\{
        \ln\left[
        \frac{\sin\left(\frac{\pi(z+X-iY)}{2W}\right)}      
             {\sin\left(\frac{\pi(z-X-iY)}{2W}\right)}\times
        \frac{\sin\left(\frac{\pi(z-X+iY)}{2W}\right)}
             {\sin\left(\frac{\pi(z+X+iY)}{2W}\right)}
        \right]
    \right\}.
\end{equation}
As required, $V(x,y)$ has singularities at
($\pm X,\pm Y$), and it satisfies the boundary conditions.


The potential at $(x,y)$ due to the images is obtained by subtracting the bare singular Green function from \eqref{eq:SemiInfiniteTotalV},
\begin{eqnarray}\label{eq:SemiInfiniteStripPotentialOnlyImages1}
    V_{\text{images}}(x,y)
    =V(x,y)-V_{\text{self}}(x,y)
    &&=\frac{1}{2\pi}
    \operatorname{Re}\left\{
        \ln\left[
            \frac{\sin\left(\frac{\pi(z+X-iY)}{2W}\right)}  
                 {\sin\left(\frac{\pi(z-X-iY)}{2W}\right)}
            \frac{\sin\left(\frac{\pi(z-X+iY)}{2W}\right)}
                 {\sin\left(\frac{\pi(z+X+iY)}{2W}\right)}
        \right]
    \right\}
    -\frac{1}{2\pi}
    \operatorname{Re}\left[
        \ln\frac{R}{z-X-iY}
    \right]\nonumber\\
    &&=\frac{1}{2\pi}
    \operatorname{Re}\left\{
        \ln\left[
            \frac{\sin\left(\frac{\pi(z+X-i Y)}{2W}\right)}
                 {\left(\frac{\pi R}{2W}\right)
                  \text{sinc}\left(\frac{\pi(z-X-iY)}{2W}\right)}
            \times
            \frac{\sin\left(\frac{\pi(z-X+iY)}{2W}\right)}
                 {\sin\left(\frac{\pi(z+X+i Y)}{2W}\right)}
        \right]
    \right\}
\end{eqnarray}
where $R$ is a cutoff length. Taking the coincident point limit of Eq. \eqref{eq:SemiInfiniteStripPotentialOnlyImages1} ($x\to X$ and $y\to Y$),
\begin{eqnarray}
    V_{\text{images}}(X,Y)
    =\lim_{\substack{x\to X\\ y\to Y}}
    V_{\text{images}}(x,y)
    &&=\frac{1}{2\pi}\operatorname{Re}\left\{
        \ln\left[
            \frac{\sin\left(\frac{\pi(X+iY+X-iY)}{2W}\right)}
                 {\left(\frac{\pi R}{2 W}\right)
                    \text{sinc}\left(\frac{\pi(X+iY-X-iY)}{2W}\right)}
                \times\
            \frac{\sin\left(\frac{\pi(X+iY-X+iY)}{2W}\right)}
                 {\sin\left(\frac{\pi(X+iY+X+iY)}{2W}\right)}
        \right]
    \right\}\nonumber\\
    &&=\frac{1}{2\pi}
    \ln\left|
        \frac{\sin\left(\frac{\pi X}{W}\right)
              \sin\left(\frac{i\pi Y}{W}\right)}
             {\left(\frac{\pi R}{2W}\right)
              \sin\left(\frac{\pi(X+iY)}{W}\right)}
    \right|.
    \label{eq:SemiInfiniteStripPotentialOnlyImages2}
\end{eqnarray}
To write Eq. \eqref{eq:SemiInfiniteStripPotentialOnlyImages2} in terms of real variables, note that 
\begin{eqnarray}
    \left|\sin(a+ib)\right|
    &&=\left|\sin a\cosh b+i\cos a\sinh b\right|
    =\sqrt{\sin^2a\cosh^2b+\cos^2a\sinh^2b}\nonumber\\
    &&=\sqrt{\sin^2a~(1+\sinh^2b)
      +(1-\sin^2a)\sinh^2b}
    =\sqrt{\sin^2a+\sinh^2b}.
\end{eqnarray}\label{}
Using $\left|\sin(a+i b)\right|=\sqrt{\sin^2a+\sinh^2b}$, Eq. \eqref{eq:SemiInfiniteStripPotentialOnlyImages2} reduces to the purely real form
\begin{equation}\label{eq:SemiInfiniteStripV_images}
    V_{\text{images}}(X,Y)=\frac{1}{2\pi}\ln\left[
    \left(\frac{2 W}{\pi R}\right)
    \frac
    {\sin\left(\frac{\pi X}{W}\right)\sinh\left(\frac{\pi Y}{W}\right)}
    {\sqrt{\sin^2\left(\frac{\pi X}{W}\right)+\sinh^2\left(\frac{\pi Y}{W}\right)}}
    \right]
\end{equation}
Equation \eqref{eq:SemiInfiniteStripV_images} provides the real-valued image potential $V_{\text{images}}(X,Y)$ for a semi-infinite strip.

The potential energy per unit length of a line charge of strength $\omega=+1$ located at $z=X+iY$ in the semi-infinite strip is obtained by evaluating the image potential at the source position.
To avoid double-counting pair interactions inherent to the method of images, a factor of $1/2$ is introduced, giving
\begin{equation}
    U=\tfrac12\omega\, V_{\text{images}}(X,Y).
\end{equation}

Restoring appropriate prefactors shows that the contribution to the vortex free energy in a semi-infinite superconducting strip from the images (i.e., due to screening by the edges) is
\begin{equation}\label{}
    F_{\text{images}}(X,Y)
    =\frac{\Phi_0^2}{2\pi\mu_0\Lambda}
    \ln\left[
        \frac{2 W}{\pi R}
        \frac{\sin\left(\frac{\pi X}{W}\right)
              \sinh\left(\frac{\pi Y}{W}\right)}
             {\sqrt{\sin^2\left(\frac{\pi X}{W}\right)+
                    \sinh^2\left(\frac{\pi Y}{W}\right)}}
    \right].
\end{equation}

Unlike the electrostatic analogue, a superconducting vortex carries a finite \emph{core} energy associated with suppression of the order parameter.
This contribution is position-independent and equals
\begin{equation}\label{}
    F_{\text{core}}
    =\frac{\Phi_0^2}{2\pi\mu_0\Lambda}\ln\left(\frac{R}{r_c}\right)
\end{equation}
where $R$ is the system-size cutoff and $r_c$ is the effective vortex core radius.

Naively adding the core term to the image contribution gives
\begin{eqnarray}\label{eq:SemiFiniteSelfEnergy_naive}
    F_{\text{self}}(X,Y)
    =F_{\text{core}}+F_{\text{images}}(X,Y)
    =\frac{\Phi_0^2}{2\pi\mu_0\Lambda}
        \ln\left[
            \frac{2 W}{\pi r_c}
            \frac{\sin\left(\frac{\pi X}{W}\right)
                  \sinh\left(\frac{\pi Y}{W}\right)}
                 {\sqrt{\sin^2\left(\frac{\pi X}{W}\right)+
                        \sinh^2\left(\frac{\pi Y}{W}\right)}}
        \right]
\end{eqnarray}
which diverges to $-\infty$ as the vortex approaches any strip boundary ($X\to 0,W$ or $Y\to 0$).
Physically, the self-energy should remain finite at the edges, where the order parameter is already suppressed.
To regularize this edge behavior while preserving the correct logarithmic scaling in the interior, we add an offset to the argument of the logarithm,
\begin{equation}\label{eq:SemiInfiniteSelfEnergy}
    \boxed{
    F_{\text{self}}(X,Y)
    =\frac{\Phi_0^2}{2\pi\mu_0\Lambda}
        \ln\left[
            \frac{2 W}{\pi r_c}
            \frac{\sin\left(\frac{\pi X}{W}\right)
                  \sinh\left(\frac{\pi Y}{W}\right)}
                 {\sqrt{\sin^2\left(\frac{\pi X}{W}\right)+
                  \sinh^2\left(\frac{\pi Y}{W}\right)}}+1
        \right].
    }
\end{equation}
This modification guarantees $F_{\text{self}}\to 0$ smoothly at all boundaries and approaches the unregularized expression away from the edges.
Its behavior is illustrated in Fig. \ref{fig:SemiInfinite_SelfEnergy_MagneticMoment}a through a 3D plot.

\subsubsection{Single-Vortex Magnetic Moment \texorpdfstring{$m$}{m} with Edge Screening}\label{sec:SemiInfiniteMagneticMoment}

\begin{summarybox}[Summary — Single-Vortex Magnetic Moment in a Semi-Infinite length Strip]
For a vortex located at $(X,Y)$ in a semi-infinite strip $0 < X < W$, $0 < Y < \infty$, 
we show that the areal magnetization is
\begin{equation}
    \mathcal{M}(x,y;X,Y)
    =\frac{\Phi_0}{\pi\mu_0 \Lambda} 
        \sum_{n=1}^{\infty} \frac{1}{n\pi} 
            \sin\left(\frac{n\pi X}{W}\right)
            \sin\left(\frac{n\pi x}{W}\right)
            \left[ 
                e^{-\tfrac{n\pi}{W}|y-Y|} 
                -e^{-\tfrac{n\pi}{W}|y+Y|} 
            \right].\nonumber
\end{equation}
Integrating $\mathcal{M}$ over the semi-infinite strip gives the total magnetic moment:
\[
    m(X,Y) = \frac{\Phi_0}{\mu_0 \Lambda}
             \left[
                   X(W-X) - 
                   \frac{8W^2}{\pi^3}
                   \sum_{n=1,3,5,\dots}^\infty\frac{1}{n^3}
                        \sin\left(\frac{n\pi X}{W}\right)
                        e^{-\tfrac{n\pi}{W}Y}
            \right].
\]
The infinite sum can be thought of as an \emph{end correction} due to the $y=0$ boundary.
If the vortex is far from the end of the strip ($Y \gg W$), the exponentials are negligible, and the magnetic moment reduces to the infinite-strip result: $m(X,Y) \approx\tfrac{\Phi_0}{\mu_0 \Lambda}X(W-X)$.

\end{summarybox}

Consider the Dirichlet Green function in an infinite strip $0<x<W$ [Eq. \eqref{eq:InfiniteStripPoisson}]. 
In Sec. \ref{sec:InfiniteStripMagneticMoment}, we obtained its explicit form, Eq. \eqref{eq:InfiniteDirectIntegration}, by direct integration method.
For the present semi-infinite strip geometry, the same strip modes govern the $x$-dependence; the only additional boundary is at $y=0$.
Thus, Eq. \eqref{eq:InfiniteDirectIntegration} serves as our starting point
\begin{eqnarray}
    G(x,y;X,0)=\sum_{n=1}^\infty\frac{1}{n\pi}
    \sin\left(\frac{n\pi X}{W}\right)
    \sin\left(\frac{n\pi x}{W}\right)
    \exp\left(-\frac{n\pi}{W}|y|\right).\nonumber
\end{eqnarray}
Now, for a semi-infinite strip ($0<x<W,~0<y<\infty$) with Dirichlet boundary at $y=0$, the corresponding Green function for a source at ($X,Y$) is obtained by the antisymmetric image construction in $y$ 
\begin{equation}\label{}
        G(x,y;X,Y)
        =\sum_{n=1}^\infty\frac{1}{n\pi}
            \sin\left(\frac{n\pi X}{W}\right)
            \sin\left(\frac{n\pi x}{W}\right)
            \left[
                \exp\left(-\frac{n\pi}{W}|y-Y|\right)
                -\exp\left(-\frac{n\pi}{W}|y+Y|\right)
            \right].
    \label{eq:SemiInfiniteStripGreenFuncitonDirectIntegration}
\end{equation}
This satisfies $G(x,0;X,Y)=0$ and reduces to Eq. \eqref{eq:InfiniteDirectIntegration} when the $y$-boundary is removed.

Therefore, the areal magnetization $\mathcal{M}$ for a vortex at $(X,Y)$ in a semi-infinite strip follows directly by multiplying Eq. \eqref{eq:SemiInfiniteStripGreenFuncitonDirectIntegration} with the prefactor $2\Phi_0/(\mu_0\Lambda)$
\begin{equation}\label{}
            \mathcal{M}(x,y;X,Y)=
            \frac{2\Phi_0}{\mu_0\Lambda}
            \sum_{n=1}^\infty\frac{1}{n\pi}
            \sin\left(\frac{n\pi X}{W}\right)
            \sin\left(\frac{n\pi x}{W}\right)
            \left[
            \exp\left(-\frac{n\pi}{W}|y-Y|\right)
            -\exp\left(-\frac{n\pi}{W}|y+Y|\right)
            \right].
\end{equation}
This representation makes explicit the subtraction of the image contribution, which enforces the boundary condition at $y=0$.

The total magnetic moment of a vortex in the semi-infinite strip is then obtained by integrating the areal magnetization over the strip geometry
\begin{equation}\label{eq:TotalMagnetic0}
    m(X,Y)
    =\int_0^W dx\int_0^\infty dy\;\mathcal{M}(x,y;X,Y).
\end{equation}

First we define an intermediate function $f(X,Y)$ for $Y<0$
\begin{eqnarray}
    f(X,Y)&&
    =\int_0^W dx \int_0^\infty dy\;
        \sum_{n=1}^\infty\frac{1}{n\pi}
            \sin\left(\frac{n\pi X}{W}\right)
            \sin\left(\frac{n\pi x}{W}\right)
            \exp\left(-\frac{n\pi}{W}|y-Y|\right)
    \nonumber\\&&
    =\int_0^W dx \int_0^\infty dy\;
    \sum_{n=1}^\infty\frac{1}{n\pi}
        \sin\left(\frac{n\pi X}{W}\right)
        \sin\left(\frac{n\pi x}{W}\right)
        \exp\left[-\frac{n\pi}{W}(y-Y)\right]
    \qquad\text{for $Y<0$, $\left|y-Y\right|=y-Y$}
    \nonumber\\&&
    =\sum_{n=1}^\infty\frac{1}{n\pi}
        \sin\left(\frac{n\pi X}{W}\right)
        \int_0^W dx\int_{-Y}^\infty du\;
            \sin\left(\frac{n\pi x}{W}\right)
            \exp\left(-\frac{n\pi}{W}u\right)
        \qquad\text{where }u=y-Y
    \nonumber\\&&
    =\sum_{n=1}^\infty\frac{1}{n\pi}
        \sin\left(\frac{n\pi X}{W}\right)
        \int_0^W dx\;\sin\left(\frac{n\pi x}{W}\right)
        \int_{-Y}^\infty du\;\exp\left(-\frac{n\pi}{W}u\right)
    \nonumber\\&&
    =\sum_{n=1}^\infty\frac{1}{n\pi}
        \sin\left(\frac{n\pi X}{W}\right)
        \left(\frac{1-\cos n\pi}{n\pi/W}\right)
        \left(\frac{\exp\left(\frac{n\pi Y}{W}\right)}{n\pi/W}\right)
    \nonumber\\&&
    =\frac{W^2}{\pi^3}
    \sum_{n=1}^\infty\frac{1}{n^3}
        \sin\left(\frac{n\pi X}{W}\right)
        \left(1-\cos (n\pi)\right)
        \exp\left(-\frac{n\pi}{W}|Y|\right).
    \label{eq:YNegative-f}
\end{eqnarray}
If $|Y|\gg W$, the sum converges exponentially fast, and only a few terms are needed. If $|Y|\ll W$, the sum still converges fairly fast because of the $1/n^3$ behavior. Thus, the above sum is amenable to numerical evaluation.

For $Y>0$, it is convenient to extend the $y$-integration to the full real line and subtract the unphysical contribution from $y<0$,
\begin{eqnarray}\label{eq:TotalMagnetic1}
    g(X,Y)&&
    =\sum_{n=1}^\infty\frac{1}{n\pi}
        \sin\left(\frac{n\pi X}{W}\right)
        \int_0^W dx\int_{-\infty}^\infty dy\;
            \sin\left(\frac{n\pi x}{W}\right)
            \exp\left(-\frac{n\pi}{W}|y-Y|\right)
    \nonumber\\&&
    -\sum_{n=1}^\infty\frac{1}{n\pi}
        \sin\left(\frac{n\pi X}{W}\right)
        \int_0^Wdx\int_{-\infty}^0 dy\;
            \sin\left(\frac{n\pi x}{W}\right)
            \exp\left(-\frac{n\pi}{W}|y-Y|\right).
\end{eqnarray}
We extend the $y$-integration to the entire real line to exploit symmetry, and subtract the unphysical contribution from $y<0$.
For the infinite strip (setting $Y=0$), the first double integral of Eq. \eqref{eq:TotalMagnetic1} simplifies using the Fourier identity to the compact quadratic form [see Sec. \ref{sec:InfiniteStripMagneticMoment}, Eqs. \eqref{eq:InfiniteStripMagneticMomentSeries}, \eqref{eq:InfiniteStripFourierSine1}, and \eqref{eq:InfiniteStripFourierSine2}]
\begin{eqnarray}\label{}
    \sum_{n=1}^\infty\frac{1}{n\pi}
        \sin\left(\frac{n\pi X}{W}\right)
        \int_0^W dx\int_{-\infty}^\infty dy\;
            \sin\left(\frac{n\pi x}{W}\right)
            \exp\left(-\frac{n\pi}{W}|y|\right)
    &&=2W^2\sum_{n=1}^\infty
        \frac{1-\cos (n\pi)}{(n\pi)^3}
        \sin\left(\frac{n\pi X}{W}\right)
    \nonumber\\&&
    =2W^2\frac{X}{4W}\left(1-\frac{X}{W}\right)
    =\frac{X}{2}\left(W-X\right).
    \label{eq:SemiInfiniteQuadraticTerm}
\end{eqnarray}
Using the quadratic dependence from Eq. \eqref{eq:SemiInfiniteQuadraticTerm}, we rewrite $g(X,Y)$ as the infinite-strip contribution minus the correction from the unphysical region $y<0$,
\begin{eqnarray}\label{eq:YPositive-g}
    g(X,Y)&&=\frac{X(W-X)}{2}
        -\sum_{n=1}^\infty\frac{1}{n\pi}
            \sin\left(\frac{n\pi X}{W}\right)
            \int_0^W dx\int_{-\infty}^0 dy\;
                \sin\left(\frac{n\pi x}{W}\right)
                \exp\left(-\frac{n\pi}{W}|y-Y|\right)
    \nonumber\\&&
    =\frac{X(W-X)}{2}
        -\sum_{n=1}^\infty\frac{1}{n\pi}
            \sin\left(\frac{n\pi X}{W}\right)
            \int_0^W dx\int_{-\infty}^0 dy\;
                \sin\left(\frac{n\pi x}{W}\right)
                \exp\left(+\frac{n\pi}{W}(y-Y)\right)
    \qquad\text{since}\;y-Y<0
    \nonumber\\&&
    =\frac{X(W-X)}{2}
        -\sum_{n=1}^\infty\frac{1}{n\pi}
            \sin\left(\frac{n\pi X}{W}\right)
            \int_0^W dx\int_Y^\infty du\;
                \sin\left(\frac{n\pi x}{W}\right)
                \exp\left(+\frac{n\pi}{W}u\right)
    \qquad\text{where}\;u=Y-y
    \nonumber\\&&
    =\frac{X(W-X)}{2}
        -\sum_{n=1}^\infty\frac{1}{n\pi}
            \sin\left(\frac{n\pi X}{W}\right)
            \int_0^W dx\;
                \sin\left(\frac{n\pi x}{W}\right)
            \int_Y^\infty du\;
                \exp\left(-\frac{n\pi}{W}u\right)
    \nonumber\\&&
    =\frac{X(W-X)}{2}-f(X,-Y).
\end{eqnarray}

To summarize, from the $Y<0$ evaluation, we obtained Eq. \eqref{eq:YNegative-f}
\begin{eqnarray}\label{}
        f(X,Y)
        =\frac{W^2}{\pi^3}\sum_{n=1}^\infty\frac{1}{n^3}
            \sin\left(\frac{n\pi X}{W}\right)
            \left(1-\cos (n\pi)\right)
            \exp\left(-\frac{n\pi}{W}|Y|\right).\nonumber
\end{eqnarray}
Note that $f(X,Y)$ depends only on $|Y|$ (i.e., it is even in $Y$).
Then, we derived Eq. \eqref{eq:YPositive-g}
\begin{eqnarray}\label{}
    g(X,Y)=\frac{X(W-X)}{2}-f(X,-Y)
    \qquad\qquad(Y>0).\nonumber
\end{eqnarray}
Because $f$ is even in $Y$, $f(X,-Y)=f(X,Y)$.
For $Y<0$, the same manipulations give
\begin{eqnarray}\label{}
    g(X,Y)=+f(X,-Y)
    \qquad\qquad(Y<0).\nonumber
\end{eqnarray}
So, picewise
\begin{eqnarray}
    g(X,Y)=
    \begin{cases}
        \frac{X(W-X)}{2}-f(X,-Y),&Y>0,\\
        ~~~~~~~~~~~+f(X,Y),&Y<0.
    \end{cases}
    \nonumber
\end{eqnarray}
Introducing the Heaviside and sign functions
\begin{eqnarray}
    \mathrm{H}(Y)=
        \begin{cases}
            1,&Y>0\\
            0,&Y\leq0
        \end{cases},
    \qquad
    \mathrm{sgn}(Y)=
        \begin{cases}
            1,&Y>0\\
            0,&Y=0\\
            -1,&Y<0.
        \end{cases}\nonumber
\end{eqnarray}
Combining the two branches [Eqs. \eqref{eq:YNegative-f} and \eqref{eq:YPositive-g}], we arrive
\begin{eqnarray}\label{eq:h}
    h(X,Y)=\frac{X(W-X)}{2}\mathrm{H}(Y)  
    -\frac{W^2}{\pi^3}\sum_{n=1}^\infty\frac{1}{n^3}
    \sin\left(\frac{n\pi X}{W}\right)
    \left(1-\cos n\pi\right)
    \exp\left(-\frac{n\pi}{W}|Y|\right)\mathrm{sgn}(Y).
\end{eqnarray}
The Heaviside step function $\mathrm{H}(Y)$ accounts for the additional contribution $X(W-X)/2$ that appears only for $Y>0$, while the sign function $\mathrm{sgn}(Y)$ ensures the series flips sign across $Y=0$.
The exponential factor $\exp\left(-\frac{n\pi}{W}|Y|\right)$ guarantees rapid convergence by suppressing contribution at large $|Y|$.

\begin{figure}[h]
    \centering
    \begin{minipage}[]{0.5\textwidth}\vspace{0pt}
        \panel{(a)}{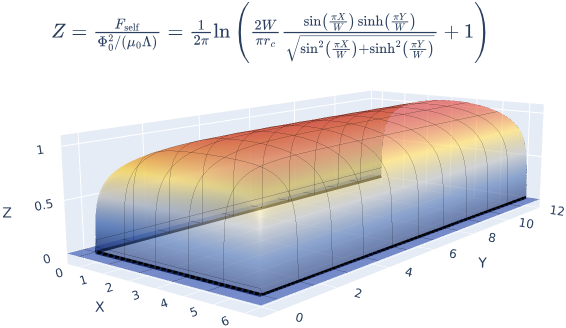}
    \end{minipage}\hfill
    \begin{minipage}[]{0.5\textwidth}\vspace{0pt}
        \panel{(b)}{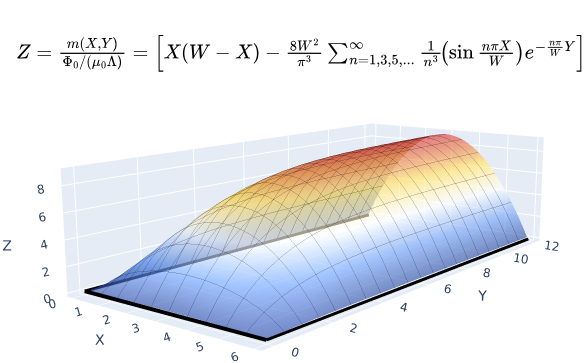}
    \end{minipage}
    \par\vspace{1ex}
    \caption{\justifying\\
    a) 3D visualization of the vortex self-energy $F_{\text{self}}(X,Y)/\left[\Phi_0^2/\left(\mu_0\Lambda\right)\right]$ [Eq. \eqref{eq:SemiInfiniteSelfEnergy}] in a semi-infinite strip of width $W=6$, with cutoff $r_c=0.005$. 
    The potential decays toward zero away from the edges and diverges logarithmically at $X=0,W$, and $Y=0$; values are clipped in the plot for clarity.\\
    b) 3D map of the vortex magnetic moment $m(X,Y;W)/\left[\Phi_0/(\mu_0\Lambda)\right]$ as a function of the vortex position $(X,Y)$ in a \emph{semi-infinite strip} with $W=6$ [Eq. \eqref{eq:SemiInfiniteStrip_m}]. 
    The magnetic moment exhibits a quadratic dependence $X(W-X)$ in the transverse ($X$) direction (vanishing at $X=0,W$) and an exponential suppression $\propto e^{-n\pi Y/W}$ with distance $Y$ from the boundary due to edge screening. 
    This highlights the combined effects of geometric confinement and boundary-induced screening on vortex magnetization.
    }
    \label{fig:SemiInfinite_SelfEnergy_MagneticMoment}
\end{figure}

Now, we wish to find Eq. \eqref{eq:TotalMagnetic0},
\begin{equation}\label{eq:SemiInfiniteMagneticMoment_h}
    m(X,Y)=\frac{2\Phi_0}{\mu_0\Lambda}\left[h(X,Y)-h(X,-Y)\right],
    \qquad\text{where }Y>0.
\end{equation}
Substituting Eq. \eqref{eq:h} into Eq. \eqref{eq:SemiInfiniteMagneticMoment_h}, and evaluating $h(X,Y)$ for $Y>0$ and $h(X,-Y)$ for $-Y<0$, we find that the step and sign functions contribute a quadratic term $\frac{X(W-X)}{2}$ together with a doubled exponential series.
This lead to
\begin{eqnarray}\label{eq:SemiInfiniteMagneticMoment_h2}
    m(X,Y)&&
    =\frac{2\Phi_0}{\mu_0\Lambda}
        \left\{
            \frac{X(W-X)}{2}
            -\frac{W^2}{\pi^3}
                \sum_{n=1}^\infty\frac{1}{n^3}
                    \sin\left(\frac{n\pi X}{W}\right)
                    \left(1-\cos n\pi\right)
                    \exp\left(-\frac{n\pi}{W}Y\right)
        \right.\nonumber\\&&\left.\qquad\qquad\qquad\qquad~~~
            -\frac{W^2}{\pi^3}
                \sum_{n=1}^\infty\frac{1}{n^3}
                    \sin\left(\frac{n\pi X}{W}\right)
                    \left(1-\cos n\pi\right)
                    \exp\left(-\frac{n\pi}{W}Y\right)\right\},
    \nonumber
\end{eqnarray}
\begin{eqnarray}\label{eq:SemiInfiniteMagneticMoment_h3}    
    \therefore\qquad
    m(X,Y)
    =\frac{\Phi_0}{\mu_0\Lambda}
        \left[ 
            X(W-X)
            -\frac{4W^2}{\pi^3}
            \sum_{n=1}^\infty\frac{1}{n^3}
                \sin\left(\frac{n\pi X}{W}\right)
                \left(1-\cos n\pi\right)
                \exp\left(-\frac{n\pi}{W}Y\right) 
        \right].
\end{eqnarray}
Using $1-\cos(n\pi)=2$ for odd $n$ (and $0$ for even $n$), Eq. \eqref{eq:SemiInfiniteMagneticMoment_h3} reduces to (see Fig. \ref{fig:SemiInfinite_SelfEnergy_MagneticMoment}b)
\begin{eqnarray}\label{eq:SemiInfiniteStrip_m}
    \boxed{
    m(X,Y)=\frac{\Phi_0}{\mu_0\Lambda}\left[X(W-X)-\frac{8W^2}{\pi^3}
    \sum_{n=1,3,5,...}^\infty\frac{1}{n^3}
    \sin\left(\frac{n\pi X}{W}\right)
    \exp\left(-\frac{n\pi}{W}Y\right)\right].
    }
\end{eqnarray}

\subsubsection{Vortex–Vortex Interaction \texorpdfstring{$F_{vv}$}{Fvv} with Edge Screening}\label{app:SemiInfiniteVortexVortex}

\begin{summarybox}[Summary — Vortex–Vortex Interaction in a Semi-Infinite Length Strip]
The interaction free energy $F_{vv}(x,y;X,Y)$ of two vortices at $(x,y)$ and $(X,Y)$ inside a semi-infinite strip $0<x<W$, $y>0$ is
\begin{equation}
    F_{vv}(x,y;X,Y) 
    =\frac{\Phi_0^2}{2\pi\mu_0\Lambda} 
        \ln\left[
            \frac{\cos\left(\tfrac{\pi(x+X)}{W}\right)-
                  \cosh\left(\tfrac{\pi(y-Y)}{W}\right)}
                 {\cos\left(\tfrac{\pi(x-X)}{W}\right)-
                  \cosh\!\left(\tfrac{\pi(y-Y)}{W}\right)}
            \;\times\;
            \frac{\cos\left(\tfrac{\pi(x-X)}{W}\right)-
                  \cosh\left(\tfrac{\pi(y+Y)}{W}\right)}
                 {\cos\!\left(\tfrac{\pi(x+X)}{W}\right)-
                  \cosh\!\left(\tfrac{\pi(y+Y)}{W}\right)}
        \right].\nonumber
\end{equation}
The interaction free energy $F_{vv}$ is always repulsive and symmetric under exchange of the two vortices.  It diverges logarithmically when the vortices coincide ($x\to X$ and $y\to Y$).  It tends to zero as either $(x,y)$ or $(X,Y)$ approaches any edge of the strip.
This expression captures the combined effects of geometric confinement and edge screening on vortex-vortex interactions in a semi-infinite length strip.
\end{summarybox}

\begin{figure}[h]
    \centering
    \begin{minipage}{0.55\textwidth} 
        \includegraphics[width=\linewidth]{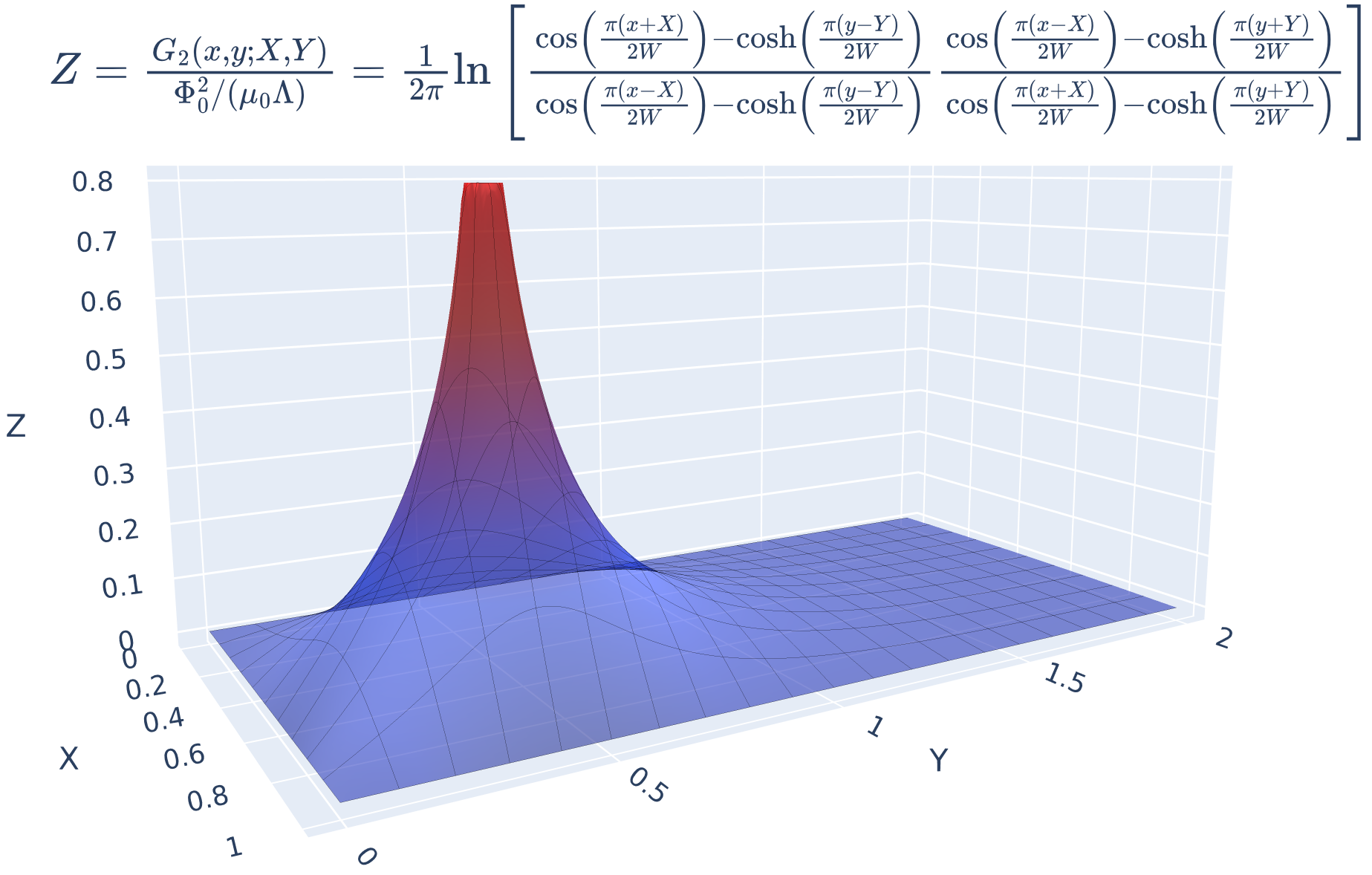}
    \end{minipage}
    \hfill
    \begin{minipage}{0.40\textwidth}
        \caption{\justifying
        3D visualization of the pairwise interaction free energy $F_{vv}(X,Y;x,y)/\left[\Phi_0^2/(\mu_0\Lambda)\right]$, Eq. \eqref{eq:SemiInfiniteStripFvv}, in a semi-infinite strip of width $W=1$.
        The plot shows the interaction energy as a function of vortex position $(x,y)$ for a reference vortex at fixed position  $(X,Y)=(0.7,0.4)$.
        The logarithmic divergence near vortex coincidence is evident, while the interaction decays smoothly with increasing separation, consistent with the expected repulsive behavior of Pearl vortex.
        }
        \label{fig:SemiInfiniteStripFvv}
    \end{minipage}
\end{figure}

Recall from Eq. \eqref{eq:SemiInfiniteTotalV} that the potential in a semi-infinite strip ($0<x<W,~0<y$) can be constructed using the method of images to enforce vanishing boundary conditions at $x=0,W$ and $y=0$.
For a vortex located at $(X,Y)$, the resulting potential at an observation point $(x,y)$ is
\begin{equation}\label{}
    V(x,y)
    =\frac{1}{2\pi}
    \operatorname{Re}\left\{
        \ln\left[
            \frac{\sin\left(\frac{\pi(z+X-iY)}{2W}\right)}  
                 {\sin\left(\frac{\pi(z-X-iY)}{2W}\right)}
            \times
            \frac{\sin\left(\frac{\pi(z-X+iY)}{2W}\right)}
                 {\sin\left(\frac{\pi(z+X+iY)}{2W}\right)}
        \right]
    \right\},\qquad\text{where }z=x+iy.\nonumber
\end{equation}
The interaction energy of a second vortex at position $(x,y)$ with the original vortex at $(X,Y)$ can be obtained by analogy with 2D electrostatics.
For line charges, $U=\tfrac{\omega^2}{2\pi\epsilon_0}\ln\frac{1}{|\textbf{S}_1-\textbf{S}_2|}$, while in superconducting films the vortex interaction takes the form $F_{vv}=\frac{\Phi_0^2}{\pi\mu_0\Lambda}\ln\frac{1}{|\textbf{S}_1-\textbf{S}_2|}$.
Comparing prefactors shows that
\begin{equation}\label{eq:SemiInfiniteStripFvv0}
    F_{vv}(x,y;X,Y)
    =\frac{\Phi_0^2}{\pi\mu_0\Lambda}
        \ln\left|
            \frac{\sin\left(\frac{\pi(z+X-iY)}{2W}\right)}
                 {\sin\left(\frac{\pi(z-X-iY)}{2W}\right)}
            \times
            \frac{\sin\left(\frac{\pi(z-X+iY)}{2W}\right)}
                 {\sin\left(\frac{\pi(z+X+iY)}{2W}\right)}
        \right|.
\end{equation}
In terms of real-valued quantities only (see Fig. \ref{fig:SemiInfiniteStripFvv}),
\begin{eqnarray}\label{eq:SemiInfiniteStripFvv}
    \boxed{
    F_{vv}(x,y;X,Y)
    =\frac{\Phi_0^2}{2\pi\mu_0\Lambda}
        \ln\left[
            \frac{\cos\left(\frac{\pi(x+X)}{W}\right)-
                  \cosh\left(\frac{\pi(y-Y)}{W}\right)}
                 {\cos\left(\frac{\pi(x-X)}{W}\right)-
                  \cosh\left(\frac{\pi(y-Y)}{W}\right)}
            \times
            \frac{\cos\left(\frac{\pi(x-X)}{W}\right)-
                  \cosh\left(\frac{\pi(y+Y)}{W}\right)}
                 {\cos\left(\frac{\pi(x+X)}{W}\right)-
                  \cosh\left(\frac{\pi(y+Y)}{W}\right)}
        \right].
    }
\end{eqnarray}

\subsection{Vortex Lattice (VL) Without Pinning in Infinite Length Strips}\label{app:MinimizeGibbs}

In this section, we minimize the Gibbs free energy for an infinite-length strip as in Bronson et al. \cite{Bronson2006}, using the previously derived expressions for the self-energy $F_{\text{self}}$, magnetic moment $m$, and vortex-vortex interaction $F_{vv}$ [Eqs. \eqref{eq:InfiniteStripF_Self2}, \eqref{eq:InfiniteStripMagneticMoment}, and \eqref{eq:InfiniteStripFvv}, respectively].
The analysis considers varying numbers of vortices, fully accounting for their mutual interactions while neglecting pinning.

To simplify the problem, we rescale all relevant quantities into a dimensionless form.
Specifically, we define:
\begin{itemize}
    \item length unit: $L_0=W$
    \item magnetic field unit: $B_0=\Phi_0/W^2$
    \item energy unit: $E_0=\Phi_0^2/\left(\mu_0W\right)$.
\end{itemize}
With these definitions, we introduce dimensionless coordinates $\tilde{X}=X/W$, $\tilde{Y}=Y/W$, and other length scales such as the dimensionless vortex core radius $\tilde{r}_c=r_c/W$.
Likewise, the external field is expressed in terms of the dimensionless magnetic field $\tilde{B}=B/B_0$.
This rescaling exposes the universal structure of the vortex lattice problem, with the strip width $W$ providing the fundamental geometric and energetic scale.

In terms of these definitions, the Gibbs free energy of a vortex $G_v(\tilde{X};\tilde{B})$ and the two-vortex interaction free energy $F_{vv}(\tilde{x},\tilde{y};\tilde{X},\tilde{Y})$ can be written as 
\begin{eqnarray}
    &&G_v(\tilde{X};\tilde{B})=
    \frac{E_0}{\tilde{\Lambda}}
    \left\{
        \frac{1}{2\pi}
        \ln\left[
            \frac{2}{\pi\tilde{r}_c}
            \sin{\left(\pi\tilde{X}\right)}+1
        \right]  
        -\tilde{X}\left(1-\tilde{X}\right)\tilde{B}
    \right\},
    \\&&
    F_{vv}(\tilde{x},\tilde{y};\tilde{x},\tilde{Y})
    =\frac{E_0}{2\pi\tilde{\Lambda}}
    \ln\left\{
    \frac{\cos{\left[\pi\left(\tilde{x}+\tilde{X}\right)\right]}-
    \cosh{\left[\pi\left(\tilde{y}-\tilde{Y}\right)\right]}}
    {\cos{\left[\pi\left(\tilde{x}-\tilde{X}\right)\right]}-
    \cosh{\left[\pi\left(\tilde{y}-\tilde{Y}\right)\right]}}
    \right\}.
\end{eqnarray}
Because both energies carry a common prefactor $E_0/\tilde{\Lambda}$, we define the dimensionless quantities $\tilde{G}_v=(\tilde{\Lambda}/E_0)G_v$ and $\tilde{F}_{vv}=(\tilde{\Lambda}/E_0)F_{vv}$,
\begin{eqnarray}
    \boxed{
    \tilde{G}_v(\tilde{X};\tilde{B})=\frac{1}{2\pi}
    \ln\left[\frac{2}{\pi\tilde{r}_c}\sin{\left(\pi\tilde{X}\right)}+1\right]
    -\tilde{B}\tilde{X}\left(1-\tilde{X}\right),}
    \label{eq:TildeGv}
\end{eqnarray}
\begin{eqnarray}
    \boxed{
    \tilde{F}_{vv}(\tilde{x},\tilde{y};\tilde{X},\tilde{Y})=
    \frac{1}{2\pi}\ln\left\{
    \frac{
    \cos{\left[\pi\left(\tilde{x}+\tilde{X}\right)\right]}-
    \cosh{\left[\pi\left(\tilde{y}-\tilde{Y}\right)\right]}
    }
    {
    \cos{\left[\pi\left(\tilde{x}-\tilde{X}\right)\right]}-
    \cosh{\left[\pi\left(\tilde{y}-\tilde{Y}\right)\right]}
    }
    \right\}.}
    \label{eq:TildeFvv}
\end{eqnarray}
These forms are independent of $\Lambda$ and depend only on the scaled coordinates (and $\tilde{r}_c,\tilde{B}$).

In the absence of pinning, the equilibrium vortex arrangement minimizes the total Gibbs free energy
\begin{eqnarray}
    G_{\text{tot}}=G_v+F_{vv},
\end{eqnarray}
where $G_v$ denotes the single-vortex confinement energy (self contribution plus coupling to the external field, and edge effects), and $F_{vv}$ is the vortex-vortex interaction.

At low magnetic fields, $G_v$ is minimized for a \textbf{single vortex column} along the center of the strip ($X=\tfrac12$).
In this regime, the inter-vortex spacing along the column is dictated by $F_{vv}$ and decreases monotonically with increasing field.
At higher fields, vortex density increases and repulsion becomes more severe, and the balance between $F_{vv}$ and confinement from $G_v$ drives a transition from a single column to a \textbf{multi-column vortex lattice} spanning the strip width \cite{Bronson2006}.

\begin{figure}[h]
    \centering
    \begin{minipage}{0.25\textwidth} 
        \includegraphics[
        width=\linewidth,
        ]{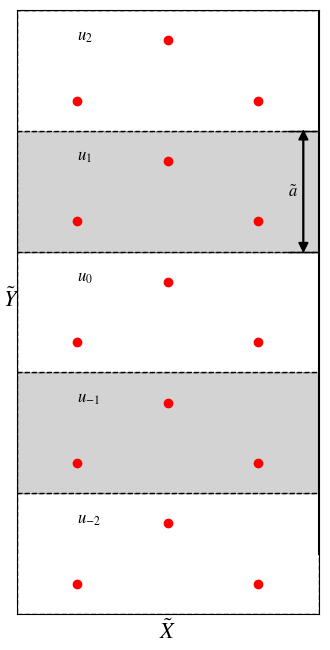}
    \end{minipage}
    \hfill
    \begin{minipage}{0.7\textwidth}
        \caption{\justifying
        Illustration of the numbering of periodic unit cells of a three-column vortex configuration.
        The strip is replicated along $\tilde{Y}$ into cells labeled by $u\in\mathbb{Z}$ with lattice spacing $\tilde{a}$; the shaded rectangle indicates the zeroth unit cell ($u=0$), which we call the motif.
        This construction underlines multi-column vortex lattices in the infinite strip.
        The first sume in Eq. \eqref{eq:InfinitStripGFEperUnitLength} accounts for the single-vortex (confinement) energy of the vortices inside the motif.
        The triple sum in Eq. \eqref{eq:InfinitStripGFEperUnitLength} runs over all vortex pairs in the motif and all of their periodic images (indexed by $u$).\\
        Figure \ref{fig:InfiniteStripOneToFourColumn}(a-d) shows explicit parametrizations of unit cells with multiple vortex columns.
        Each unit cell extends from $\tilde{Y}=0$ to $\tilde{Y}=\tilde{a}$ (shaded in Fig. \ref{fig:InfiniteStripOneToFourColumn}) and tiles periodically along $\tilde{Y}$ with spacing $\tilde{a}$.
        A cell contains $c$ vortices whose vertical positions are\\
         $~~~~~~~~~~~~~~~~~~~~~~~~~~~~~~~~~~~~~
            \tilde{Y}_i=\frac{\tilde{a}}{2}+(-1)^i\frac{\tilde{a}}{4},
            \qquad i=1,\dots,c,
        $\\
        depending on $\tilde{a}$.
        The horizontal positions within each motif are $\tilde{X}_1$, $\tilde{X}_2$, ..., $\tilde{X}_c$.
        By mirror symmetry about $\tilde{X}=\frac{1}{2}$, only $\tilde{X}_1$, $\tilde{X}_2$, ..., $\tilde{X}_{\left\lfloor{c/2}\right\rfloor}$ are independent (see Fig. \ref{fig:InfiniteStripOneToFourColumn}).
        }
        \label{fig:InfiniteStripSlice}
    \end{minipage}
\end{figure}

From $\tilde{G}_v$ and $\tilde{F}_{vv}$ [Eqs. \eqref{eq:TildeGv} and \eqref{eq:TildeFvv}], we obtain the dimensionless Gibbs free energy density of a $c$-column lattice, $\tilde{g}_c$, as the Gibbs free energy of a unit cell divided by the unit cell area ($\mathcal{A}=a\times W$).  In terms of the dimensionless lattice spacing is $\tilde{a}=a/W$,
\begin{eqnarray}\label{eq:InfinitStripGFEperUnitLength}
    \tilde{g}_c
        \left(\tilde{a};\{\tilde{X}_i\},\{\tilde{Y}_i\}\right)
    =\frac{1}{\tilde{a}}
    \left(
        \sum_{i=1}^c\tilde{G}_v(\tilde{X}_i;\tilde{B})+
        \frac{1}{2}
        \sum_{i=1}^c\sum_{j=1}^c\sum_{u\in\mathbb{Z}}
            \tilde{F}_{vv}
                \left(\tilde{X}_i,\tilde{Y}_i;\tilde{X}_j,\tilde{Y}_j+u\tilde{a}\right)
    \right).
\end{eqnarray}
Here $u\in\mathbb{Z}$ enumerates periodic images (unit cells) along $y$; the $j$-th vortex is shifted by $u\tilde{a}$ in the argument $\tilde{Y}_j+u\tilde{a}$.
The $1/2$ prefactor avoids for double-counting, and the self-term $(i=j,u=0)$ is excluded (see Fig. \ref{fig:InfiniteStripSlice}).

\begin{figure}[h]
  \centering
  \begin{minipage}[t]{0.2\textwidth}\vspace{0pt}
    \panel{(a)}{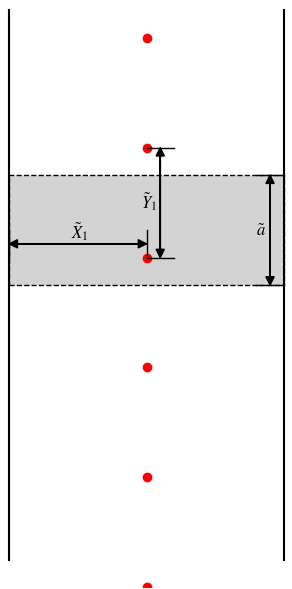}
  \end{minipage}\hfill
  \begin{minipage}[t]{0.2\textwidth}\vspace{0pt}
    \panel{(b)}{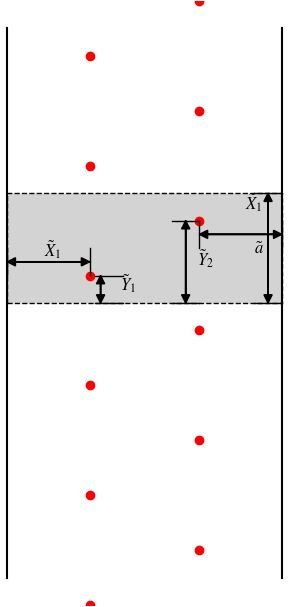}
  \end{minipage}\hfill
  \begin{minipage}[t]{0.2\textwidth}\vspace{0pt}
    \panel{(c)}{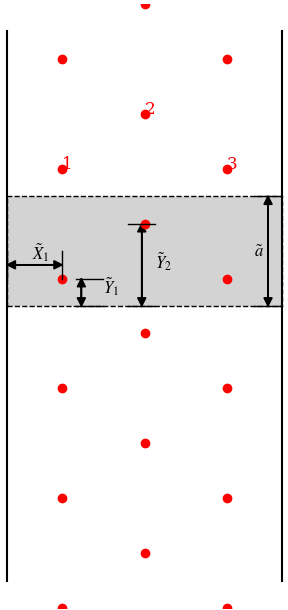}
  \end{minipage}\hfill
  \begin{minipage}[t]{0.2\textwidth}\vspace{0pt}
    \panel{(d)}{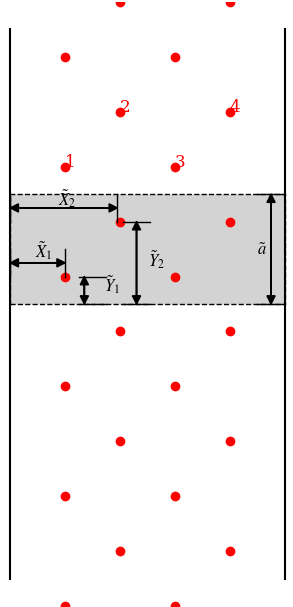}
  \end{minipage}
  \par\vspace{1ex}
  \caption{\justifying
  Parametrization of vortex positions in infinite-strip unit cells.
  Panels (a–d) illustrate motifs with (a) one ($c=1$), (b) two ($c=2$), (c) three ($c=3$), and (d) four ($c=4$) vortex columns.
  The shaded rectangle denotes the fundamental unit cell of height $\tilde{a}$, which tiles periodically along $\tilde{Y}$.
  Within each cell, vortices are specified by horizontal coordinates $\tilde{X}_i$ and vertical coordinates $\tilde{Y}_i$.
  By mirror symmetry across the strip midline, only half of the $\tilde{X}_i$ are independent; the remaining positions are fixed by reflection.
  The parametrization forms the basis for constructing multi-column vortex lattice in the infinite strip.
  }
  \label{fig:InfiniteStripOneToFourColumn}
\end{figure}

For computational efficiency, the triple sum in Eq. \eqref{eq:InfinitStripGFEperUnitLength} is decomposed into contributions from vortices within the \emph{same cell} and from vortices in \emph{different cells}.
This eliminates redundant evaluations and reduces the computational effort.
The resulting expression for the dimensionless Gibbs free energy per unit cell of a $c$-column lattice is
\begin{mdframed}[roundcorner=10pt]
\begin{align}\label{eq:SameCellDifferentCell}
    &\tilde{g}_c
        \left(\tilde{a};\{\tilde{X}_1, \tilde{X}_2, ...,\tilde{X}_{\lfloor c/2\rfloor}\}\right)
    \nonumber\\&\qquad\qquad
    =\frac{1}{\tilde{a}}
    \Bigg(
        \sum_{i=1}^c\tilde{G}_v(\tilde{X}_i;\tilde{B})+
        \sum_{i=1}^{c-1}\sum_{j=i+1}^c\tilde{F}_{vv}
            \left(\tilde{X}_i,\tilde{Y}_i;\tilde{X}_j,\tilde{Y}_j\right)
    +\sum_{i=1}^c\sum_{j=1}^c\sum_{u=1}^{\infty}
        \tilde{F}_{vv}
        \left(\tilde{X}_i,\tilde{Y}_i;\tilde{X}_j,\tilde{Y}_j+u\tilde{a}\right)
    \Bigg).
\end{align}
\end{mdframed}
The first term represents the confinement energy of individual vortices, the second accounts for pairwise interactions within the same unit cell (with $j>i$ to avoid double-counting), and the third term accounts for interactions with vortices in periodic images shifted by integer multiples of $\tilde{a}$.
Evaluation of Eq. \eqref{eq:SameCellDifferentCell} can be further optimized by omitting insignificant terms.
If $\left| u\tilde{a} + \tilde{Y}_j - \tilde{Y}_i \right| \gg 1$,
then $\tilde{F}_{vv} \sim \exp\left(-\pi (u\tilde{a} + \tilde{Y}_j - \tilde{Y}_i) \right)$.
Thus, if $u\tilde{a} > 40$ or $|\tilde{Y}_j-\tilde{Y}_i| > 40$, then $\tilde{F}_{vv} \lesssim 10^{-16}$.

To investigate vortex behavior beyond the threshold magnetic field---where the first vortex enters at the strip center---we minimize the dimensionless Gibbs free energy per unit cell $\tilde{g}_c$ with respect to the lattice spacing $\tilde{a}$ and column positions $\tilde{X}_i$.
The minimization is carried out using the L-BFGS-B optimization algorithm for infinite superconducting strips with width $W=\SI[parse-numbers=false]{7,\,10,\,\text{and }20}{\micro\meter}$ at fixed applied field $\tilde{B}$.

The numbers of vortex columns is determined by minimizing the Gibbs free energy, which decreases monotonically with increasing magnetic field.
To identify transitions between successive column numbers, we compare the Gibbs free energies $\tilde{g}_c$ for adjacent values of $c$.
The first column $(c=1)$ appears when $\tilde{g}_1<0$, defining the threshold field $B_L$.
As the field increases, we compare $\tilde{g}_1$ and $\tilde{g}_2$; if $\tilde{g}_2 < \tilde{g}_1$, the system favors two columns. 
Continuing in this manner, we locate each transition from $c\to c+1$ according to the criterion $\tilde{g}_{c+1}<\tilde{g}_{c}$.

Through the minimization, we compute the following vortex density and the total number of vortices in the strip:
\begin{itemize}
    \item From the optimized unit length $\tilde{a}$, the dimensionless unit cell density is
    \begin{eqnarray}
        n_a=\frac{1}{aW}\;\Longrightarrow\;
        \tilde{n}_{\tilde{a}} = \frac{1}{\tilde{a}} = \frac{W}{a}.
    \end{eqnarray}
    \item Accounting for $c$ columns across the strip width, the dimensionless vortex density becomes
    \begin{eqnarray}
        n_c=\frac{c}{aW}\;\Longrightarrow\;
        \tilde{n}_{c} = \frac{c}{\tilde{a}} = \frac{Wc}{a}.
    \end{eqnarray}
    $\tilde{n}_c$ coincides with the number of vortices in an segment of length $W$ or an area of $W^2$ in the strip and can therefore be computed from experimental data. $\tilde{n}_c$ is used in the main text to compare results from strips with different $W$ on the same rescaled axes (see Fig. 5). 
    \item For a finite strip of width $W$ and length $L$, the total number of vortices is then
    \begin{eqnarray}
        n_v = \tilde{n}_{c} \, \frac{L}{W}.
    \end{eqnarray}
\end{itemize}
\noindent Here, $\tilde{n}_c$ represents the dimensionless vortex number per unit cell, while the factor $L/W$ rescales it to the actual vortex count for a finite strip of width $W = \SI[parse-numbers=false]{7,\,10,\,20}{\micro\meter}$ and length $L = \SI{200}{\micro\meter}$. 
This step connects the theoretical density to experimentally measurable vortex counts.

\begin{figure}[h]
    \centering
    \includegraphics[width=0.95\textwidth]{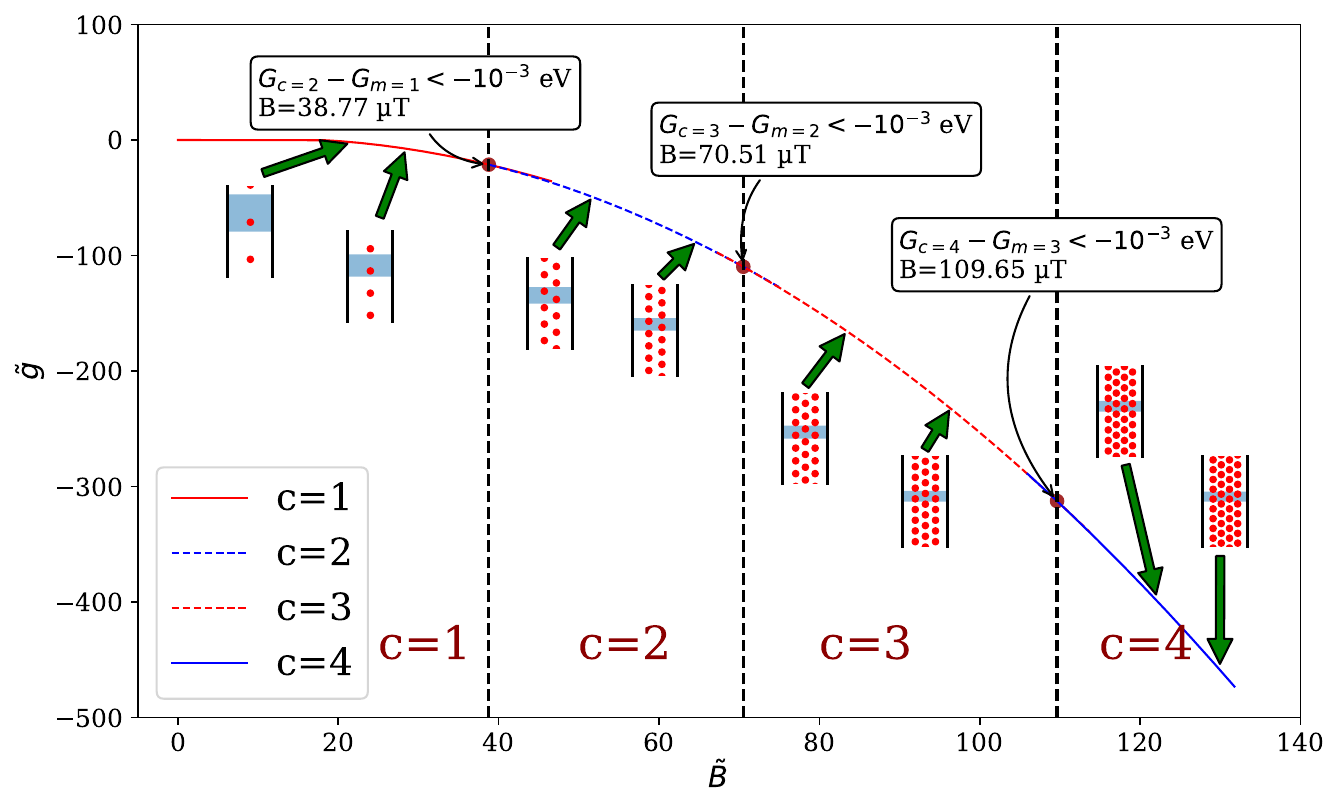}
    \caption{\justifying
    Gibbs free energy per unit cell, $G_c=\tilde{g}_cE_0$, as a function of applied magnetic field $B$ for an infinite superconducting strip of width $W=\SI{20}{\micro\meter}$ and vortex core radius $r_c=\SI{60}{\nano\meter}$ ($\tilde{r}_c=0.003$).
    The natural energy and field scale are $E_0=\Phi_0^2/(\mu_0 W)=1.062\,$eV and $B_0=\SI{5.170}{\micro\tesla}$. 
    Each curve corresponds to a configuration with $c$ vortex columns across the strip width. 
    As the field increases, free energy crossings $G_{c+1}<G_c$ mark successive transitions from $c\to c+1$, illustrated by the insets.
    These transitions reflects the competition between edge confinement and vortex-vortex repusion.}
    \label{fig:MinimizeGibbs}
\end{figure}

\begin{figure}[h]
    \centering
    \includegraphics[width=\textwidth]{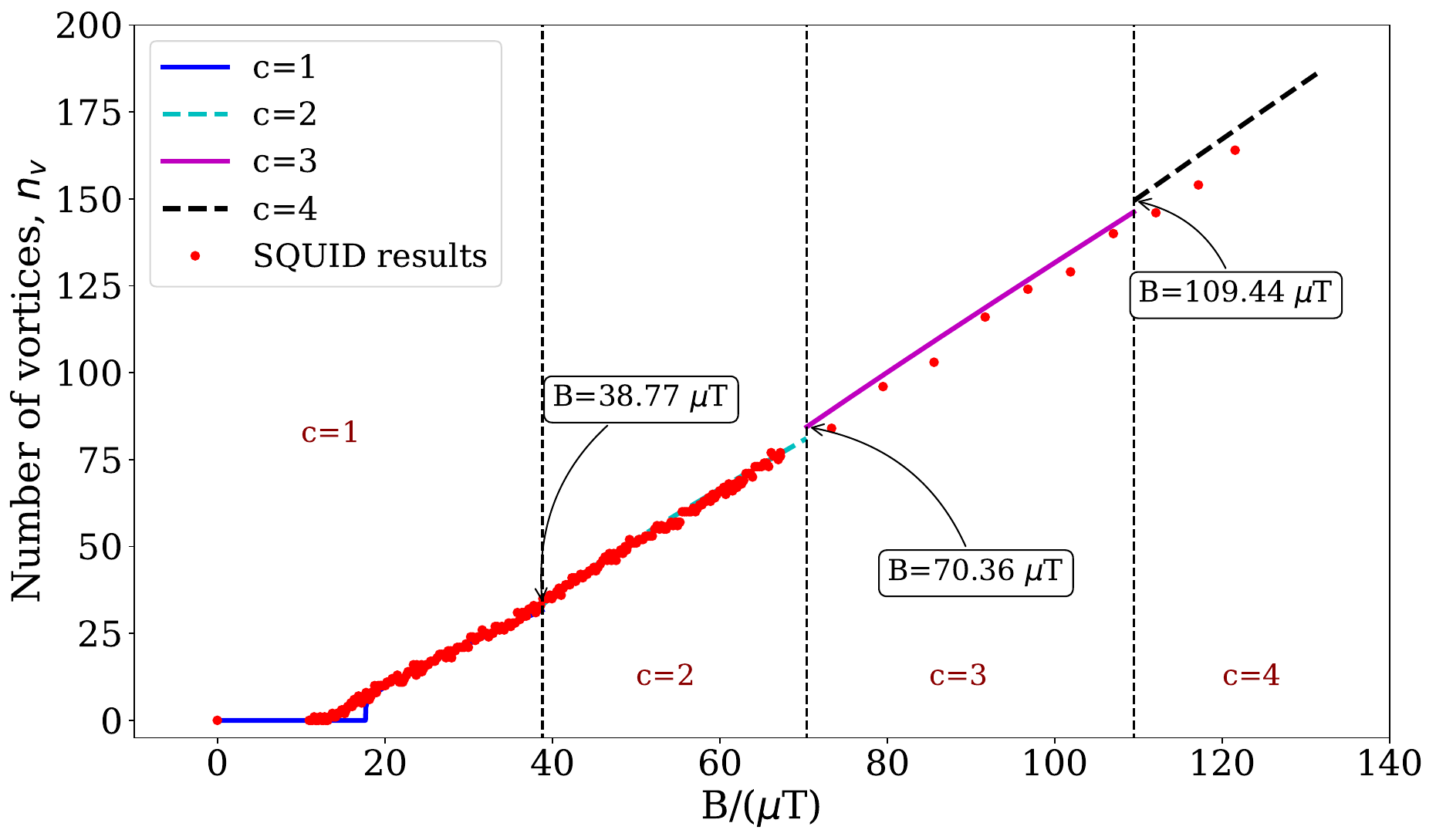}
    \caption{\justifying
    Simulated number of vortices $n_v=\tilde{n}_c\tfrac{L}{W}$ as a function of applied magnetic field $B$, compared with experimental SQUID measurements (red dots). 
    Simulations are performed for an infinite strip with $W=\SI{20}{\micro\meter}$, $L=\SI{200}{\micro\meter}$, $r_c=\SI{60}{\nano\meter}$ ($\tilde{r}_c=0.003$), and $B_0=\SI{5.170}{\micro\tesla}$.
    Distinct slope changes in the calculated curves correspond to successive transitions from $c\to c+1$ vortex columns, occurring at the transition fields marked by vertical dashed lines.
    The strong agreement between simulation and experiment confirms that vortex column transitions are governed by the balance between edge confinement and vortex-vortex repulsion.}
    \label{fig:SingleFiveRow_g_n}
\end{figure}

Figure \ref{fig:MinimizeGibbs} illustrates the minimization procedure for an infinite strip with $\tilde{r}_c = 0.003$, and the predicted vortex count for a $W\times L$ region is shown in Fig. \ref{fig:SingleFiveRow_g_n}, alongside measurements on a \SI{20}{\micro\meter}-wide strip.

\subsection{One-Dimensional Monte Carlo Simulated Annealing (SA) with Pinning in Semi-Infinite Length Strips}\label{app:MonteCarlo}

\subsubsection{Overview}
 

In the presence of defects, the strip no longer has translational symmetry.  A few-parameter vortex lattice ansatz such as Fig.~\ref{fig:InfiniteStripSlice} is no longer applicable.  Instead, one must minimize the free energy with respect to the positions of all vortices.  The free energy landscape has multiple local minima, corresponding to metastable vortex configurations.  Local optimization algorithms (such as conjugate gradient and BFGS) may become trapped in local minima.  Here we use global optimization algorithm (simulated annealing) with the intention of finding the global minimum of the free energy.  (We note that simulated annealing is not guaranteed to find the true global minimum, but neither do experiments necessarily attain the global minimum.)

At a fixed magnetic field, we minimize the total Gibbs free energy using simulated annealing (Metropolis Monte Carlo with exponential cooling), in order to approach thermodynamic equilibrium vortex configurations. 
Following Bronson \textit{et al.} \cite{Bronson2006}, we adopt an effective one-dimensional model, in which the strip length is much greater than the strip width ($L \gg W$), and vortices are restricted to occupy $N$ equally spaced discrete sites along the centerline. 
The site spacing $L/N$ is chosen much smaller than the characteristic inter-vortex separation in the single-column regime, i.e.\ $L/N \ll W$. 
This guarantees that each segment of length $W$ contains many available sites, ensuring that discretization does not artificially restrict the vortex positions. 
Each site is randomly assigned a pinning energy $E_{\mathrm{pin}}$. Most sites have $E_{\mathrm{pin}}=0$, while a fraction have $E_{\mathrm{pin}}<0$, representing defects energetically favorable for trapping vortices.

For a vortex at position $(X,Y)$ within a semi-infinite strip ($0<X<W;~0<Y<\infty$), we used Eq. \eqref{eq:SemiInfiniteSelfEnergy} for the vortex self-energy $F_{\mathrm{self}}$ and Eq. \eqref{eq:SemiInfiniteStrip_m} for the vortex magnetic moment $m$. 
The Gibbs free energy of a single vortex in an applied magnetic field $B$ is therefore
\begin{equation}
  G_v(X,Y;B) = F_{\mathrm{self}} - F_{mB} = F_{\mathrm{self}} - m(X,Y)B.
\end{equation}
The vortex–vortex interaction free energy $F_{vv}$ is computed by Eq. \eqref{eq:SemiInfiniteStripFvv}.  
The total Gibbs free energy, including self-energy, pinning and vortex-vortex interaction, is
\begin{equation}
  G_{\mathrm{tot+pin}}(X,Y;B) = G_v(X,Y;B) + E_{\mathrm{pin}} + F_{vv}(x,y;X,Y) .
\end{equation}

This one-dimensional model is applied to fields below the threshold at which vortex–vortex interactions promote the formation of a second vortex column. 
In narrower strips, this single-column regime persists up to relatively high fields, whereas in wider strips (e.g., \SI{20}{\micro\meter}) the second column forms at lower fields. 
Therefore, our simulations focus on strips of width \SI{7}{\micro\meter} and \SI{10}{\micro\meter}.

Monte Carlo updates consist of three possible moves: (i) adding a vortex to an empty site, (ii) removing a vortex from an occupied site, or (iii) shifting a vortex to an neighboring unoccupied site. 
Each move is accepted with a probability determined by the change in Gibbs free energy (including $G_v$, $F_{vv}$, and $E_{\mathrm{pin}}$), under an exponential annealing schedule that gradually cools the system. 
Flowcharts \ref{alg:I} and \ref{alg:II} summarize the algorithm and numerical procedure.

We adopt dimensionless units by setting $l_0=\SI{1}{\micro\meter}$, $\Phi_0 = \mu_0 = k_B = 1$, and defining the magnetic field scale $B_1=\Phi_0/l_0^2=\SI{2.1}{\milli\tesla}$. 
Energies are expressed in units of $\Phi_0^2/(\mu_0 l_0)$, lengths ($W,L,d,r_c$) are scaled by $l_0$, and magnetic fields by $B_1$. 
With this normalization, equilibrium vortex configurations in the absence of pinning are independent of the Pearl length $\Lambda$, provided $\Lambda\gg W$.
In contrast, pinning energies $E_{\mathrm{pin}}$ remain correlated with the natural energy scale $\Phi_0^2/(\mu_0 \Lambda)$.

\RestyleAlgo{ruled}
\begin{algorithm}[h]
\caption{Initialization of the  simulated annealing procedure.}
\label{alg:I}
Choose system parameters: $W,\, L,\, r_c,\, c$.\\
Set number of sites: $N=100$ or $200$.\\
Define field array: $\mathbf{B}=\{B_1,\dots,B_{\max}\}$.\\
Set maximum MC runs: $t_{\text{max}}{=}8000$.\\
Initialize storage: $n_v$ (array of vortex counts), 
$E_{\text{pin}}$ (random pinning energies for $N$ sites).\\
\For{$B\in \mathbf{B}$}{
  Initialize vortex occupancy array: $n_y \leftarrow \text{zeros}(N)$.\\
  Set initial temperature: $kT \leftarrow 0.5$.\\
  \For{$Y \text{ in } N \text{ sites}$}{
     Calculate 1D array $G_v(W,L,r_c,B,W/2,Y)$.\\
  }
  Add elementwise $E_{\text{pin}}$ to $G_v$.\\
  \For{$Y_1 \text{ in } N \text{ sites}$}{
    \For{$Y_2 \text{ in } N \text{ sites}$}{
        Calculate 2D array $F_{vv}(W,L,W/2,Y_1,W/2,Y_2)$.\\
     }
  }
  \For{$t=0,\ldots,t_{\max}$}{
     Perform Metropolis update using $G_v, F_{vv}, n_y$ \tcp{Do Metropolis step}
     Monte Carlo simulation shown in Algorithm~\ref{alg:II}.\\
     Update temperature: $kT \gets kT \times 0.997$.\\
     Record vortex number: $n_v(B)=\sum_{i=1}^N n_{y_i}$.\\
  }
}
\end{algorithm}

\begin{algorithm}[h]
    \caption{Metropolis update($G_v,F_{vv},n_y$)}\label{alg:II}
    $RAND\equiv\text{Uniform random distribution in a range (0,1)}$\\
    \eIf(\tcp*[h]{Add or Remove a vortex}){$RAND<0.5$}
        {
            $y_i \gets$ a random site from $N$\\
            \eIf(\tcp*[h]{Add a vortex}){Site is empty, $n_{y_i}=0$}
                {
                    $\Delta G=G_v[y_i]+(F_{vv}[y_i]\times n_y)$\\
                    \If{$RAND<\exp{(-\frac{\Delta G}{kT})}$}{$n_y[y_i]=1$}
                }(\tcp*[h]{Remove a vortex})
                {
                    $\Delta G=-G_v[y_i]-(F_{vv}[y_i]\times n_y)$\\
                    \If{$RAND<\exp{(-\frac{\Delta G}{kT})}$}{$n_y[y_i]=0$}
                }
        }(\tcp*[h]{Move a vortex to the adjacent site})
        {
            \If{$\sum_{i=1}^N n_{y_i}=0$}
            {
                Break the loop\\
            }
            \While{$n_y[y_i]=0$}{$y_i \gets$ a random site from $N$}
            Choose Randomly direction to move (1 or -1)\\
            $y_f=y_i+(1~\text{or}~-1)$\\
            \If{$0\leq y_f< y_{max}~~~and~~~n_y[y_f]=0$}
                {
                    $\Delta G=\left(G_v[y_f]-G_v[y_i]\right)
                    +\left[\left(F_{vv}[y_f]-F_{vv}[y_i]\right)\times n_y)\right]
                    -F_{vv}[y_f][y_i]$
                }
            \If{$RAND<\exp{(-\frac{\Delta G}{kT})}$}
                {
                    $n_y[y_i]=0$\\
                    $n_y[y_f]=1$\\
                }
        }
\end{algorithm}

\subsubsection{Choice of parameters for simulations}

For a given strip width $W$, the simulated vortex-density curves $n_v(B)$ are set by two inputs: the vortex core radius $r_c$ and the pinning-energy distribution $E_{\mathrm{pin}}$.
We specify the latter by choosing a discrete set of pinning energies $\{E_i\}\le0$ and assigning probabilities $\{p_i\}$ such that $\sum_i p_i=1$. Varying $r_c$ primarily shifts the curve horizontally, through the $r_c$-dependence of the single-vortex threshold field $B_L$ (see Fig. \ref{fig:MC_rc}). Changing $\{E_i,p_i\}$ primarily affects the curve shape near threshold and has less impact once vortex-vortex spacings approach $W$. Pinning can only lower site energies, it therefore reduces the threshold field relative to the case without pinning.

For each strip, we choose the vortex core radius $r_c$ before adding pinning sites to the simulation. $r_c$ is chosen by matching the experimental data at higher magnetic fields above the shoulder that are not strongly affected by pinning to the simulated curve without pinning.
For example, $r_c=\SI{40}{\nano\meter}$ reproduces the behavior of the \SI{7}{\micro\meter} strip (Fig. \ref{fig:MC_rc}) in magnetic fields greater than \SI{140}{\micro\tesla}.
Once $r_c$ is fixed, the pinning energy distributions is adjusted to capture the shape of the data at lower fields, particularly its low-field tail (Fig. \ref{fig:MC_parameters}).
This two-step procedure separates the roles of $r_c$ (onset) and $E_\mathrm{pin}$ (broadening), providing a systematic comparison between simulation and experiment.

For each strip, we performed simulations with 30 independent realizations of pinning energy configurations, all generated from the same probability distribution. 
Figure \ref{fig:MC_std} presents the results for the \SI{7}{\micro\meter} strip from S2.

\subsubsection{Additional experimental support for the presence of pinning: shallow tails observed for S1 strips}
\begin{figure}
    \centering
    \includegraphics[width = 0.5\textwidth]{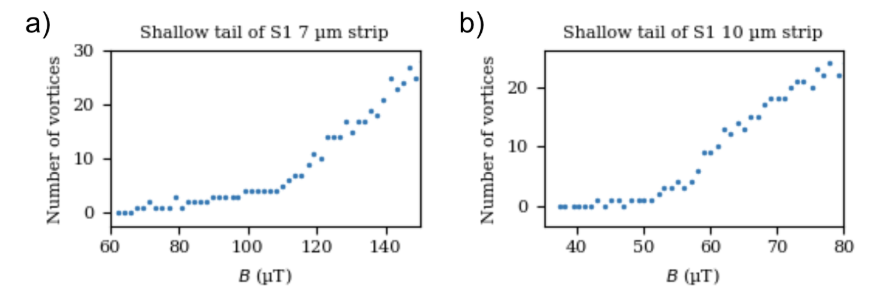}
    \caption{Zoom-in on the shallow tail observed in the 7 µm and 10 µm strips from S1. This data is the same as in Fig. 4 in the manuscript reporting absolute vortex numbers rather than scaling the vertical axis to report vortex density.}
    \label{fig:S1_tail}
\end{figure}
The previous sections show that by introducing pinning sites to the strips in numerical simulation, we can reproduce the shallow tails seen in the experimental data. Here in this section, we provide experimental evidence showing that such shallow tails are indeed caused by strong pinning sites on the strips.

\begin{figure}
    \centering
    \includegraphics[width = 0.9\textwidth]{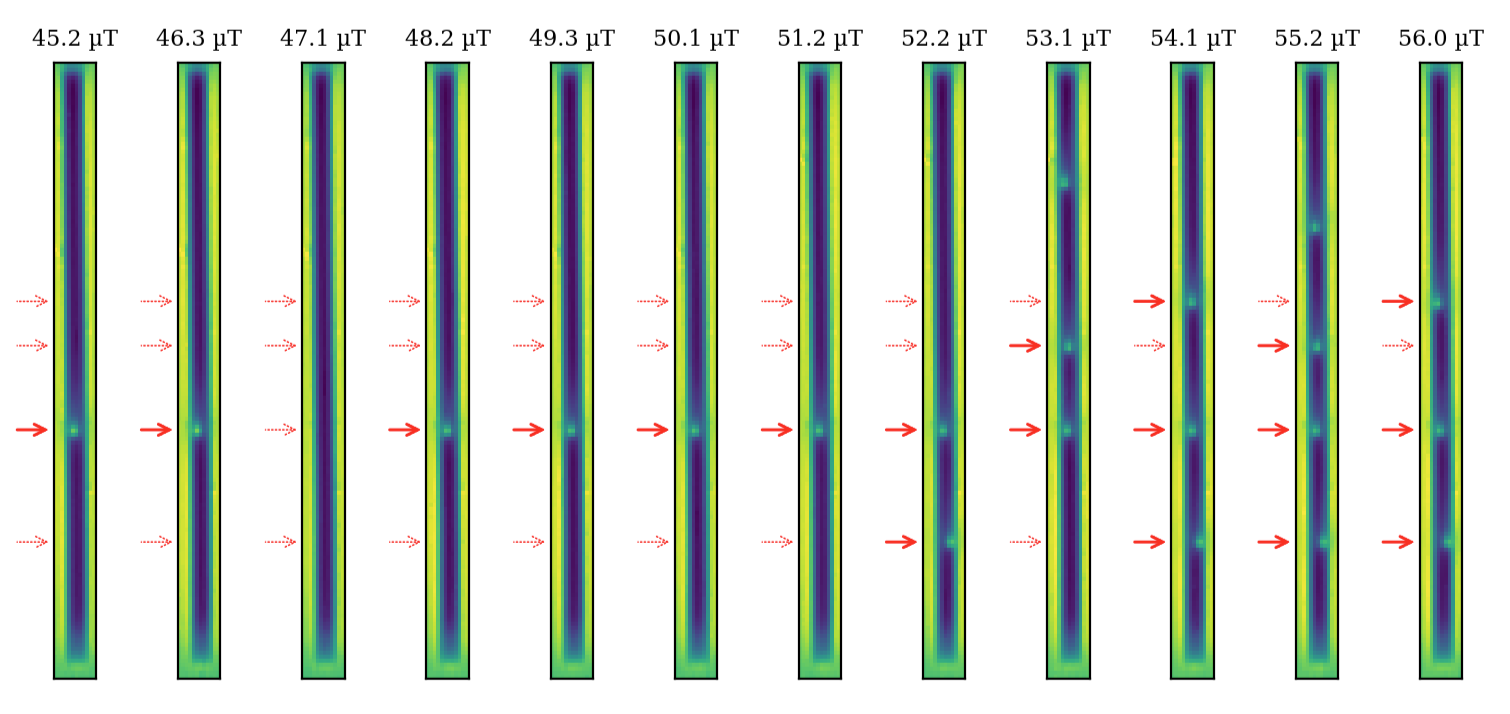}
    \caption{Scanning SQUD images of the 10 µm strip on S1 cooled under magnetic fields corresponds to the shallow tail in Fig. 5. Four locations where vortices repeatedly appear are labelled with red arrows. Such locations are the pinning sites that lower the threshold field of the strip and induce the shallow tail in the n vs B plot.}
    \label{fig:S1_10um_tail}
\end{figure}
Fig.~\ref{fig:S1_tail} zooms in to the n vs B plot's shallow tails of the \SI{7}{\micro\meter} and \SI{10}{\micro\meter} strips. The discrete steps show individual vortices populating the strips at low field. Upon inspecting the images associated with thses data points, as shown in Fig.~\ref{fig:S1_10um_tail} for the \SI{10}{\micro\meter} strip on S1, we find that the vortices repeatedly occur at a few select locations within the strips. We associate these locations with strong pinning sites.

\begin{figure}[h]
    \centering
    \begin{minipage}{0.55\textwidth} 
        \includegraphics[width=\linewidth]{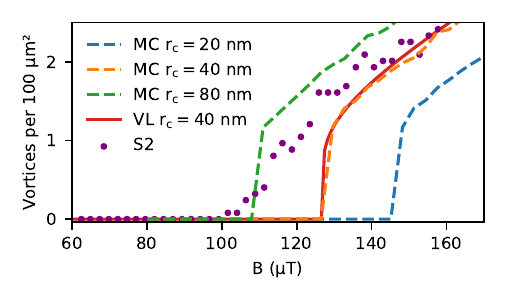}
    \end{minipage}
    \hfill
    \begin{minipage}{0.4\textwidth}
        \caption{\justifying
        Effect of vortex core radius $r_c$ on the simulated vortex number $n_v(B)$ for a strip of width $W=\SI{7}{\micro\meter}$. 
        Dotted lines show Monte Carlo simulated annealing (SA) results without pinning for $r_c=\SI[parse-numbers=false]{20,\,40,\,80}{\nano\meter}$}.
        Increasing $r_c$ lowers the threshold field $B_L$, shifting the curves to the left.
        The solid line shows vortex lattice (VL) deterministic free energy minimization with $r_c =\SI{40}{\nano\meter}$, which closely matches the SA simulation for the same $r_c$. 
        Points are experimental data for the \SI{7}{\micro\meter} strip from S2.
        The agreement demonstrates consistency between SA and VL approaches and confirms the role of $r_c$ in setting $B_m$.
        \label{fig:MC_rc}
    \end{minipage}
\end{figure}

\begin{figure}[h]
    \centering
    \begin{minipage}[]{0.5\textwidth}\vspace{0pt}
        \panel{(a)}{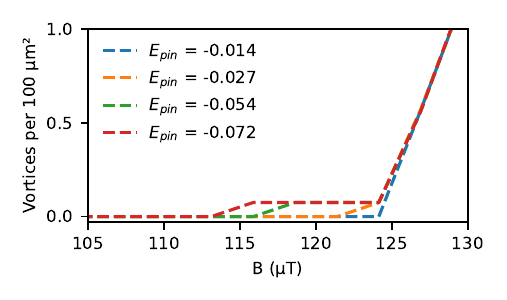}
    \end{minipage}\hfill
    \begin{minipage}[]{0.5\textwidth}\vspace{0pt}
        \panel{(b)}{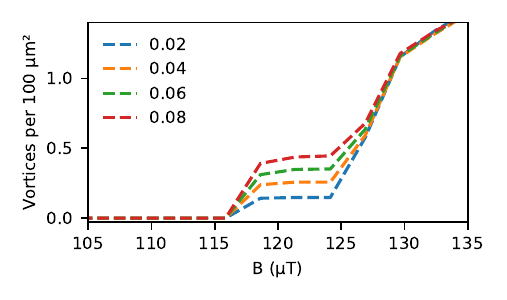}
    \end{minipage}
    \par\vspace{1ex}
    \caption{\justifying
    Influence of pinning on the simulated vortex number $n_v(B)$ in a \SI{7}{\micro\meter} width, \SI{200}{\micro\meter} length strip with $r_c=\SI{40}{\nano\meter}$}, obtained from SA simulations.
    a) Effect of varying pinning energy $E_{\mathrm{pin}}$ at fixed probability 0.01 per site.
    Increasing the magnitude of $E_{\mathrm{pin}}$ shifts the onset of vortex entry to lower fields.
    b) Effect of varying pinning probability for a fixed pinning energy $E_{\text{pin}} = -0.054$.
    Larger probabilities enhance the overall impact of pinning, broadening the tail of the vortex-field curve.
    Pinning energies are expressed in units of $\Phi_0^2/(\mu_0 \Lambda)$.
    \label{fig:MC_parameters}
\end{figure}

\begin{figure}[h]
    \centering
    \begin{minipage}{0.55\textwidth} 
        \includegraphics[width=\linewidth]{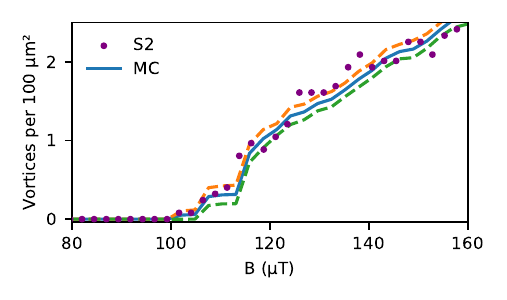}
    \end{minipage}
    \hfill
    \begin{minipage}{0.4\textwidth}
    \caption{\justifying
        Simulated annealing (SA) results averaged over 30 independent runs with the same pinning energy probability distribution for a \SI{200}{\micro\meter} long, \SI{7}{\micro\meter} width strip containing 100 sites. 
        The solid line denotes the ensemble average, while the dotted lines indicate the average $\pm$ one standard deviation at each field. 
        Experimental data for the \SI{7}{\micro\meter} strip from S2 are shown as purple points.}
    \label{fig:MC_std}
    \end{minipage}
\end{figure}

\subsection{Pinning Energy: Gibbs Free Energy of a Vortex at a Circular Hole}

To connect the abstract pinning energies in the simulation with physical pinning centers in superconducting films, we model defects as circular holes of radius $r_h$.
A vortex located at such a hole experiences a reduction in Gibbs free energy relative to a uniform film, which defines the effective pinning energy.
This reduction arises from two contributions: a change in the vortex self-energy, $\Delta F_{\text{self}}=F_{\text{self}}^{\text{hole}}-F_{\text{self}}^{\text{film}}$, and a change in the vortex-field interaction energy, $\Delta F_{mB}=B\left(m_{\text{self}}^{\text{hole}}-m_{\text{self}}^{\text{film}}\right)$.

Closed-form approximations for the Gibbs free energy reduction at a circular hole are obtained from variational Ginzburg-Landau arguments (a detailed derivation will be presented elsewhere). 
The analytical limits for both small $(r_h\lesssim \xi)$ and large hole radii $(\xi\lesssim r_h\ll \Lambda)$ are:
\begin{eqnarray}\label{eq:FselfPin}
    \Delta F_{\text{self}}=\frac{\Phi_0^2}{2\pi\mu_0 \Lambda}
    \left\{
    \begin{array}{cc}
      -\left(\dfrac{1}{8}+\alpha^2\right) \left(\frac{r_h}{\xi}\right)^2,   & 0 < r_h \lesssim \xi, \\[1em]
      -\left(\dfrac{1}{8}+\alpha^2\right) - \ln\left(\dfrac{r_h}{\xi}\right), & \xi \lesssim r_h \ll \Lambda,
    \end{array}
    \right.
\end{eqnarray}

\begin{eqnarray}\label{eq:FmBPin}
    \Delta F_{mB}=\frac{\Phi_0}{\mu_0 \Lambda}
    \left\{
    \begin{array}{cc}
      -\dfrac{1}{4}\alpha \left(\frac{r_h^4}{\xi^2}\right)B,   & 0 < r_h \lesssim \xi, \\[1em]
      -\left[\dfrac{1}{4}\alpha \xi^2+\frac{1}{2}\left(r_h^2-\xi^2\right)\right]B, & \xi \lesssim r_h \ll \Lambda
    \end{array}
    \right.
\end{eqnarray}
Here, $\alpha=0.4123$ denotes the near-core slope of the order parameter in the limit $\frac{\Lambda}{\xi}\to \infty$, and the vortex core radius is related to the coherence length through $r_c=0.966\,\xi$. 
For $r_h\lesssim \xi$, the pinning energy grows quadratically with hole size; while for $r_h\gg \xi$ the dependence becomes logarithmic.
We see that when $\xi\sim r_h$, the contribution of $\Delta F_{mB}$ is negligible. 
In both cases, the total Gibbs free energy reduction $G_{\text{pin}}=\Delta F_{\text{self}}-\Delta F_{m B}$ quantifies the strength of the pinning site.

Thus, the pinning energies used in our simulated annealing can be interpreted as the effective free-energy reductions associated with vortices trapped at circular holes of radius $r_h$.
This correspondence establishes a quantitative link between the abstract simulation parameters and experimentally realizable defect sizes in superconducting thin films.

%

\end{document}